\documentclass[lettersize,journal]{IEEEtran}
\usepackage{amsmath,amsfonts,amssymb}
\usepackage{algorithmic}
\usepackage{algorithm}
\usepackage{array}
\usepackage[caption=false,font=footnotesize,labelfont=sf,textfont=sf]{subfig}
\usepackage{textcomp}
\usepackage{stfloats}
\usepackage{url}
\usepackage{verbatim}
\usepackage{graphicx}
\usepackage{cite}
\usepackage{xcolor}
\usepackage{pgfplots}
\usepackage{hyperref}
\pgfplotsset{compat=1.18}
\usepackage[printonlyused]{acronym}

\begin{document}
\newacro{6g}[6G]{sixth-generation}
\newacro{5g}[5G]{fifth-generation}
\newacro{dof}[DoF]{degree of freedom}
\newacro{ris} [RIS] {reconfigurable intelligent surface}
\newacro{starris}[STAR-RIS]{simultaneously transmitting and reflecting RIS}
\newacro{bdris}[BD-RIS]{beyond diagonal RIS}
\newacro{mimo} [MIMO] {multiple-input multiple-output}
\newacro{ummimo} [UM-MIMO] {ultra-massive multiple-input multiple-output}
\newacro{irs} [IRS] {intelligent reflecting surface}
\newacro{elaa}[ELAA]{extremely large antenna arrays}
\newacro{xlmimo}[XL-MIMO]{extremely large-scale MIMO}
\newacro{thz}[THz]{terahertz}
\newacro{mmwave}[mmWave]{millimeter-wave}
\newacro{aosa}[AoSA]{array-of-sub-arrays}
\newacro{daosa}[DAoSA]{dynamic AoSA}
\newacro{gosa}[GoSA]{group-of-sub-arrays}
\newacro{hmimo}[H-MIMO]{holographic MIMO}
\newacro{dma}[DMA]{dynamic metasurface antennas}
\newacro{capmimo}[CAP-MIMO]{continuous aperture MIMO}
\newacro{fas}[FAS]{fluid antenna systems}
\newacro{sim}[SIM]{stacked intelligent metasurfaces}
\newacro{rhs}[RHS]{reconfigurable holographic surface}
\newacro{lis}[LIS]{large intelligent surfaces}
\newacro{fim}[FIM]{flexible intelligent metasurfaces}
\newacro{ma}[MA]{movable antenna}

\newacro{ae}[AE]{array element}
\newacro{pwm}[PWM]{planar wave model}
\newacro{swm}[SWM]{spherical wave model}
\newacro{hspwm}[HSPWM]{hybrid
spherical planar wave model}
\newacro{sa}[SA]{sub-array}
\newacro{sas}[SAs]{sub-arrays}
\newacro{arvs}[ARVs]{array response vectors}
\newacro{em}[EM]{electromagnetic}
\newacro{isac}[ISAC]{integrated sensing and communications}
\newacro{ai}[AI]{artificial intelligence}
\newacro{rf}[RF]{radio frequency}
\newacro{los}[LoS]{line-of-sight}
\newacro{nlos}[NLoS]{non-line-of-sight}
\newacro{csi}[CSI]{channel state information}
\newacro{cs}[CS]{compressed sensing}
\newacro{sca}[SCA]{successive convex approximation}
\newacro{omp}[OMP]{orthogonal matching pursuit}
\newacro{mc} [MC] {mutual coupling}
\newacro{ber} [BER] {bit error rate}

\newacro{upa} [UPA] {uniform planar array}
\newacro{ula} [ULA] {uniform linear  array}
\newacro{uca} [UCA] {uniform circular arrays}
\newacro{spd} [SPD] {spatially discrete}
\newacro{cap} [CAP] {continuous aperture}
\newacro{1d} [1D] {one-dimensional}
\newacro{2d} [2D] {two-dimensional}
\newacro{3d} [3D] {three-dimensional}

\newacro{rsls} [RS-LS] {reduced-subspace least square }
\newacro{ls} [LS] {least square }
\newacro{zf} [ZF] {zero-forcing }
\newacro{lmmse} [LMMSE] {linear minimum mean square error }
\newacro{mmse} [MMSE] {minimum mean square error }

\newacro{grand} [GRAND] {guessing random additive noise decoding }
\newacro{snr} [SNR] {signal-to-noise ratio }
\newacro{ttd} [TTD] {true-time-delay }

\newacro{pcb} [PCB] {printed circuit board }

\newacro{omp}[OMP]{orthogonal matching pursuit}
\newacro{sbl}[SBL]{sparse Bayesian learning}
\newacro{anm}[ANM]{atomic norm minimization}
\newacro{lista}[LISTA]{learned iterative shrinkage-thresholding algorithm}
\newacro{wmmse}[WMMSE]{weighted minimum mean square error}
\newacro{mrt}[MRT]{maximum ratio transmission}
\newacro{ml}[ML]{maximum likelihood}
\newacro{llr}[LLR]{log-likelihood ratio}
\newacro{ldpc}[LDPC]{low-density parity-check}
\newacro{scl}[SCL]{successive cancellation list}
\newacro{bch}[BCH]{Bose–Chaudhuri–Hocquenghem}
\newacro{bler}[BLER]{block error rate}
\newacro{nmse}[NMSE]{normalized mean square error}
\newacro{ofdm}[OFDM]{orthogonal frequency-division multiplexing}
\newacro{qpsk}[QPSK]{quadrature phase shift keying}
\newacro{sinr}[SINR]{signal-to-interference-plus-noise ratio}
\newacro{mse}[MSE]{mean square error}

\newacro{psi}[PSI]{pseudo-soft information}

\newacro{kaust}[KAUST]{King Abdullah University of Science and Technology}
\newacro{ucsd}[UCSD]{University of California San Diego}
\newacro{aub}[AUB]{American University of Beirut}

\newcommand{\DicRed}{\mathrm{DR}}
\newcommand{\TT}{\mathsf{T}}
\newcommand{\HH}{\mathsf{H}}
\newcommand{\nth}[1]{{#1}{\text{th}}}
\newcommand{\abs}[1]{\left|{#1}\right|}
\newcommand{\norm}[1]{\left\|{#1}\right\|}
\newcommand{\diagopr}[1]{\mathrm{diag}\left(#1\right)}
\newcommand{\invdiag}[1]{\mathrm{diag}^{-1}\left(#1\right)}
\newcommand{\vect}[1]{\mathrm{vec}\left(#1\right)}
\newcommand{\invvec}[1]{\mathrm{vec}^{-1}\left(#1\right)}
\newcommand{\mbf}[1]{\mathbf{#1}}

\newcommand{\red}[1]{{\color{red}{#1}}} 
\newcommand{\blue}[1]{{\color{blue}{#1}}}
\newcommand{\cyan}[1]{{\color{cyan}{#1}}}
\newcommand{\green}[1]{{\color{green}{#1}}} 
\newcommand{\yellow}[1]{{\color{yellow}{#1}}} 
\newcommand{\magenta}[1]{{\color{magenta}{#1}}}
\newcommand{\orange}[1]{{\color{orange}{#1}}}
\newcommand{\brown}[1]{{\color{brown}{#1}}} 
\newcommand{\simon}[1]{\small{\color{blue}{[Simon: #1]}}} 
\newcommand{\hadi}[1]{\small{\color{red}{[DrHadi: #1]}}}

\newacro{profTareq}[\cyan{Group1: Prof. Tareq}]{G1}
\newacro{profHakan}[\green{Group2: Prof. Hakan}]{G2}
\newacro{profRobert}[\yellow{Group3: Prof. Robert}]{G3}
\newacro{profHadi}[\magenta{Group4: Prof. Hadi}]{G4}

\newfont{\bb}{msbm10 scaled 1100}
\newcommand{\PP}{\mbox{\bb P}}
\newcommand{\EE}{\mbox{\bb E}}
\newcommand{\av}{{\bf a}}
\newcommand{\bv}{{\bf b}}
\newcommand{\cv}{{\bf c}}
\newcommand{\dv}{{\bf d}}
\newcommand{\ev}{{\bf e}}
\newcommand{\fv}{{\bf f}}
\newcommand{\gv}{{\bf g}}
\newcommand{\hv}{{\bf h}}
\newcommand{\iv}{{\bf i}}
\newcommand{\jv}{{\bf j}}
\newcommand{\kv}{{\bf k}}
\newcommand{\lv}{{\bf l}}
\newcommand{\mv}{{\bf m}}
\newcommand{\nv}{{\bf n}}
\newcommand{\ov}{{\bf o}}
\newcommand{\pv}{{\bf p}}
\newcommand{\qv}{{\bf q}}
\newcommand{\rv}{{\bf r}}
\newcommand{\sv}{{\bf s}}
\newcommand{\tv}{{\bf t}}
\newcommand{\uv}{{\bf u}}
\newcommand{\wv}{{\bf w}}
\newcommand{\xv}{{\bf x}}
\newcommand{\yv}{{\bf y}}
\newcommand{\zv}{{\bf z}}
\newcommand{\zerov}{{\bf 0}}
\newcommand{\onev}{{\bf 1}}
\newcommand{\avr}{\av_\text{R}}
\newcommand{\Am}{{\bf A}}
\newcommand{\Bm}{{\bf B}}
\newcommand{\Cm}{{\bf C}}
\newcommand{\Dm}{{\bf D}}
\newcommand{\Em}{{\bf E}}
\newcommand{\Fm}{{\bf F}}
\newcommand{\Gm}{{\bf G}}
\newcommand{\Hm}{{\bf H}}
\newcommand{\Id}{{\bf I}}
\newcommand{\Jm}{{\bf J}}
\newcommand{\Km}{{\bf K}}
\newcommand{\Lm}{{\bf L}}
\newcommand{\Mm}{{\bf M}}
\newcommand{\Nm}{{\bf N}}
\newcommand{\Om}{{\bf O}}
\newcommand{\Pm}{{\bf P}}
\newcommand{\Qm}{{\bf Q}}
\newcommand{\Rm}{{\bf R}}
\newcommand{\Sm}{{\bf S}}
\newcommand{\Tm}{{\bf T}}
\newcommand{\Um}{{\bf U}}
\newcommand{\Wm}{{\bf W}}
\newcommand{\Vm}{{\bf V}}
\newcommand{\Xm}{{\bf X}}
\newcommand{\Ym}{{\bf Y}}
\newcommand{\Zm}{{\bf Z}}
\newcommand{\Onem}{{\bf 1}}
\newcommand{\Zerom}{{\bf 0}}
\newcommand{\At}{{\rm A}}
\newcommand{\Bt}{{\rm B}}
\newcommand{\Ct}{{\rm C}}
\newcommand{\Dt}{{\rm D}}
\newcommand{\Et}{{\rm E}}
\newcommand{\Ft}{{\rm F}}
\newcommand{\Gt}{{\rm G}}
\newcommand{\Ht}{{\rm H}}
\newcommand{\It}{{\rm I}}
\newcommand{\Jt}{{\rm J}}
\newcommand{\Kt}{{\rm K}}
\newcommand{\Lt}{{\rm L}}
\newcommand{\Mt}{{\rm M}}
\newcommand{\Nt}{{\rm N}}
\newcommand{\Ot}{{\rm O}}
\newcommand{\Pt}{{\rm P}}
\newcommand{\Qt}{{\rm Q}}
\newcommand{\Rt}{{\rm T}}
\newcommand{\St}{{\rm S}}
\newcommand{\Tt}{{\rm T}}
\newcommand{\Ut}{{\rm U}}
\newcommand{\Vt}{{\rm V}}
\newcommand{\Wt}{{\rm W}}
\newcommand{\Xt}{{\rm X}}
\newcommand{\Yt}{{\rm Y}}
\newcommand{\Zt}{{\rm Z}}

\newcommand{\alphav}{\hbox{\boldmath$\alpha$}}
\newcommand{\betav}{\hbox{\boldmath$\beta$}}
\newcommand{\gammav}{\hbox{\boldmath$\gamma$}}
\newcommand{\deltav}{\hbox{\boldmath$\delta$}}
\newcommand{\etav}{\hbox{\boldmath$\eta$}}
\newcommand{\lambdav}{\hbox{\boldmath$\lambda$}}
\newcommand{\kappav}{\hbox{\boldmath$\kappa$}}
\newcommand{\epsilonv}{\hbox{\boldmath$\epsilon$}}
\newcommand{\nuv}{\hbox{\boldmath$\nu$}}
\newcommand{\muv}{\hbox{\boldmath$\mu$}}
\newcommand{\zetav}{\hbox{\boldmath$\zeta$}}
\newcommand{\phiv}{\hbox{\boldmath$\phi$}}
\newcommand{\varphiv}{\hbox{\boldmath$\varphi$}}
\newcommand{\psiv}{\hbox{\boldmath$\psi$}}
\newcommand{\thetav}{\hbox{$\boldsymbol\theta$}}
\newcommand{\varthetav}{\hbox{$\boldsymbol\vartheta$}}
\newcommand{\tauv}{\hbox{\boldmath$\tau$}}
\newcommand{\omegav}{\hbox{\boldmath$\omega$}}
\newcommand{\xiv}{\hbox{\boldmath$\xi$}}
\newcommand{\sigmav}{\hbox{\boldmath$\sigma$}}
\newcommand{\piv}{\hbox{\boldmath$\pi$}}
\newcommand{\rhov}{\hbox{\boldmath$\rho$}}

\newcommand{\Gammam}{\hbox{\boldmath$\Gamma$}}
\newcommand{\Lambdam}{\hbox{\boldmath$\Lambda$}}
\newcommand{\Deltam}{\hbox{\boldmath$\Delta$}}
\newcommand{\Sigmam}{\hbox{\boldmath$\Sigma$}}
\newcommand{\Phim}{\hbox{\boldmath$\Phi$}}
\newcommand{\Pim}{\hbox{\boldmath$\Pi$}}
\newcommand{\Psim}{\hbox{\boldmath$\Psi$}}
\newcommand{\psim}{\hbox{\boldmath$\psi$}}
\newcommand{\chim}{\hbox{\boldmath$\chi$}}
\newcommand{\omegam}{\hbox{\boldmath$\omega$}}
\newcommand{\Thetam}{\hbox{\boldmath$\Theta$}}
\newcommand{\Omegam}{\hbox{\boldmath$\Omega$}}
\newcommand{\Xim}{\hbox{\boldmath$\Xi$}}

\mathchardef\mhyphen="2D

\title{Physically Consistent Channel Modeling and Signal Processing for Reconfigurable Wireless Systems}

\author{Ahmad~Dkhan,~\IEEEmembership{Student Member,~IEEE,}
    Simon~Tarboush,~\IEEEmembership{Student Member,~IEEE,}
        Hadi Sarieddeen,~\IEEEmembership{Senior Member,~IEEE,}
        Robert W. Heath Jr.,~\IEEEmembership{Fellow,~IEEE,}
        Hakan Bagci,~\IEEEmembership{Senior Member,~IEEE,}
        and~Tareq~Y. Al-Naffouri,~\IEEEmembership{Fellow,~IEEE}
\thanks{Ahmad~Dkhan and Hadi~Sarieddeen are with the Department of Electrical and Computer Engineering, American University of Beirut (AUB), Beirut 1107 2020, Lebanon (amd53@mail.aub.edu, hadi.sarieddeen@aub.edu.lb). Simon Tarboush is with the Communications and Information Theory Group (CommIT), Technical University of Berlin, 10587 Berlin, Germany (simon.tarboush@tu-berlin.de). The contributions of S. Tarboush to this work were made during his research at KAUST. Robert W. Heath Jr. is with the Department of Electrical and Computer Engineering, University of California San Diego, La Jolla, CA 92093, USA (rwheathjr@ucsd.edu). Hakan Bagci and Tareq Y. Al-Naffouri are with the Department of Computer, Electrical and Mathematical Sciences and Engineering (CEMSE), King Abdullah University of Science and Technology (KAUST), Thuwal 23955-6900, Kingdom of Saudi Arabia (hakan.bagci@kaust.edu.sa, tareq.alnaffouri@kaust.edu.sa). This work was supported by the AUB University Research Board (URB) and Vertically Integrated Projects (VIP) program and the KAUST Office of Sponsored Research (OSR) under Award No. ORFS-CRG12-2024-6478.}
}

\maketitle

\begin{abstract}

Reconfigurable antennas are increasingly integrated into multi-antenna communication systems to exploit large apertures while reducing the hardware complexity, energy consumption, and implementation costs of classical massive arrays. Their reconfigurable electromagnetic (EM) properties, including dynamically varying radiation patterns and state-dependent mutual coupling, challenge the fixed-antenna and decoupled-port assumptions of conventional channel models. This motivates a physically consistent framework connecting Maxwell's equations, circuit theory, and information theory. In this tutorial, we develop a unified framework spanning three coupled dimensions: (i) reconfigurable antenna and transceiver architectures, (ii) physically consistent channel modeling, and (iii) physically consistent signal processing. We first establish a taxonomy covering tunable antennas, reconfigurable transceivers, and emerging array architectures, highlighting their reconfiguration mechanisms and hardware-performance trade-offs. We then develop modeling approaches based on Maxwell's equations, wavenumber-domain representations, multiport network theory, and computational electromagnetics, and use them to construct end-to-end channel and noise models that capture near-field propagation, mutual coupling, and circuit-level impairments. Building on these models, we examine architecture-aware channel estimation, beamforming, data detection, and channel decoding, emphasizing how physical structure reshapes algorithm design and performance-complexity trade-offs. Overall, the tutorial treats physical architecture, channel and noise models, and communication algorithms as coupled components of an end-to-end design, providing a unified foundation for physically consistent reconfigurable wireless systems.

\end{abstract}


\def\BState{\State\hskip-\ALG@thistlm}

\section{Introduction}

Future generations of wireless communications must significantly enhance data rates, reduce latency, and improve reliability in densely connected environments~\cite{Zhang20196G,Dang2020Should,Giordani2020Toward,You2021Towards}. They must also accommodate new use cases and functionalities, facilitating ubiquitous connectivity and accurate localization and sensing~\cite{Letaief2019Roadmap,Yang20196G,Akyildiz20206G}. These advances necessitate radical changes in physical-layer designs, where integrating physically consistent \ac{em} propagation laws into modeling and system design becomes crucial for understanding the limits and opportunities of future wireless networks~\cite{Bjornson2024Towards,Di2024Electromagnetic,Zhu2024Electromagnetic,Wang2024Electromagnetic}. 

The required shift in physical-layer design points toward three tightly coupled research directions: first, \textbf{designing} reconfigurable wireless architectures that introduce new degrees of freedom for software-controlled manipulation of \ac{em} wave propagation; second, \textbf{developing} physically consistent channel modeling frameworks that accurately capture \ac{em} behavior, near-field effects, and hardware-induced impairments; and third, \textbf{devising} signal processing techniques that explicitly account for both architectural constraints and propagation characteristics. The remainder of this tutorial develops each direction in turn.


Meeting the ambitious requirements of next-generation wireless systems demands a simultaneous expansion of both spectral and \acp{dof}. While higher-frequency bands, from the upper mid-band to sub-\ac{thz} and \ac{thz} frequencies~\cite{Bjornson2024Enabling,Rappaport2019Wireless,Sarieddeen2021Overview}, offer abundant bandwidth, their increased propagation losses make large-aperture arrays essential for achieving sufficient link budgets. Spatial \acp{dof} can thus be enhanced through large-aperture \ac{ummimo} systems, also referred to as \ac{elaa}, \ac{xlmimo}, or \ac{hmimo}~\cite{Hu2018Beyond,Prather2017Optically,Bjornson2019Massive,Huang2020Holographic,Lu2021Communicating,Li2023Modular,Deng2023Reconfigurable,Wang2024Tutorial,Gong2024Holographic,Bjornson2024Towards}. While \ac{ris} can shape the radio propagation environment in real time~\cite{Basar2019Wireless,Wu2019Towards,Wu2019Intelligent,Di2020Smart,Bjornson2022Reconfigurable,Pan2022An,Cheng2024Degree}, the focus of this tutorial is on active transceiver-side arrays rather than passive environmental reflectors. Achieving spatial gains at scale through such active arrays, however, significantly increases hardware cost, signal processing complexity, and energy consumption~\cite{You2025Next}, especially under conventional fully digital or fully connected hybrid beamforming architectures~\cite{Zhang2005Variable,Alkhateeb2014MIMO,Heath2016Overview}. This tension between desired performance and practical implementation motivates a fundamental rethinking of active array architectures, which constitutes the first research direction.

\textbf{1) Reconfigurable wireless architectures.} With the deployment of large \ac{mimo} systems and the increasing demands for energy efficiency, computational power, and low-cost hardware, novel \ac{mimo} array architectures are required~\cite{Sarieddeen2021Overview,Wang2024Tutorial,Liu2024Near,Liu2024Densifying}. As array size and bandwidth expand, conventional fully digital and fully connected hybrid beamforming structures become impractical due to the large number of \ac{rf} chains, phase shifters, and analog circuits~\cite{Alkhateeb2014MIMO,Heath2016Overview}. A promising alternative lies in reconfigurable array architectures, encompassing software-tunable antennas and surfaces, that enable \ac{em} wave manipulation with reduced hardware overhead~\cite{Cui2014Coding,Wang2020Metantenna,Katwe2024Overview,Cui2024Roadmap,Fu2025Fundamentals}. Recent research in this direction addresses two key aspects: first, acknowledging the unpredictable nature of the wireless environment, which calls for innovative reconfigurable arrays and surfaces to enhance propagation control~\cite{Basar2019Wireless,Di2020Smart,Wu2019Towards,Wu2019Intelligent}; second, deploying numerous antenna arrays with large physical or electrical apertures~\cite{Liu2024Densifying,Wei2024Electromagnetic,You2025Next,Gong2024Holographic}. These two aspects jointly drive the evolution from conventional fixed hardware toward programmable and scalable \ac{em} systems.

\textbf{2) Physically-consistent channel modeling.} Adapting large-aperture arrays at high frequencies demands fundamental changes to conventional channel models. As apertures grow and wavelengths shrink, classical far-field assumptions become invalid~\cite{stutz2026near}, and system behavior becomes inherently wave-based rather than angular. Modeling and design must align with physical laws governed by Maxwell's equations in addition to the theoretical bounds of Shannon's information theory. Consequently, integrating tools from \ac{em} theory, circuit theory, and information theory is essential~\cite{Di2024Electromagnetic,Zhu2024Electromagnetic,Bjornson2024Towards,Wang2024Electromagnetic,Mezghani2023Reincorporating,Ivrlavc2010Toward}.

Several modeling approaches have been developed, progressing from simple extensions of far-field models to full \ac{em} formulations. A first approach extends the conventional \ac{pwm} by incorporating the \ac{swm}, which accounts for both distance and angle properties between each transmit-receive antenna pair. The resulting near-field model captures non-uniform path gains and non-linear phase variations across antenna elements~\cite{Lu2021Communicating,Cui2022Near,Li2023Modular,Liu2024Near,Wang2024Tutorial,Cui2022Channel}, while power variations remain negligible beyond the Björnson distance~\cite{Bjornson2021Primer}. For near-field \ac{nlos} conditions, stochastic correlated Rayleigh fading better captures small-scale fading and spatial correlation across \ac{ummimo} dimensions~\cite{Dong2022Near,Demir2022Channel,Haghshenas2023New,Demir2024Spatial}.

For continuous-aperture systems, a more rigorous representation is derived directly from Maxwell's equations~\cite{Bjornson2024Towards,Di2024Electromagnetic,Zhu2024Electromagnetic,Wang2024Electromagnetic,Wei2024Electromagnetic}. In a linear time-invariant setting, Maxwell's equations link the electric current density at the transmitter to the resulting electric field at the receiver via the Green function operator. The \ac{em} channel can thus be interpreted as a continuous vector wave field excited by the transmitted current density~\cite{Wei2024Electromagnetic}. Modeling efforts based on this perspective include using the Dyadic Green's function for line-of-sight channels in both far- and near-field conditions~\cite{Gong2023Generalized,Gong2024HolographicJSAC,Gong2024Near,Wei2024Electromagnetic}, as well as Fourier plane-wave expansions for stochastic far-field non-line-of-sight channels~\cite{Pizzo2020Spatially,Pizzo2022Fourier,Pizzo2022Spatial}. Recent work has extended these methods to near-field non-line-of-sight stochastic channels using non-stationary Gaussian random fields~\cite{Wan2024Near} and stochastic Green's functions~\cite{Lin2023Predicting}. Green's function methods have also been applied to \ac{ris}-assisted channel modeling~\cite{Agrawal2024Towards,Danufane2021Path}.

With an infinite number of antennas in continuous-aperture \ac{hmimo}, the asymptotic limits of \ac{ummimo} can be explored by integrating \ac{em} and information theories, leading to analyses of \acp{dof}~\cite{Dardari2020Communicating,Pizzo2020Degrees,Pizzo2022Nyquist,Yuan2021Electromagnetic,Ruiz2023Degrees,Svarvare2024Spatial,Wang2024Analytical,Kosasih2024Roles}, mutual information, and channel capacity~\cite{Wan2023Mutual,Zhu2024Electromagnetic,Zhu2024MIMO}. However, solving Maxwell's equations for large-scale, strongly coupled antenna arrays is computationally intractable for iterative system design. Circuit theory offers a powerful and practical alternative. By modeling the transceiver and its antennas as a multi-port network, characterized by impedance, admittance, or scattering parameters~\cite{Ivrlavc2014Multiport,Pozar2011Microwave}, we retain essential physical consistency, accurately capturing mutual coupling, impedance mismatch, and noise correlation~\cite{Akrout2022Achievable,Shyianov2021Achievable}, while working with voltage and current pairs rather than continuous field distributions~\cite{Bjornson2024Towards,Akrout2023Super}. This approach inherently respects fundamental physical limits, such as Chu's antenna Q-factor trade-off~\cite{Chu1948,Bjornson2024Towards,Ivrlavc2010Toward} and the Bode-Fano bandwidth-impedance matching limit~\cite{Fano1950,Mezghani2023Reincorporating}. It is far more tractable than full-wave EM simulations, making it an ideal bridge between the complexity of Maxwell's equations and the abstract models of information theory~\cite{Ivrlavc2010Toward,Ivrlavc2014Multiport,Mezghani2023Reincorporating,Bjornson2024Towards}, while still enabling analytical models for both transceiver arrays and \ac{ris}-assisted designs~\cite{Sun2022Characteristics,Akrout2022Achievable,Akrout2023Super,Gradoni2021End,Abrardo2023Design,Di2023Modeling}. Consequently, circuit theory enables end-to-end system-level models that incorporate the physical limitations of the transceiver hardware, including the noise characteristics of RF chain components such as amplifiers, filters, and matching networks~\cite{Mezghani2023Reincorporating,Bjornson2024Towards,Akrout2022Achievable}.

\textbf{3) Tailored signal processing algorithms.} The shift toward reconfigurable hardware architectures and physically consistent channel models fundamentally reshapes signal processing. Efficient channel estimation, beamforming, and data detection must now account for near-field spherical wavefronts, mutual coupling, and circuit-level impairments. For instance, channel estimation algorithms originally designed for far-field planar wavefronts fail to capture the distance-dependent phase variations inherent in \ac{ummimo} systems, necessitating new parametric or compressed-sensing approaches. Beamforming designs must balance the enhanced spatial resolution offered by large apertures against the increased sensitivity to hardware imperfections and coupling. Data detection schemes must operate reliably under the non-stationary, correlated fading statistics predicted by \ac{em}-consistent models. The unifying principle is clear: signal processing for reconfigurable wireless systems must be co-designed with the underlying hardware architectures and propagation physics. This contrasts with traditional approaches that treat the channel as a given, abstracted quantity independent of the antenna and circuit implementation.

Despite significant progress, existing surveys and tutorials typically address reconfigurable elements/architectures, \ac{em}-based channel modeling, and signal processing methodologies in isolation~\cite{Bjornson2024Towards,Di2024Electromagnetic,Zhu2024Electromagnetic,Gong2024Holographic}. As a result, they do not provide a unified treatment that jointly connects physically consistent \ac{em} propagation models with circuit-theoretic impairments, end-to-end reconfigurable communication system design, and signal processing. More specifically, physics-driven studies of \ac{ummimo} systems emphasize \ac{em}-consistent modeling and near-field effects but do not fully incorporate emerging multi-layer architectures or tri-hybrid transceiver designs~\cite{Bjornson2024Towards}. \ac{em} information-theoretic frameworks provide fundamental insights into wave-domain communication limits, yet they often abstract away circuit-level constraints such as impedance mismatch, mutual coupling, and noise correlation, and do not systematically address architecture-aware signal processing~\cite{Di2024Electromagnetic,Zhu2024Electromagnetic}. Similarly, surveys on holographic and large-aperture \ac{mimo} systems focus on antenna surface design and wave manipulation, but do not unify these models with parasitic arrays, fluid antenna systems, or pinching antenna architectures, nor do they provide a consistent link to estimation and detection algorithms under hardware constraints~\cite{Gong2024Holographic}.

This tutorial establishes a unified framework that integrates \ac{em} theory, circuit theory, and information theory to enable end-to-end physically consistent modeling of reconfigurable wireless systems. This integration bridges propagation physics, hardware constraints, and signal processing design within a single coherent perspective. Table~\ref{tab:comprehensive_comparison} summarizes this distinction by comparing representative prior works with the proposed tutorial along key dimensions, including architectural coverage, physical modeling fidelity, and signal processing integration.

Building on the three core thrusts of reconfigurable architectures, physically consistent channel modeling, and architecture-aware signal processing, this tutorial makes the following key contributions:

\begin{enumerate}
    \item A \textbf{comprehensive and hierarchical taxonomy} of reconfigurable wireless systems, spanning reconfigurable antennas, transceiver architectures, and emerging antenna array paradigms. The tutorial highlights the underlying reconfiguration mechanisms and compares alternative architectures in terms of hardware complexity, power consumption, and beamforming flexibility.

    \item A \textbf{unified physically consistent modeling framework} grounded in Maxwell's equations, wavenumber-domain channel representations, multiport network theory, and computational \ac{em} methods. The framework captures near-field spherical-wave propagation, mutual coupling, and circuit-level effects, and develops \textit{end-to-end channel and noise models} tailored to emerging architectures.

    \item A \textbf{systematic treatment of architecture-aware signal processing} for reconfigurable wireless systems, covering channel estimation, beamforming, data detection, reliability extraction, and channel decoding under practical physical and hardware constraints, and highlighting how physically consistent modeling influences algorithm design and performance.
\end{enumerate}

\begin{figure}[!t] \centering \includegraphics[width=0.48\textwidth]{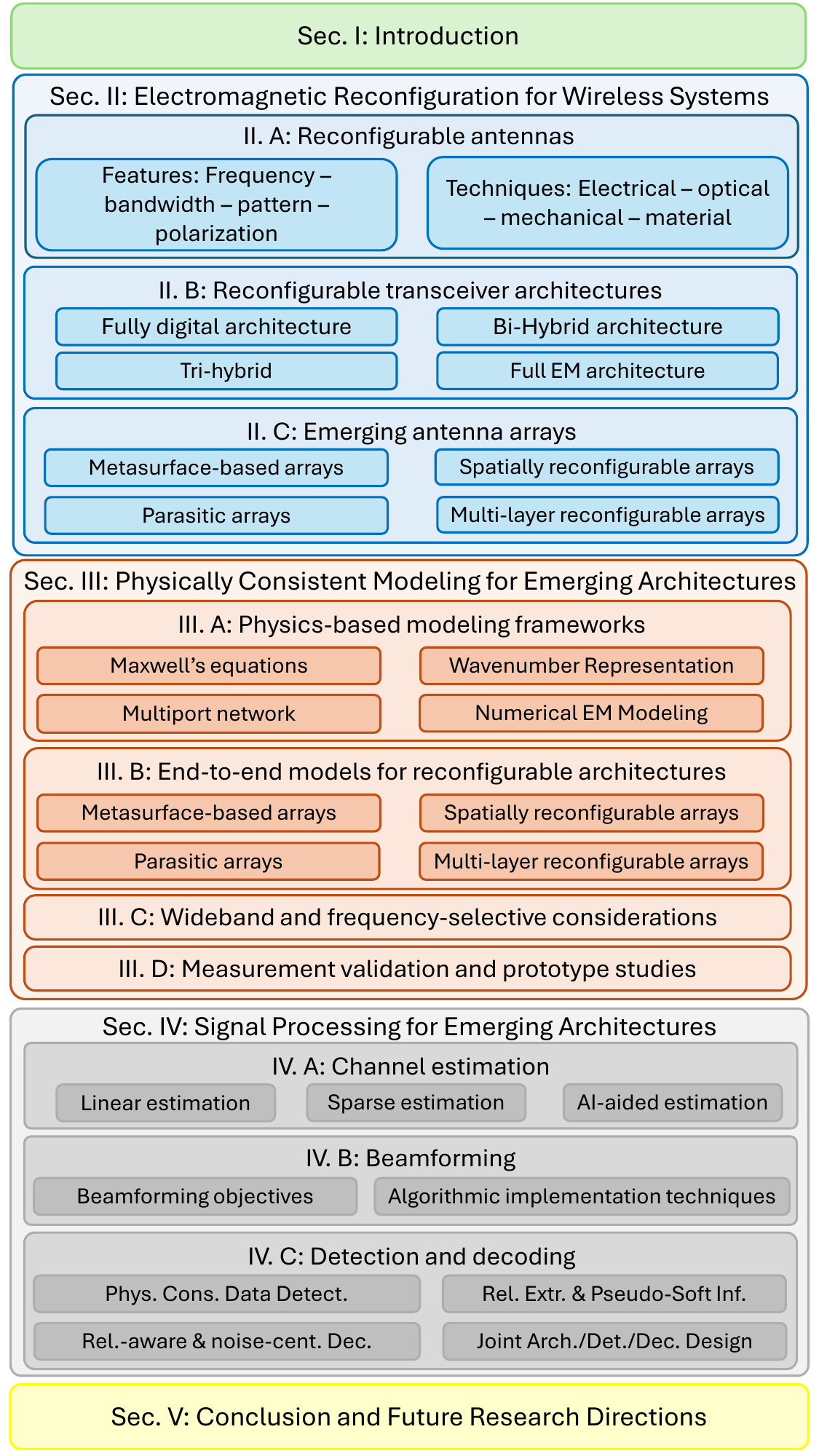} \caption{Tutorial organization. Sec.~II reviews \ac{em} reconfiguration for antennas and transceiver architectures. Sec.~III presents physically consistent channel modeling for emerging architectures. Sec.~IV covers signal processing techniques, including channel estimation, beamforming, and data detection. Sec.~V concludes and discusses future directions.} \label{fig:organization} \end{figure}

The overall scope and organization of the tutorial are illustrated in Fig.~\ref{fig:organization}. Sec.~\ref{Sec:II} reviews \ac{em} reconfiguration techniques for antennas and transceiver architectures. Sec.~\ref{Sec:III} presents the physically consistent modeling framework and develops end-to-end channel and noise models for emerging architectures. Sec.~\ref{Sec:IV} discusses architecture-aware signal processing, including channel estimation, beamforming, detection, reliability extraction, and decoding. Finally, Sec.~\ref{sec:conclusion} concludes the tutorial and outlines future research directions.

\begin{table*}

\centering
\caption{comparison of tutorials and surveys on next-generation wireless technologies, focusing on their technical coverage.}
\label{tab:comprehensive_comparison}
\begin{tabular}
{|p{1.3cm}|p{1.5cm}|p{1.5cm}|p{1.5cm}|p{1.5cm}|p{1.5cm}|p{1.5cm}|p{1.3cm}|p{1cm}|p{1cm}|}
\hline
\textbf{Ref.} & \textbf{Main focus} & \textbf{Array architecture} & \textbf{Channel modeling} & \textbf{Signal processing} & \textbf{Reconfig. tech.} & \textbf{\ac{mimo} tech.} & \textbf{Near field} & \textbf{Hardware array} & \textbf{\ac{em} theory} \\
\hline
\cite{Bjornson2024Towards} & \ac{ummimo}: physics-driven design & \checkmark (\ac{ummimo}) & \checkmark & \checkmark (channel est.) & $\odot$ (RIS mention) & \ac{ummimo} & \checkmark (radiative NF) & $\odot$ (array design) & \checkmark \\
\hline
\cite{Di2024Electromagnetic} & ESIT fundamentals & $\odot$ & \checkmark (NF LoS) & $\odot$  & $\odot$ (RIS)  & (\ac{mimo}, RIS) & \checkmark & $\times$ & \checkmark \\
\hline
\cite{Gong2024Holographic} & \ac{hmimo} & \checkmark  & \checkmark & \checkmark & $\odot$ & \ac{hmimo} & \checkmark & \checkmark & \checkmark  \\
\hline
\cite{Zhu2024Electromagnetic} & EIT fundamentals & $\times$  & \checkmark (\ac{em}) & $\times$ & $\times$ & General & $\odot$ & $\times$ & \checkmark   \\
\hline
\cite{Han2022Terahertz} & THz channels & $\times$ & \checkmark & $\times$ & $\times$ & $\odot$ (\ac{ummimo}) & $\odot$ & $\times$ & $\times$ \\
\hline
\cite{Wang2023Extremely} & \ac{xlmimo} & $\times$  & \checkmark & $\odot$ & $\times$ & \ac{xlmimo} & \checkmark & \checkmark  & $\odot$ \\
\hline
\cite{An2024Near} & Near-field comm. & \checkmark & \checkmark  & \checkmark  & $\odot$ & General & \checkmark & $\odot$ & $\times$ \\
\hline
\cite{Katwe2024Overview}  & Intelligent metaSurfaces & $\times$ & $\times$ & $\odot$ & \checkmark & RIS-aided \ac{mimo} & $\times$ & $\times$ & $\times$ \\
\hline

\cite{Wang2024Tutorial} & \ac{xlmimo} tutorial & \checkmark & \checkmark & \checkmark & $\times$ & \ac{xlmimo} & \checkmark & \checkmark & \checkmark   \\
\hline
\cite{Wang2024Electromagnetic} & EIT for 6G systems & $\times$ & $\checkmark$ & $\times$ & $\odot$ (RIS) & General & $\odot$ & $\times$ & \checkmark  \\
\hline
\cite{Wei2024Electromagnetic} & EIT for \ac{hmimo} & $\odot$  & \checkmark  & $\odot$ & $\odot$ (HoloS) & \ac{hmimo} & \checkmark & \checkmark  & \checkmark \\
\hline
\cite{You2025Next} & Next-gen transceivers & \checkmark  & $\odot$ & \checkmark & $\odot$ (RIS) & \ac{hmimo} & \checkmark & \checkmark & $\checkmark$ \\
\hline
\cite{Liu2025Near} & NFC survey & \checkmark & \checkmark & $\odot$ & $\odot$ & General & \checkmark & \checkmark & $\odot$  \\
\hline
\cite{Bidabadi2025Physically} & \ac{em}-aware RIS Opt. & $\odot$ (RIS) & \checkmark (multiport, MC,) & $\odot$ (optim.) & \checkmark (RIS, BD-RIS) & $\odot$ (RIS-aided) & $\odot$ & $\odot$ & \checkmark \\
\hline
\cite{Liu2024Near} & NFC Survey & \checkmark & \checkmark & \checkmark &  \checkmark & General & \checkmark & \checkmark & $\odot$ \\
\hline
Our tutorial & Reconfig. archit., phys. consistent model., signal process. & \checkmark  & \checkmark  & \checkmark  & \checkmark  & Reconfig. \ac{mimo} & \checkmark & \checkmark  & \checkmark  \\
\hline
\end{tabular}
\\[2mm]
\footnotesize
Legend: 
\checkmark : Primary focus; $\odot$ : Mentioned or secondary topic; $\times$ : Not covered

\end{table*}

Throughout this tutorial, scalars $(a, A)$, vectors $(\mathbf{a})$, and matrices $(\mathbf{A})$ are represented by non-bold, bold lowercase, and bold uppercase letters, respectively. $\mathbf{I}_M$ denotes an $M \times M$ identity matrix. The operators $(\cdot)^{\TT}$, $(\cdot)^{\HH}$, and $(\cdot)^{-1}$ stand for the transpose, Hermitian (conjugate transpose), and inverse operators, respectively. The Euclidean norm of a vector is denoted by $\|\cdot\|$, and the Frobenius norm of a matrix by $\|\cdot\|_{\mathrm{F}}$. $\mathbb{E}[\cdot]$ denotes the expectation operator. $\operatorname{diag}(a_1, a_2, \dots, a_N)$ represents an $N \times N$ diagonal matrix with diagonal entries $\{a_1, a_2, \dots, a_N\}$. The sets of real and complex numbers are denoted by $\mathbb{R}$ and $\mathbb{C}$, respectively.

\section{Electromagnetic reconfiguration for wireless systems}
\label{Sec:II}

This section covers reconfigurable antennas and arrays at three levels. Sec.~\ref{sec:reconfigurable_antennas} classifies antennas by tunable properties and mechanisms. Sec.~\ref{sec:reconfigurable_transceiver_architectures} scales to transceiver architectures, from digital to full-EM. Sec.~\ref{sec:emerging_arrays} examines metasurface-based, spatially reconfigurable, parasitic, and multi-layer arrays. We highlight trade-offs among hardware complexity, power consumption, and beamforming flexibility.

\subsection{Reconfigurable antennas}
\label{sec:reconfigurable_antennas}
\subsubsection{Features}

Based on the operational properties dynamically adjusted, e.g., frequency of operation, radiation pattern, polarization, or a combination of any of these properties, reconfigurable antennas can be classified as follows~\cite{Castellanos2025Embracing}: 

\paragraph{Frequency reconfigurable antennas} 
Frequency reconfigurable antennas can dynamically adjust their operating frequency, making them suitable for multi-band and multi-service wireless systems~\cite{Patriotis2021Millimeter,Tang2021Frequency}. Reconfiguration is typically achieved by tuning the antenna’s electrical properties or modifying the matching network, allowing operation over a range of frequencies or switching between discrete bands. 

\paragraph{Bandwidth reconfigurable antennas} 
Bandwidth reconfigurable antennas adjust their operational bandwidth to cover both narrow-band and wideband requirements~\cite{Alqurashi2019Liquid,Tran2023Flexible}. This is usually accomplished via reconfigurable feed-line networks or modifications to the antenna structure, enabling impedance bandwidths ranging from 22\% to 78\%. 
These antennas can switch between multi-resonance and wideband/ultra-wideband (UWB) modes, allowing coexistence with systems such as fourth-generation (4G) cellular networks, wireless local area networks (WLANs), worldwide interoperability for microwave access (WiMAX) systems, and satellite communications, thereby increasing flexibility for multi-mode wireless applications.

\paragraph{Pattern reconfigurable antennas} 
Pattern reconfigurable antennas can intentionally modify their radiation characteristics, particularly the spherical distribution of the radiation pattern, making them essential for applications requiring dynamic beam steering, in communications, radar, and \ac{em} imaging~\cite{Cano2022Pattern}. These antennas achieve reconfigurability through two primary mechanisms: movable or rotatable structures, such as metasurfaces, which can physically reorient to adjust the radiation pattern, allowing dynamic modification without relying on electronic components; and switchable reactive-loaded capacitive elements, which electronically control the current distribution on the antenna surface, enabling rapid and precise adjustments of radiation characteristics~\cite{Piazza2010Experimental}.

The discrete pattern approach provides a practical implementation of pattern reconfigurable antennas, typically using positive-intrinsic-negative (PIN) diodes or electronic switches to toggle between a few distinct radiation patterns~\cite{Stroh2021Pattern}. A control signal adjusts the switch state, though some insertion loss occurs along the switched path. Designers usually design these antennas so that each discrete pattern points in a different direction, enabling partial beam steering with a single element. More recent designs employ fluid-based reconfigurable structures, such as radiation pattern reconfigurable fluid antenna systems (RPR-FAS)~\cite{Li2025Radiation}, which provide adaptive beam shaping and improved interference resilience.

\paragraph{Polarization reconfigurable antennas} 
Polarization-reconfigurable antennas can effectively mitigate multipath fading by dynamically switching between different polarization states~\cite{Wu2017Wideband}. Most studies in this area focus on switching between right-hand circular polarization (RHCP) and left-hand circular polarization (LHCP) at a desired frequency~\cite{Lin2018Reconfigurable}, while linear polarizations such as vertical and horizontal modes are also explored~\cite{Wang2016Reconfigurable}. 

Structural modifications, such as slots, slits, parasitic elements, or truncated corners in the main radiator combined with active components, commonly achieve polarization reconfigurability~\cite{Cai2016Compact}. Alternatively, active metasurfaces and adaptive feeding networks have been adopted to enable polarization switching~\cite{Zhu2014Design,Qin2025High}. Other approaches include the use of reconfigurable external polarizers or phase shifters~\cite{Row2014Design}. 

\paragraph{Multi-feature reconfigurable antennas} 
Compound reconfigurable antennas are capable of independently tuning two or more characteristics, such as frequency, bandwidth, radiation pattern, and polarization~\cite{Suryapaga2024Review}. Reported designs include frequency–bandwidth switching using reconfigurable band-pass filters or parasitic structures~\cite{Ojaroudi2019Recent,Kingsly2018Multiband}, frequency–pattern combinations that integrate active elements with switchable parasitics for beam steering~\cite{Nguyen2017Dual,Zainarry2018Frequency}, and frequency–polarization designs that exploit an active \ac{em} band gap (EBG) or metasurface structures for polarization diversity~\cite{Nguyen2015Frequency}. Radiation pattern and polarization diversity has also been achieved using parasitic elements and reconfigurable feeding networks~\cite{Narbudowicz2014Omnidirectional}. Researchers have investigated fully compound designs that simultaneously reconfigure frequency, radiation pattern, and polarization, with pixel-surface techniques providing an effective implementation approach~\cite{Rodrigo2014Frequency}.

The reconfigurable features discussed above can be realized through different physical implementation mechanisms. The choice of mechanism determines the achievable tuning speed, insertion loss, power consumption, fabrication complexity, and reliability. The following subsections review the principal implementation approaches (techniques).

\subsubsection{Techniques}

Antenna reconfiguration can be achieved using various mechanisms that alter the radiation characteristics, frequency response, or impedance of the antenna. The main mechanisms include electrical, optical, mechanical, and material-based approaches~\cite{Ojaroudi2020Reconfigurable,Ojaroudi2019Recent}. The choice of mechanism affects both the reconfiguration speed and the complexity of the antenna design.

\paragraph{Electrical control}
Electrical reconfiguration relies on integrating switching components into the antenna structure to alter surface current distributions or radiating edges~\cite{Ojaroudi2019Recent,Piazza2010Experimental}. Common components include PIN diodes, which enable fast on/off switching, providing nanosecond-scale reconfiguration and high dynamic tuning capability. Similarly, varactor diodes allow continuous tuning of antenna reactance via bias voltage, achieving smooth frequency or phase changes~\cite{Gutierrez2007Design}. Another widely used option is radio frequency micro-electro-mechanical systems (RF-MEMS) switches, which provide low-loss, high-linearity switching using mechanical movement, with switching speeds ranging from 1--200~$\mu$s, making them suitable for applications with moderate speed requirements. In addition, field-effect transistors (FETs) are also employed for continuous or discrete tuning.

\paragraph{Optical control}
Optical reconfiguration employs photoconductive switches or semiconductor materials activated by laser light, offering high isolation due to the absence of a biasing network and nanosecond-scale switching speeds, while also enabling easy integration with optical backhauls~\cite{Patron2014Optical}. Common materials include doped silicon, gallium arsenide, and organic semiconductors such as poly(3-hexylthiophene) (P3HT)~\cite{Patriotis2021Millimeter}. For instance, photoconductive silicon switches can shift a patch antenna's resonance from 18--19 GHz to 12 GHz upon laser activation, while P3HT-based designs have demonstrated beam steering and frequency sweeping under variable optical intensity~\cite{Patriotis2021Millimeter,Alqurashi2019Liquid}. Despite these benefits, practical deployment faces challenges such as the need for precise optical alignment, sufficient laser power, and increased system complexity, which largely restrict optical reconfiguration to infrastructure nodes or specialized scenarios where fiber-based fronthauls already exist.

\paragraph{Mechanical adjustment} 
Mechanical reconfiguration modifies the antenna geometry or the position of radiating elements using actuators, providing an exceptionally wide tuning range while keeping the RF signal path free of semiconductor-induced distortion and power limitations~\cite{Costantine2014Reconfigurable}. Modern implementations leverage MEMS actuators, electro-active polymers, shape-memory alloys, and origami-inspired folding mechanisms to achieve continuous frequency or pattern tuning~\cite{Costantine2014Reconfigurable,Piazza2010Experimental}. For example, MEMS-based electrostatic comb actuators enable wide frequency tuning with superior isolation and linearity compared to PIN diodes~\cite{Piazza2010Experimental}. However, the actuation speed remains orders of magnitude slower than electrical or optical methods (milliseconds to seconds), and moving parts introduce reliability concerns, increased form factor, and sensitivity to environmental factors~\cite{Costantine2014Reconfigurable}. Consequently, mechanical approaches are best suited for quasi-static applications such as satellite ground stations or one-time base-station adjustments, rather than real-time beamforming for high-mobility scenarios.

\paragraph{Material property tuning}  
\label{Material_property}

Metamaterials enable control over \ac{em} properties by engineering their subwavelength unit cells and, for reconfigurable designs, by exploiting tunable intrinsic material properties. By tailoring the geometry, orientation, and arrangement of these meta-atoms, designers can tune the effective permittivity ($\varepsilon_{\text{eff}}$) and permeability ($\mu_{\text{eff}}$), achieving values not found in natural materials~\cite{Noginov2011Tutorials}. This tuning allows for dynamic, reconfigurable manipulation of wave behavior, particularly at \ac{rf} and microwave frequencies, and in more limited forms at \ac{thz} and optical bands. Depending on their \ac{em} response, operating frequency, and implementation method, metamaterials can be grouped into several categories:

\begin{itemize}
\item Linear and nonlinear metamaterials:  
Linear metamaterials are characterized by field-independent constitutive parameters $\varepsilon$ and $\mu$. In nonlinear metamaterials, these parameters depend on the field amplitudes, for example, $\varepsilon = \varepsilon(|\mathbf{E}|)$ or $\mu = \mu(|\mathbf{H}|)$, which enables tunable responses and harmonic generation.  

Nonlinear behavior can be achieved using inclusions such as varactor-loaded split-ring resonators or Kerr-type dielectrics. Liquid crystals are another example of tunable media, whose effective permittivity can be reconfigured by an applied bias voltage to reorient their molecules~\cite{Liu2008Liquid}.  

Ferrite materials also offer reconfigurability, as their permeability changes under an applied static magnetic field. Additional tunable options include barium-strontium-titanate (BST), which enables electric-field tuning of permittivity, while ferrites such as yttrium iron garnet (YIG) enable magnetic-field tuning of permeability and gyrotropic response.~\cite{Zhu2014Design}.

\item Microwave metamaterials:
At microwave frequencies, metamaterials are commonly realized using metallic inclusions such as split-ring resonators (SRRs), complementary SRRs, and wire arrays. Dielectric resonator-based metamaterials have also been demonstrated at microwave frequencies using high-permittivity ceramic resonators. Metallic implementations remain the most widely studied due to their ease of fabrication using printed-circuit techniques and their ability to demonstrate negative-index propagation, compact filters, and impedance-matching surfaces~\cite{Hussain2023Metamaterials}. Their relatively low loss and geometric scalability make them ideal testbeds for metamaterial verification and antenna integration.

\item Dielectric metamaterials:
Employ high-permittivity inclusions to support electric and magnetic Mie-type resonances, avoiding the ohmic losses associated with metallic inclusions~\cite{Cheben2023Recent}.
Their all-dielectric nature enables high efficiency and, depending on the design, potentially wider bandwidth than plasmonic counterparts, while remaining compatible with photonic platforms.
Representative implementations include dielectric resonator arrays, photonic-crystal slabs, and gradient-index lenses.

\item Gain-assisted and active metamaterials:
Loss compensation is critical across the \ac{em} spectrum, particularly in plasmonic and metallic metamaterials where ohmic losses degrade performance~\cite{Pacheco2021Temporal,Guan2026High}. Incorporating optical gain within the metamaterial host can counteract metallic absorption. 
Gain-assisted metamaterials use doped polymers, semiconductor quantum dots, or laser dyes embedded in the matrix to partially compensate absorption losses and improve transmission, with amplification demonstrated in selected experimental configurations. 
These active systems enable tunable metasurfaces and reconfigurable negative-index layers.

\item Anisotropic and hyperbolic metamaterials:
Strongly anisotropic metamaterials possess tensorial permittivity or permeability where at least one principal component differs in sign, producing hyperbolic dispersion~\cite{Lee2022Hyperbolic,Shekhar2014Hyperbolic}. 
Such hyperbolic metamaterials (HMMs) support propagating modes with large wave vectors (high-$k$), a large photonic density of states, and sub-diffraction imaging.
Practical realizations include metal-dielectric multilayers and nanowire composites.

\item Bianisotropic and chiral metamaterials:
Bianisotropic metamaterials exhibit magnetoelectric coupling between electric and magnetic fields. Let $\mathbf{D}$ and $\mathbf{B}$ denote the electric displacement and magnetic flux density, respectively, and let $\boldsymbol{\xi}$ and $\boldsymbol{\zeta}$ be the coupling tensors that mix electric and magnetic effects. The constitutive relations are then
\begin{equation}
\mathbf{D} = \boldsymbol{\varepsilon}\mathbf{E} + \boldsymbol{\xi}\mathbf{H}, \quad
\mathbf{B} = \boldsymbol{\zeta}\mathbf{E} + \boldsymbol{\mu}\mathbf{H}.
\end{equation}

Chiral metamaterials, a subset of bianisotropic media, employ geometrically handed inclusions to control polarization and optical activity. These materials produce circular dichroism, polarization rotation, and polarization-dependent phase control.

\item Metafluidic metamaterials:
Metafluidic metamaterial is a metamaterial the optical response of which is dependent on fluid contributed metamolecules. The dependence originates either from a fluid background coupling to the metamolecule or from the resonance in a liquid structured metamolecule. Different liquid materials including water, liquid crystal, and liquid metals are applied to realize the metafluidic metamaterial. Sophisticated technologies like electric bias and microfluidic system have been used for active control of metafluidic metamaterials which provide a new platform for \ac{em} wave manipulation and metadevice realization~\cite{Holloway2012Overview,Zhu2013Frequency,Memon2018Microfluidic,Wu2022Active,Prakash2025Microfluidic,Gonzalez2019Microfluidic}. 

\end{itemize}

When the unit cell dimensions are not sufficiently smaller than the operating wavelength, the local medium approximation fails, and the response becomes wave-vector dependent. 
This spatial dispersion requires nonlocal constitutive modeling that captures the dependence of $\varepsilon_{\text{eff}}$ and $\mu_{\text{eff}}$ on the wave vector $\mathbf{k}$. 
Accurate retrieval of material parameters under such conditions remains a major research challenge.

Fabrication methods vary by spectral regime~\cite{Ji2023Recent,Feng2025Fabrication}. 
Microwave metamaterials are typically manufactured using printed-circuit or waveguide processes, whereas optical metamaterials employ electron-beam lithography, focused-ion-beam milling, or nanoimprint lithography. 
In all regimes, losses, dispersion, and limited bandwidth constrain performance. 
Hybrid metal-dielectric composites and two-dimensional analogues, known as metasurfaces~\cite{Ataloglou2023Metasurfaces}, are promising paths toward scalable and tunable metamaterial devices. Placing a metasurface atop a patch antenna and rotating it modifies the equivalent relative permittivity, thereby shifting the resonant frequency~\cite{Zhu2014Design}.

\subsection{Reconfigurable transceiver architectures}
\label{sec:reconfigurable_transceiver_architectures}

The integration of reconfigurable components into transceiver architectures represents a significant departure from conventional wireless system design, enabling novel hardware and arrays that dynamically adapt to varying traffic demands, interference conditions, and propagation environments, thereby supporting highly efficient, resilient, and adaptable communications. In this section, we present a combination of traditional and emerging architectures that leverage tunable antennas and metasurfaces~\cite{Liu2025Reconfigurable,Li2025Tri} (as summarized in Table~\ref{tab:TXRX_architectures}), extending capabilities well beyond conventional digital and hybrid designs.

\begin{figure*}[t]
 \centering
 \subfloat[Bi-hybrid digital-analog.]{\label{fig:D_A} \includegraphics[width=0.48\linewidth]{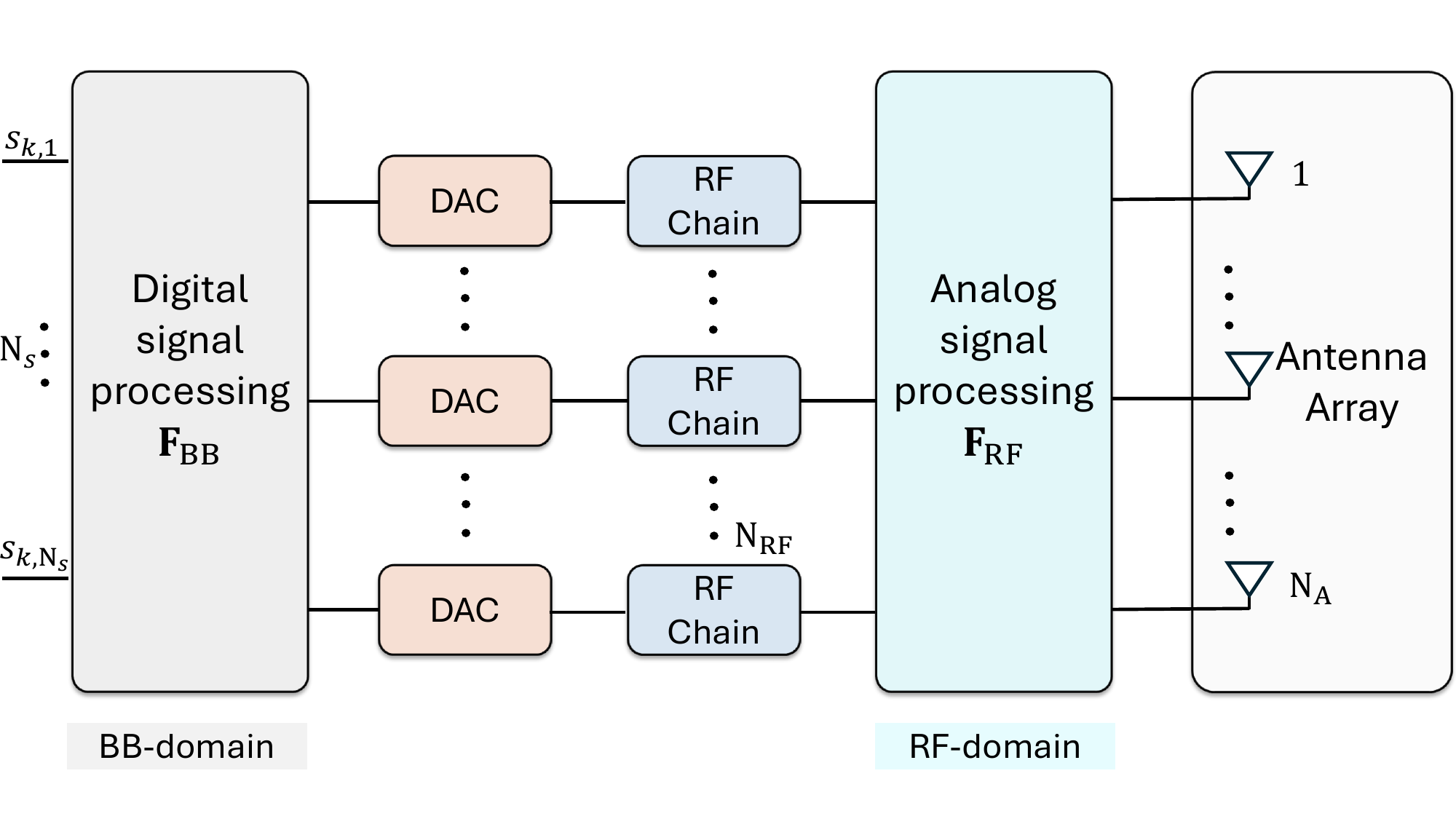}}%
  \hfill
 \subfloat[Bi-hybrid digital-\ac{em}.]{\label{fig:D_EM} \includegraphics[width=0.48\linewidth]{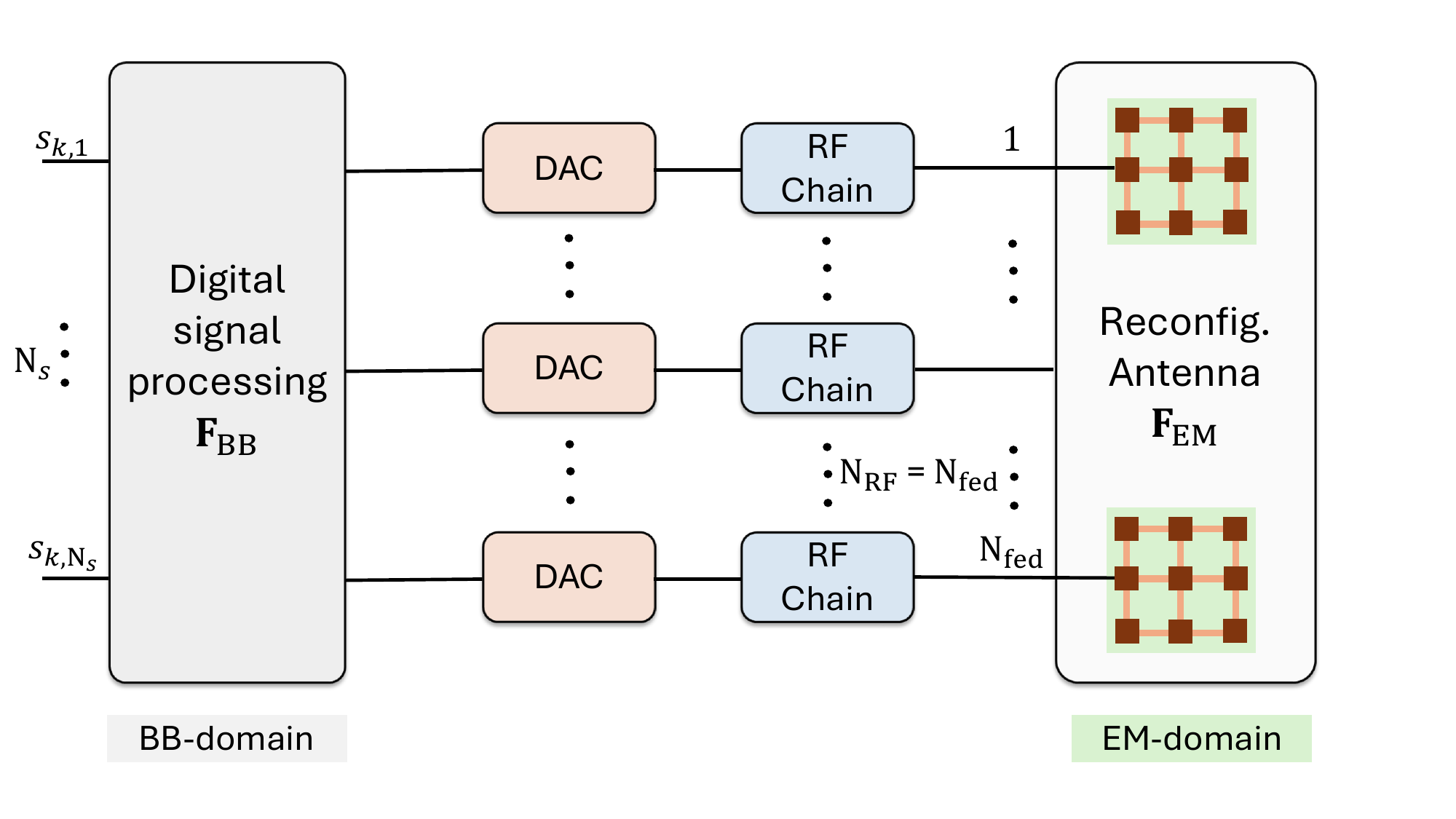}}%
  \hfill
 \subfloat[Tri-hybrid digital-analog-\ac{em}.]{\label{fig:D_A_EM} \includegraphics[width=0.48\linewidth]{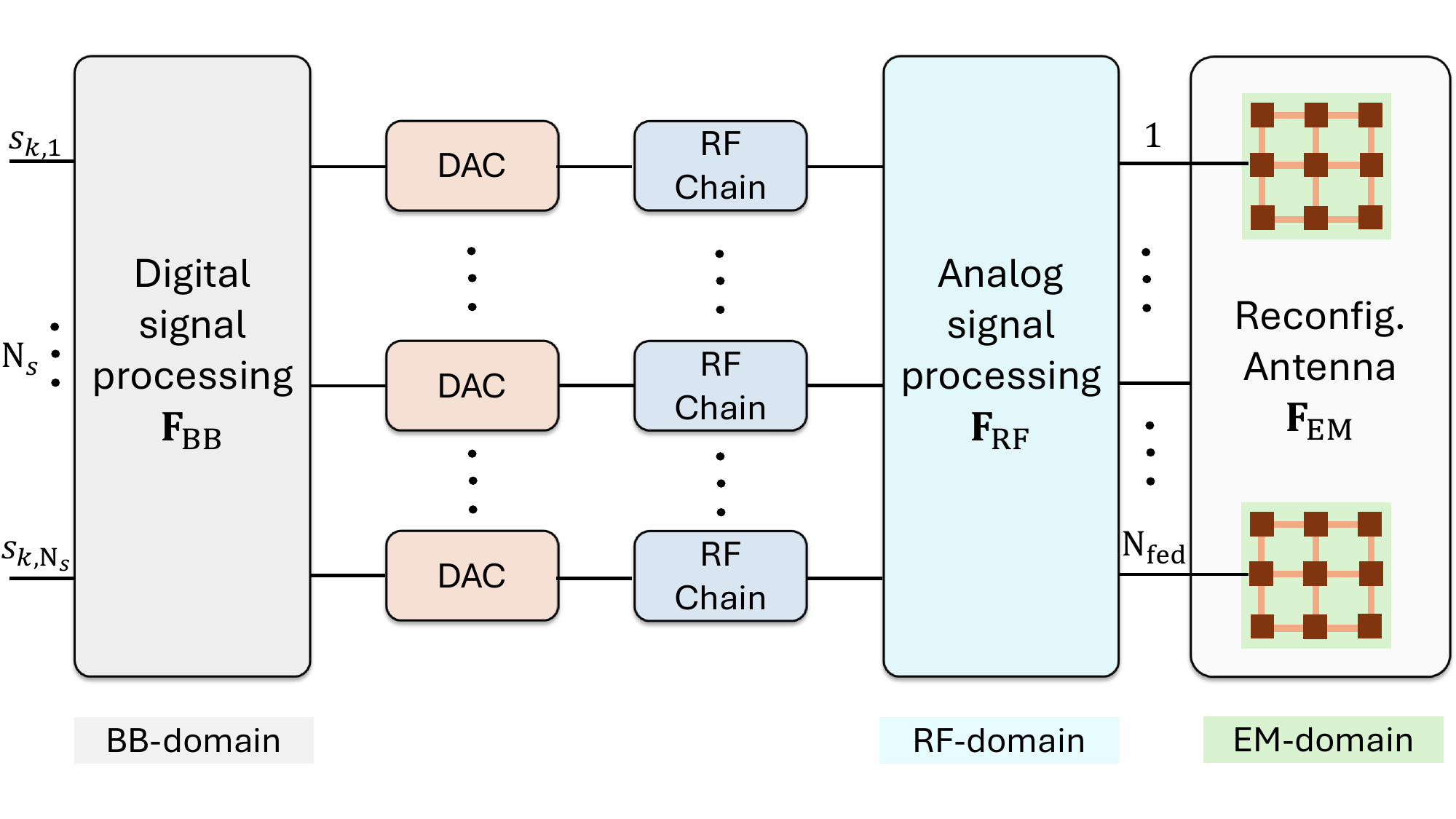}}%
  \vspace{0.2cm}
\caption{Illustration of hybrid transceiver architectures: (a) bi-hybrid digital-analog architecture, where analog phase shifters perform \ac{rf}-domain beamforming; (b) bi-hybrid digital-\ac{em} architecture, where the analog precoder is replaced by a reconfigurable antenna or metasurface operating in the \ac{em} domain; and (c) tri-hybrid digital-analog-\ac{em} architecture, combining digital processing, analog \ac{rf} beamforming, and \ac{em}-domain wave manipulation.}
 \label{fig:hybrid_arch}
\end{figure*}

To establish a unified mathematical foundation for comparing the following architectures, we consider a wideband orthogonal frequency-division multiplexing (OFDM) system with \(K\) subcarriers. Let $N_{\mathrm{in}}$ denote the number of effective input ports to the antenna system (e.g., RF chains, driven elements, or guided feeds). The transmitted symbol vector is $\mathbf{s}_k \in \mathbb{C}^{N_S}$, and $\mathbf{F}_k \in \mathbb{C}^{N_{\mathrm{in}} \times N_S}$ denotes the baseband precoder (which may be digital, analog, or hybrid) applied after the \ac{em} layer.

The reconfigurable \ac{em} layer is characterized by a configuration-dependent matrix $\mathbf{F}_{\mathrm{EM},k}$, which maps the $N_{\mathrm{in}}$ input ports to the radiating aperture. Consequently, the resulting end-to-end channel shapes the overall propagation based on the EM domain configurations. We denote this channel as $\mathbf{H}_{\mathrm{eff},k}(\mathbf{F}_{\mathrm{EM},k}) \in \mathbb{C}^{U \times N_{\mathrm{in}}}$.

The received signal vector at the $U$ users is expressed as
\begin{equation}
\mathbf{y}_k = \mathbf{H}_{\mathrm{eff},k}(\mathbf{F}_{\mathrm{EM},k}) \, \mathbf{F}_k \, \mathbf{s}_k + \mathbf{n}_k,
\label{eq:unified_system_model}
\end{equation}
where $\mathbf{n}_k$ is the additive noise.

This factorization separates the physics-based EM processing from the conventional signal processing, unifying all reconfigurable architectures.

\begin{table*}[t]
    \centering
    \caption{Comparison of transceiver \ac{mimo} Array Architectures \footnotesize{In wideband operation, \(\mathbf{F}_{\mathrm{BB}}\) and \(\mathbf{F}_{\mathrm{EM}}\) are generally frequency-selective (per subcarrier), while \(\mathbf{F}_{\mathrm{RF}}\) is frequency-flat (shared across subcarriers) unless TTD units are used. Here, \(N_{\text{fed}}\) denotes the number of guided feeds for the EM layer.}}
    \begin{tabular}{|p{4.2cm}|p{2.2cm}|p{2.2cm}|p{2.2cm}|p{2.2cm}|p{2.6cm}|}
    \hline
    \textbf{Feature / architecture} & \textbf{Fully digital} & \textbf{Hybrid (digital–analog)} & \textbf{Hybrid (digital–EM)} & \textbf{Tri-hybrid (digital–analog-EM)} & \textbf{Full EM } \\
    \hline
    \textbf{RF chain to antenna ratio / data streams} 
    & \boldmath$N_S \leq N_{\text{RF}} = N_\mathrm{A}$  
    & \boldmath$N_S \leq N_{\text{RF}} \ll N_\mathrm{A}$  
    & \boldmath$N_S \leq N_{\text{RF}} = N_{\text{fed}} \ll N_\mathrm{A}$  
    & \boldmath$N_S \leq N_{\text{RF}} \leq N_{\text{ana}} \leq N_\mathrm{A}$  
    & \boldmath$N_S = N_{\text{RF}} = N_\text{fed} \ll N_\mathrm{A}$  
     \\
    \hline
    \textbf{Beamforming domain} & Digital baseband & Digital + analog & Digital + EM & Digital + analog + EM & EM domain only \\
    \hline
    \textbf{Hardware complexity} & Very high & Moderate & Low–moderate & Moderate & Low (minimal baseband DSP; no digital precoder) \\
    \hline
    \textbf{Power consumption} & Very high & Moderate & Low & Low & low \\
    \hline
    \textbf{Beamforming flexibility} & Maximum & Moderate–high & High & Very high & Very high \\
    \hline
    \textbf{Scalability to large arrays} & Limited & Moderate & High & Very high & Very high \\
    \hline
    \textbf{End-to-end latency} & High (DSP-limited) & Moderate (hybrid delays) & Low (passive switching) & Low & Ultra-low (near propagation limit) \\
    \hline
    \end{tabular}
    \label{tab:TXRX_architectures}
\end{table*}

\subsubsection{Fully digital architecture}

\begin{figure*}[t]
 \centering
 \subfloat[Fully Digital.]{\label{fig:Full_D} \includegraphics[width=0.45\linewidth]{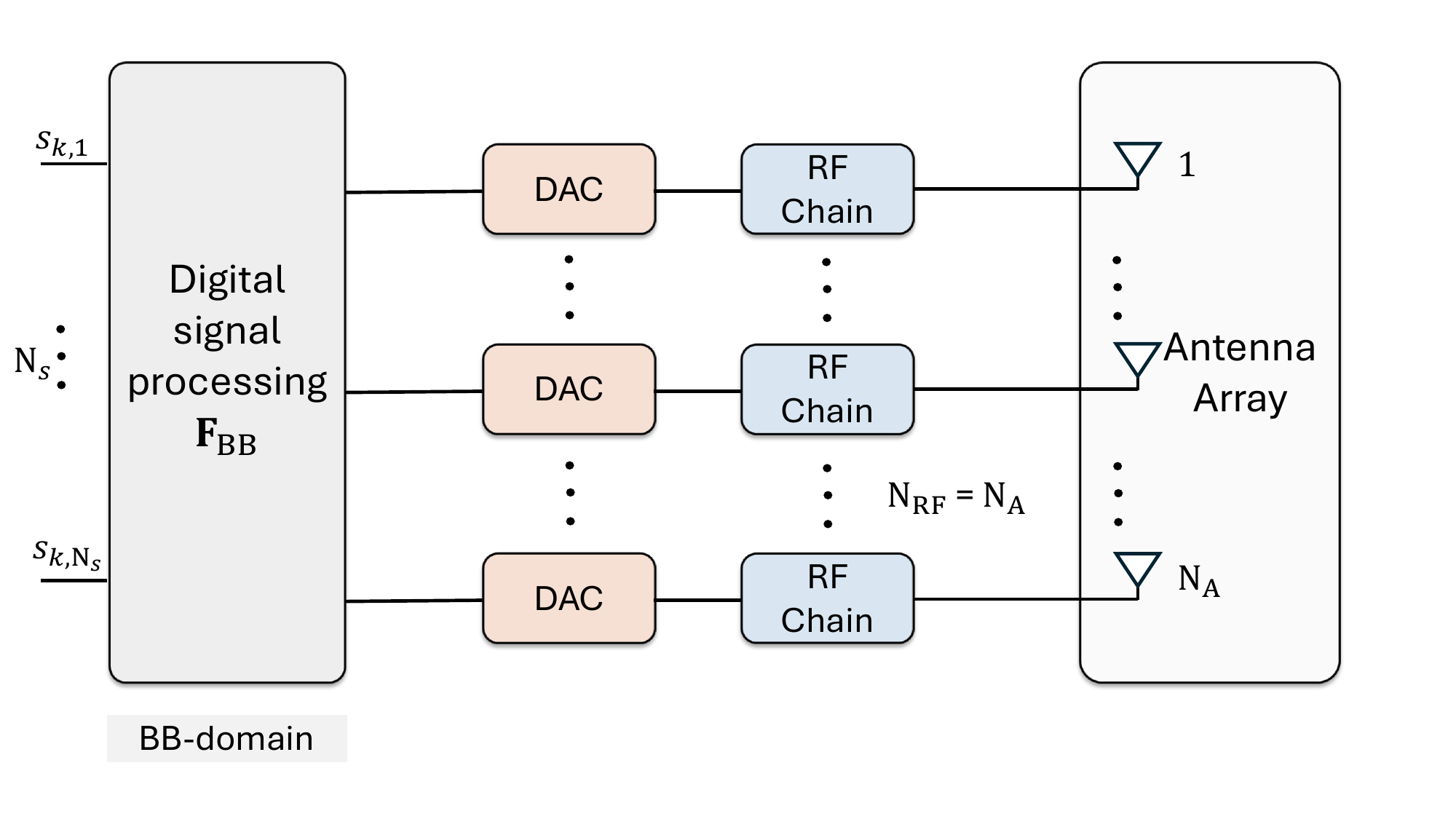}}%
 \subfloat[Fully EM.]{\label{fig:Full_EM} \includegraphics[width=0.45\linewidth]{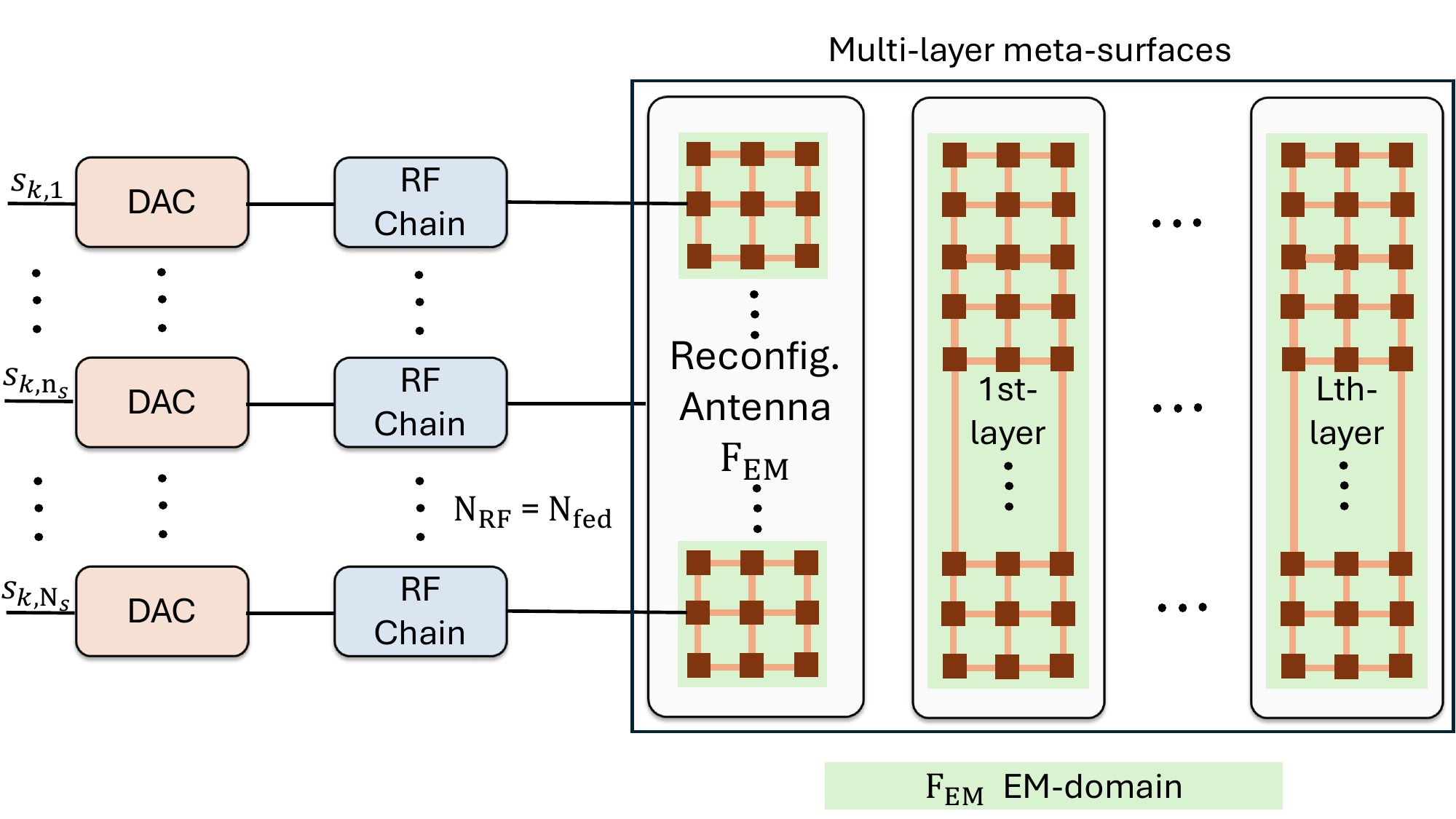}}%
\caption{Illustration of full architectures: (a) fully digital architecture with $N_{\text{RF}} = N_{\mathrm{A}}$ RF chains, and (b) fully \ac{em} architecture where precoding is performed in the \ac{em} domain using programmable metasurfaces.}
\label{fig:Full_arch}
\end{figure*}

In a fully digital transceiver architecture, as shown in Fig.~\ref{fig:Full_D}, each antenna element is connected to a dedicated \ac{rf} chain and high-speed analog-to-digital/digital-to-analog converter (ADC/DAC), enabling per-element baseband processing. The overall precoder at subcarrier \(k\) is simply the digital baseband precoder, i.e., \(\mathbf{F}_k = \mathbf{F}_{\mathrm{BB}, k}\), with \(\mathbf{F}_{\mathrm{BB}, k} \in \mathbb{C}^{N_{\mathrm{RF}} \times N_S}\), where \(N_{\mathrm{RF}} = N_{\mathrm{A}}\). Since each subcarrier can have its own precoding matrix, this architecture provides maximal frequency-selective beamforming flexibility. It allows for the generation of arbitrary beam patterns in both near-field and far-field regimes, offering high spatial resolution and adaptability~\cite{Heath2016Overview}. The benefits of fully digital processing, however, come at the cost of increased hardware complexity and power consumption, with cost scaling linearly with the number of antennas, i.e., \(N_{\text{RF}} = N_\mathrm{A}\)~\cite{Gao2016Energy}.

\subsubsection{Bi-hybrid architecture}

We use the term Bi-hybrid architecture to refer to transceiver designs that integrate two distinct layers of signal processing, typically combining digital baseband processing with analog or \ac{em} processing. This category includes two architectures

\begin{itemize}
    \item {Hybrid digital–analog architecture:} This architecture refers to traditional hybrid designs that aim to balance the high performance of a fully digital structure with the reduced hardware complexity of analog solutions, as illustrated in Fig.~\ref{fig:D_A}. The hybrid digital–analog structure employs a two-stage precoder. The overall precoder at subcarrier \(k\) is factorized as \(\mathbf{F}_k = \mathbf{F}_{\mathrm{RF}} \mathbf{F}_{\mathrm{BB}, k}\), where \(\mathbf{F}_{\mathrm{RF}} \in \mathbb{C}^{N_\mathrm{A} \times N_{\mathrm{RF}}}\) denotes the frequency-flat analog precoder implemented with phase shifters and is shared across all subcarriers, and \(\mathbf{F}_{\mathrm{BB}, k} \in \mathbb{C}^{N_{\mathrm{RF}} \times N_S}\) is the frequency-selective baseband digital precoder. This sharing of \(\mathbf{F}_{\mathrm{RF}}\) across subcarriers is the root cause of beam-squint in wideband operation. Two primary configurations exist for the analog precoder: the fully-connected and sub-connected architectures.

    In the fully-connected structure, each \ac{rf} chain is connected to all antenna elements via a network of phase shifters, offering high beamforming flexibility and spatial resolution. This comes, however, at the cost of increased hardware complexity and power consumption due to the large number of required phase shifters and combiners~\cite{Heath2016Overview}.
    
    Conversely, the sub-connected (or partially-connected) architecture reduces hardware complexity by connecting each \ac{rf} chain to a distinct subset of antenna elements, thereby forming multiple subarrays. This configuration significantly lowers the number of required analog components, such as phase shifters and combiners, resulting in reduced power consumption and implementation cost. The reduced connectivity, however, limits beamforming flexibility and spatial resolution~\cite{Molisch2017Hybrid}.
    
    A notable implementation of the sub-connected structure is the \ac{aosa} architecture, which has gained increasing attention in high-frequencies and large-scale systems~\cite{Dkhan2026Hierarchical,Tarboush2021Teramimo,Sarieddeen2021Overview}. Common variants of \ac{aosa} include:
    
    \begin{itemize}
        \item Dynamic AoSA (DAoSA): allows dynamic reconfiguration of \ac{rf}-chain–to-subarray mappings to adapt to changing channel conditions~\cite{Yan2020Dynamic}.
        \item Group-of-subarrays (GoSA): clusters subarrays into groups to balance performance and scalability~\cite{Park2017Dynamic}.
        \item Widely-spaced AoSA: increases the spatial aperture by spacing subarrays further apart, enhancing angular resolution~\cite{Akyildiz2022Terahertz}.
        \item Virtual AoSA: leverages channel sparsity and switching mechanisms to emulate a larger array with fewer physical elements~\cite{Elbir2021Terahertz}.
    \end{itemize}

   To mitigate beam-squint in wideband operation, these architectures often employ \ac{ttd} elements, which provide frequency-invariant beam steering by implementing actual time delays rather than phase shifts~\cite{Myers2021Infocus,Deshpande2022Nonuniform}.
   In bandwidth-constrained or cost-sensitive designs, low-resolution phase shifters are alternatively used as a power-efficient solution, though they do not address beam-squint and are better suited for narrowband operation~\cite{Han2021Hybrid}. A broader discussion of wideband effects, including frequency‑dependent mutual coupling and amplitude‑phase‑frequency coupling, is provided in Section~\ref{subsec:wideband}.

   \item Hybrid digital–EM architecture: Recognizing the limitations of conventional analog precoding, especially in large arrays, including increased hardware complexity, elevated power consumption, limited angular resolution, and reduced adaptability to time-varying channel conditions, recent research has explored replacing traditional \ac{rf} modules with reconfigurable technologies that offer enhanced spatial control, scalability, and energy efficiency. In this paradigm, the analog precoder \(\mathbf{F}_{\mathrm{RF}}\) is replaced by a programmable \ac{em} structure that directly shapes the radiated field through spatially varying impedance, amplitude, or phase profiles. The EM-domain processing \(\mathbf{F}_{\mathrm{EM},k}\) directly affects the effective channel, and optimizing its reconfigurable parameters enables adaptive control of the end-to-end propagation response. The remaining (digital) precoder is \(\mathbf{F}_k = \mathbf{F}_{\mathrm{BB},k} \in \mathbb{C}^{N_{\mathrm{fed}} \times N_S}\). Since the response of metasurface elements is inherently frequency-dependent, \(\mathbf{F}_{\mathrm{EM},k}\) must be optimized separately for each subcarrier. Here, \(N_{\mathrm{fed}}\) denotes the number of guided feeds (e.g., microstrip or waveguide ports)~\cite{Di2020Hybrid,Huang2023Integrating,Jamali2020Intelligent,Mishra2023Transmitter,Shlezinger2021Dynamic,Zhang2022Beam}, as illustrated in Fig.~\ref{fig:D_EM}.
    
\end{itemize}

\subsubsection{Tri-hybrid / digital–analog–EM architecture}
The tri-hybrid architecture extends conventional hybrid (digital-analog) precoding by incorporating an additional \ac{em} domain enabled by reconfigurable antennas, thereby jointly exploiting digital baseband processing, analog \ac{rf} precoding, and \ac{em}-domain wavefront control, as illustrated in Fig.~\ref{fig:D_A_EM}. In this framework, transmit precoding is realized through a cascade of processing stages across the digital, \ac{rf}, and \ac{em} domains~\cite{Heath2025Tri}. The EM-domain processing \(\mathbf{F}_{\mathrm{EM},k}\) provides an additional layer of adaptability by enabling wavefront manipulation directly at the physical layer, complementing the conventional analog and digital precoding stages. The remaining precoder is the cascade of analog and digital stages: \(\mathbf{F}_k = \mathbf{F}_{\mathrm{RF}} \mathbf{F}_{\mathrm{BB},k}\), where \(\mathbf{F}_{\mathrm{RF}}\) is the frequency-flat analog precoder (typically implemented with phase shifters and shared across all subcarriers) and \(\mathbf{F}_{\mathrm{BB},k}\) is the frequency-selective digital baseband precoder. If \ac{ttd} elements are employed, \(\mathbf{F}_{\mathrm{RF}}\) becomes frequency-dependent and is denoted by \(\mathbf{F}_{\mathrm{RF},k}\). This cascade satisfies the dimensionality constraint \(N_{\mathrm{RF}} \leq N_{\mathrm{ana}} \leq N_{\mathrm{A}}\), which reflects the progressively increasing degrees of freedom from the digital domain to the \ac{em} domain.

Recent studies~\cite{Castellanos2025Embracing,Castellanos2023Energy} demonstrate that the tri-hybrid architecture can improve spectral and energy efficiency while reducing hardware complexity and power consumption relative to fully digital and conventional hybrid (digital-analog) \ac{mimo} systems. At the same time, the introduction of the \ac{em} domain poses new challenges related to \ac{em} precoder design, real-time control, and system-level modeling. In particular, efficient operation requires coordinated optimization across the digital, analog, and \ac{em} domains, as well as accurate characterization of the \ac{em} channel response induced by reconfigurable antenna structures.

\subsubsection{Full electromagnetic architecture}

Full-\ac{em} architectures perform precoding, combining, and signal processing directly in the EM domain, without relying on per-symbol digital computation. In this paradigm, all processing is realized through the EM-domain transformation \(\mathbf{F}_{\mathrm{EM},k}\), with minimal baseband processing, yielding ultra-low latency at the expense of reduced per-symbol flexibility. The effective end-to-end channel is given by \(\mathbf{H}_{\mathrm{eff},k}(\mathbf{F}_{\mathrm{EM},k})\).

One realization of this paradigm is based on cascaded intelligent layers that manipulate \ac{em} waves to perform signal processing. By configuring multiple layers of programmable metasurfaces, which can act like a deep neural network, \ac{em} waves can be processed and transformed at the speed of light, replacing or augmenting complex digital processing in communication systems. This approach enables faster processing, reduces hardware complexity and energy consumption, and provides more control over wave propagation. An example of this architecture is \ac{sim}~\cite{An2024Stacked,An2023Stacked,Liu2025Stacked}.

Another approach to reducing digital processing complexity is the use of reconfigurable microwave networks that operate as analog computers, exemplified by the microwave linear analog computer (MiLAC)~\cite{Nerini2025Analogcomputing,Nerini2025Analog}. MiLAC is a programmable multiport network with tunable admittance components, capable of realizing linear operators, such as \ac{zf} and \ac{mmse} precoders and combiners, directly in the analog domain using circuit-theoretic principles, thereby significantly reducing hardware requirements. Unlike conventional digital matrix inversion, which incurs $O(N^3)$ arithmetic complexity, the number of tunable admittance components in a fully connected MiLAC topology scales as $O(N^2)$~\cite{Nerini2025Analogcomputing}. While distinct from the wave-domain processing of \ac{sim}, MiLAC offers a complementary analog-computing paradigm; the key differences between the two architectures are summarized in Table~\ref{table:MILAC_SIM}.

\begin{table*}[t]
    \caption{Comparison Between MiLAC and \ac{sim} Architectures}
    \label{table:MILAC_SIM}
    \centering
    \renewcommand{\arraystretch}{1.2}
    \begin{tabular}{|p{3.8cm}|p{6.2cm}|p{6.2cm}|}
    \hline
    \textbf{Aspect} & \textbf{MiLAC} & \textbf{\ac{sim}} \\ \hline
    \textbf{Nature} & Reconfigurable microwave network that encodes input vectors as port voltages, processes them via tunable admittance components, and reads outputs as voltages & Passive wave manipulator that shapes \ac{em} waves via stacked metasurfaces \\
    \hline
    \textbf{Key functionality} & Hardware-level matrix operations (e.g., inversion, multiplication) replace digital baseband computation & Meta-atoms apply phase shifts; wave superposition physically steers energy toward users \\
    \hline
    \textbf{Computational capability} & Supports arbitrary matrix operations (e.g., ZF, MMSE, DFT) with digital-like flexibility & Limited to transformations achievable via \ac{em} wave propagation. Suitable for fixed beamforming (e.g., DFT) \\
    \hline
    \textbf{Architecture} & Tunable admittance components (resistors, capacitors) configured as a reconfigurable microwave network & Stacked layers of programmable meta-atoms (phase shifters) \\
    \hline
    \textbf{Processing domain} & Microwave analog signal domain; matrix solutions via physical laws & \ac{em} wave domain; transformations encoded in layer-wise phase profiles \\
    \hline
    \textbf{Complexity vs. flexibility} & Higher complexity due to fine-grained analog tuning; excels in dynamic tasks & Lower complexity and power; limited by phase-shift constraints; best for static or repetitive transformations \\
    \hline
    \textbf{Energy efficiency} & High (avoids digital computations) & Very high (fully passive operation) \\
    \hline
    \textbf{Processing speed} & Near-instantaneous (speed of analog signal propagation) & Near-instantaneous (speed of light through passive structure) \\
    \hline
    \textbf{RF chains required} & Minimal (equal to number of data streams) & Minimal (equal to number of data streams) \\
    \hline
    \textbf{ADC/DAC resolution} & Low-resolution (symbols processed in analog) & Low-resolution \\
    \hline
    \textbf{ Use cases} & Real-time beamforming in massive \ac{mimo}, high-dimensional precoding (e.g., ZF, MMSE), analog computing for matrix operations & Low-complexity DFT beamforming, and \ac{isac} \\ \hline
    \end{tabular}
\end{table*}

\subsection{Emerging antenna arrays}
\label{sec:emerging_arrays}

In this section, we review key reconfigurable antenna architectures and their underlying mechanisms and implementation strategies. A comparative overview of representative reconfigurable antenna technologies is provided in Table~\ref{tab:Em_Rec_Arr}. These architectures span a range of technologies, from established designs such as parasitic arrays and fluid antennas to more recent developments like dynamic metasurface antennas and stacked intelligent metasurfaces.

\subsubsection{ Metasurface-based arrays}
\label{subsubsec:metasurface_antennas}
Among the most promising reconfigurable platforms are metasurface-based antennas, which leverage engineered subwavelength structures to enable fine-grained control of \ac{em} waves.

\paragraph{Dynamic metasurface antennas}
\ac{dma} are a class of reconfigurable antennas that use a single waveguide feed and electronically tunable elements, offering a compact, energy-efficient alternative to traditional phased arrays~\cite{Shlezinger2021Dynamic}. The antenna consists of a waveguide fed at one end and a grid of tunable radiating elements (“meta-atoms”) on top. Each meta-atom includes a low-power electronic component, typically a varactor diode, whose capacitance can be adjusted via an external bias voltage, altering the resonant frequency of the slots~\cite{Gong2024Holographic,Bjornson2024Towards}. By controlling the phase and amplitude of each meta-atom, the overall radiation pattern can be dynamically shaped, enabling electronic beam steering and low-cost, power-efficient analog beamforming~\cite{Jabbar202460,Shlezinger2019Dynamic}. Compared to conventional phased arrays, DMAs replace numerous analog circuits and phase shifters with microstrip lines embedded with tunable metamaterial cells, reducing complexity, cost, and power consumption~\cite{Huang2020Holographic,Gong2024Holographic,Bjornson2024Towards}.

Closely related to \ac{dma}, but differing in beam synthesis strategy, are \acp{rhs}. Both use guided waves coupled to tunable metasurfaces, yet \ac{dma} rely on element-wise electronic control of radiation, while \acp{rhs} employ holographic principles to convert a surface wave into a directive beam through coordinated leakage.

\paragraph{Reconfigurable holographic surface}

\acp{rhs} are a compact, ultra-thin, and energy-efficient metasurface-based antenna technology that leverages the holographic principle to generate directional beams without relying on conventional phase shifters~\cite{Deng2022Reconfigurable,Di2025Reconfigurable}. Unlike \acp{dma}, which use electronically tunable meta-atoms to control radiation from a guided-wave structure, RHSs typically employ amplitude-modulation techniques based on the interference pattern between a reference wave and a desired object wave. This holographic approach enables beamforming with simplified hardware, as each metamaterial element adjusts its radiation amplitude (and possibly phase) so that the superposition of all radiated waves forms the desired far-field beam~\cite{Gong2024Holographic,Zhang2025Holographic}.

The \ac{rhs} architecture comprises three main components: (i) feeds, which convert \ac{rf} chain signals into reference waves; (ii) a waveguide for surface propagation; and (iii) a planar array of metamaterial elements, whose local \ac{em} responses shape the radiation pattern. The \ac{rf} front-end is fully integrated into the \ac{pcb}, eliminating bulky components and enabling seamless integration with transceivers.

\acp{rhs} achieve high gain and low sidelobe levels (e.g., below -\(8\) dB) with reduced hardware complexity~\cite{Deng2022Reconfigurable}. Unlike \acp{ris}, which passively reflect incident waves and require external \ac{rf} feeds, \acp{rhs} actively generate and steer beams without relying on incoming signals. Consequently, \acp{rhs} are better suited for transceiver applications, whereas \acp{ris} typically function as passive relays. These key differences are summarized in Table~\ref{table:RHS_RIS}, which distinguishes \ac{rhs} from \ac{ris} architectures.

Recent multi-user beam training techniques for \acp{rhs} enable joint angle and range estimation in both near- and far-field regimes, reducing training overhead~\cite{Zhang2025Holographic}. Hierarchical 3D beam training approaches have been proposed to handle dense \ac{rhs} element arrays, hardware constraints, and mutual coupling effects, demonstrating robust performance across near-field and far-field deployments~\cite{Dong2025Near}. 

\begin{table*}[t]
\caption{Key Differences Between RHS and \ac{ris} Architectures}
\label{table:RHS_RIS}
\centering
\renewcommand{\arraystretch}{1.2}
\begin{tabular}{|p{3cm}|p{7cm}|p{6cm}|}
\hline
\textbf{Feature} & \textbf{RHS} & \textbf{RIS} \\
\hline
\textbf{Functionality} & Active antenna that generates and steers beams & Passive surface that reflects incident waves \\
\hline
\textbf{Beamforming method} & Holographic beamforming via leaky-wave radiation & Reflection-based phase shift control \\
\hline
\textbf{RF front-end} & Integrated into the \ac{pcb}, no external control required & Requires external \ac{rf} chains and control link \\
\hline
\textbf{Wave source} & Generates its own reference wave & Relies on external incident wave \\
\hline
\textbf{Typical role} & Used as transmit/receive antenna & Used as passive relay or reflector \\
\hline
\end{tabular}
\end{table*}

While electronically tunable metasurfaces offer fast reconfiguration with solid-state components, an alternative paradigm seeks to achieve reconfigurability through physical deformation of the antenna structure itself. These spatially or structurally reconfigurable antennas adapt their geometry, rather than just their electrical loading, to enable large-scale changes in radiation behavior, often with minimal \ac{rf} hardware.

\paragraph{Lens antennas}
Lens antennas represent another class of passive beamforming structures that leverage \ac{em} focusing to achieve high gain and directional transmission. By shaping the wavefront through a dielectric or metamaterial lens, these antennas can focus energy into narrow beams, effectively implementing a spatial Fourier transform that maps incident angles to specific antenna elements~\cite{Ouyang2022Lens,Saravanakumar2025High}. In MIMO systems, this enables beamspace processing, where the spatial channel is transformed into a sparse angular domain, allowing for the selection of only the most dominant beams and significantly reducing the number of required RF chains while maintaining high spectral efficiency~\cite{Ouyang2022Lens,Saravanakumar2025High}. Recent advances have also demonstrated reconfigurable lens antennas capable of dynamic beam steering and polarization control, further expanding their applicability in next-generation wireless systems~\cite{Yin2025Wideband}.

\subsubsection{Spatially reconfigurable arrays}

Mechanically and fluidically reconfigurable antenna arrays adapt their physical structure to achieve dynamic beam steering, spatial diversity, and enhanced reconfigurability in complex environments. These systems modify the geometry of radiating elements rather than relying solely on electronic control, enabling flexible, hardware-level adaptation\cite{Zhu2026Movable}.

\paragraph{Fluid antenna systems}
The \ac{fas}~\cite{Wong2020Fluid,Wong2023FluidI,Wong2023FluidIII,New2024Tutorial} represent a key approach, encompassing fluid-driven, deformable, and liquid-metal-based implementations~\cite{Castellanos2025Embracing}. These arrays reconfigure the aperture by adjusting the position, shape, orientation, polarization, or dimensions of the antenna elements, providing flexible control of the radiation pattern and improving interference mitigation~\cite{Khammassi2023New,Zhang2024Successive,Chen2024Joint,Lu2024Group}. Specifically, fluid-based designs manipulate radiation characteristics using liquid conductors within microfluidic channels or dielectric substrates, whereas deformable structures employ flexible materials, such as shape-memory alloys, to physically reshape the antenna surface~\cite{Wong2020Fluid,Wong2023FluidI}.

Some \ac{fas} designs support sub-wavelength positioning precision and allow continuous or quasi-continuous movement across ports, which is particularly advantageous in near-field and high-frequency applications. Nevertheless, challenges remain in fabrication complexity, mechanical durability, and long-term reliability, particularly for liquid-metal-based implementations. Recent work~\cite{Zhang2025Fluid} proposes RHS-enabled \acp{fas}, where \acp{rhs} adjust antenna positions by activating different element subsets, which also reprograms the effective array aperture. Fluid beam training employs sliding windows (i.e., different active element sets per codeword) to improve channel quality and enhance received signal strength. To avoid \ac{csi} acquisition, it adopts a hierarchical structure with narrowing beamwidth across layers.

\paragraph{Movable antenna systems}

Closely related to fluid antennas are \acp{ma}, which achieve spatial reconfiguration through mechanical actuation rather than fluidic mechanisms~\cite{Zhu2023Movable,Ning2024Movable,New2024Tutorial}. Unlike \ac{fas}, where the antenna element is typically a liquid conductor that can move between discrete ports, \acp{ma} are generally defined as physical antennas connected to the \ac{rf} chain through a flexible cable, with mobility enabled by a positioning mechanism or driver such as a stepper motor~\cite{Zhu2023Movable,Ning2024Movable,New2024Tutorial}. This mechanical approach allows \acp{ma} to adjust their position within a designated region, effectively exploiting the continuous spatial variations of the wireless channel to optimize connectivity and enhance communication performance~\cite{Zhu2023Movable,Ning2024Movable}. The concept has recently been extended to six-dimensional movable antennas (6DMA), which can flexibly adjust both the three-dimensional (3D) positions and 3D rotations of antennas or sub-arrays, offering unprecedented degrees of freedom for system design in future networks~\cite{Zhu2023Movable}. While \ac{fas} and \ac{ma} share the same underlying mathematical framework of flexible antenna positioning, they represent distinct hardware implementations and have evolved through separate research communities~\cite{Zhu2024Historical}. The choice between \ac{fas} and MAS often depends on practical considerations such as switching speed, mechanical reliability, and environmental robustness.

\paragraph{Flexible intelligent metasurfaces}
Building on this direction, \ac{fim} have been proposed as a three-dimensional extension of planar \acp{ris}. \ac{fim} consist of structurally adaptive, low-cost elements capable of individual vertical reconfiguration, known as \ac{3d} morphing, allowing real-time reshaping of the aperture surface. This design is particularly advantageous at \ac{mmwave} and \ac{thz} frequencies, as the 3D morphing provides the agility to compensate for small-scale fading. Recent studies show that joint optimization of \ac{fim} geometry and transmit beamforming can substantially reduce transmit power under user-specific \ac{sinr} constraints in downlink scenarios~\cite{An2025Flexible}.

\paragraph{Pinching-antenna systems}

The pinching-antenna concept introduces a flexible radiating structure in which small dielectric ``pinches'' (i.e., particles) are applied along a dielectric waveguide so that radiation can be emitted from any chosen point along the guide~\cite{Ding2025Flexible}. Unlike conventional flexible-antenna systems such as fluid antennas and movable antennas, which are typically constrained to small-scale position shifts within an aperture of several wavelengths, pinching-antenna systems (PASS) enable large-scale antenna reconfiguration, where antennas can be deployed at arbitrary positions along a waveguide that can be hundreds of wavelengths long~\cite{Liu2025PASS,Yang2025Pinching}. This unique mechanism allows pinching antennas to be placed close to users, thereby establishing strong \ac{los} links that mitigate large-scale path loss and significantly improve channel gain~\cite{Ding2025Flexible,Yang2025Pinching}. The system also supports activating multiple pinch elements on one waveguide to serve several users with a shared signal, making it suitable for non-orthogonal multiple-access (NOMA) or beamforming applications~\cite{Ding2025Flexible}. Furthermore, PASS is cost- and energy-efficient due to its simple structure, as its mechanism involves adding or removing dielectric materials along the waveguide~\cite{Yang2025Pinching,Ouyang2025ArrayGain}.

Despite these advantages, practical implementation of PASS faces several challenges. Determining and adjusting pinching antennas to arbitrary positions along the waveguides is a non-trivial task, and automating the pinching process remains difficult~\cite{Wang2025LCX}. Current implementations typically rely on mechanical actuation to configure the pinching elements, which introduces issues such as slow response times (on the order of milliseconds), increased weight, higher power consumption, and reduced reliability due to moving parts~\cite{Chen2026Hybrid}. As a result, practical PASS deployments often pre-install pinching antennas at uniformly spaced fixed locations along the waveguide rather than dynamically adjusting their positions in real time~\cite{PinchingISAC2025}. Overcoming these hardware limitations is a key direction for future research, with emerging approaches exploring electronic control of radiation amplitudes and phases to mitigate the drawbacks of mechanical actuation~\cite{Chen2026Hybrid}.

\subsubsection{Parasitic arrays}

Parasitic arrays employ a single driven element surrounded by one or more passive parasitic elements that are electromagnetically coupled rather than directly fed. By adjusting the reactance or geometry of these parasitic elements, they can be tuned to act as reflectors or directors, thereby shaping and steering the radiation beam. This mechanism enables low-cost, phase-shifter-free beamforming and pattern reconfiguration, although it may introduce efficiency degradation due to mutual coupling losses and impedance mismatch~\cite{Deshpande2025Beamforming}. In particular,~\cite{Zhang2004Pattern} demonstrates that switchable microstrip parasitic arrays can realize multiple radiation patterns over a shared bandwidth, highlighting practical pattern reconfigurability for compact antenna systems.

To further enhance spatial \acp{dof} and \ac{em} processing capability, recent work has extended reconfigurable metasurfaces into the third dimension through multi-layer architectures. These stacked structures enable depth-based wave manipulation, mimicking optical or neural systems in the \ac{rf} domain.

\subsubsection{Multi-layer reconfigurable arrays}

\paragraph{Stacked intelligent metasurfaces}

\ac{sim} is a multi-layer architecture composed of programmable metasurfaces stacked to form a unified \ac{em} processing structure. Recently, \ac{sim} has emerged as a promising approach to~\ac{hmimo} implementation, enabling $N_{\text{RF}} = N_{S} \ll N_\mathrm{A}$. By stacking multiple metasurface layers, the architecture performs information processing, transmit precoding, and receive combining directly at the~\ac{em} level~\cite{An2024Stacked,An2023Stacked,An2024Hybrid,Liu2025Stacked,Yao2024Channel,An2024Two}, eliminating the need for digital precoding/combining~\cite{An2023Stacked,An2024Near,An2024Stacked}.

Inspired by diffractive multi-layer neural networks, \ac{sim} processes signals through passive wave propagation and diffraction across its metasurface layers, achieving complex signal processing tasks at the speed of light with ultra-low latency~\cite{Liu2022Programmable,An2023Stacked,An2024Near,An2024Stacked}. Recent work has further extended the concept by integrating meta-fibers, enabling two-layer structures that achieve comparable flexibility to conventional multi-layer \acp{sim} while significantly reducing the number of meta-atoms~\cite{Niu2026Introducing}. This design reduces structural complexity and computational overhead, thereby enhancing energy efficiency and simplifying the alternating optimization procedure for system configuration~\cite{Niu2026Introducing}.

The energy efficiency of multi-layer \ac{sim} has been extensively studied, demonstrating that careful hybrid precoding design with 2–5 layers maximizes energy efficiency while maintaining high spectral efficiency using alternating optimization, semidefinite programming, and projected gradient methods~\cite{Shi2025Energy}. A recent survey provides a comprehensive overview of these advances~\cite{Di2025State}.

\paragraph{Stacked flexible intelligent metasurface}
Stacked flexible intelligent metasurfaces (SFIMs) extend conventional \acp{sim} by allowing each metasurface layer to be physically deformable, providing additional adaptability compared to rigid stacked architectures~\cite{Magbool2025Stacked}. Simulation results in~\cite{Magbool2025Stacked} demonstrate that the proposed SFIM-based system significantly outperforms its rigid \ac{sim} counterpart, achieving a 47\%–58\% improvement in sum rate and saving approximately 9~dBm of transmit power for a target rate of 7~bps/Hz (bits per second per Hertz), highlighting the potential of mechanical reconfigurability in multi-layer metasurface-assisted wireless systems.

Collectively, these emerging array architectures illustrate a fundamental change from static, feed-driven antennas toward intelligent, adaptive, and often software-defined \ac{em} surfaces capable of co-designing hardware and signal processing for next-generation wireless systems.

\vspace{0.5cm}

\begin{table*}[t]
\centering
\scriptsize
\caption{Comparison of emerging reconfigurable antenna and surface technologies}
\label{tab:Em_Rec_Arr}
\begin{tabular}{|p{2cm}|p{1.8cm}|p{2.2cm}|p{2cm}|p{2.5cm}|p{2.8cm}|p{0.5cm}|}
\hline
\textbf{Technology} & \textbf{Core principle} & \textbf{Beamforming mechanism} & \textbf{Control mechanism} & \textbf{Key advantage} & \textbf{Primary limitation} & \textbf{ Refs.} \\
\hline
\textbf{RIS} & Reflective mode only & Phase control of reflected signals & PIN diodes or varactors on passive elements & Ultra-low power, scalable deployment & Requires external illuminator, limited gain & \cite{Huang2019Reconfigurable, Wu2024Intelligent} \\
\hline
\textbf{BD-RIS} & Reflective, transmissive, or hybrid mode & Phase and amplitude control via multi-port scattering matrix & Multi-port reconfigurable impedance network & Supports multiple modes; generalization of conventional \ac{ris} & Higher hardware complexity & \cite{Xu2021Star, Li2023Beyond} \\
\hline
\textbf{STAR-RIS} & Hybrid (simultaneous reflection/transmission) & Phase and amplitude control of transmitted and reflected signals & Cell-wise single-connected impedance network & Full-space (360°) coverage; specialized case of \ac{bdris} & Higher design complexity, mutual coupling, power split between reflection and transmission & \cite{Xu2021Star, Li2023Beyond} \\
\hline
\textbf{RHS} & Leaky-wave radiation with amplitude control & Amplitude-controlled holographic pattern & Amplitude control of each element using PIN diodes or tunable materials & Ultra-thin structure, integrable with transceivers, high beampattern gain & Real-domain amplitude constraints & \cite{Deng2022Reconfigurable, Di2025Reconfigurable} \\
\hline
\textbf{DMA} & Guided Wave Radiation & Modulation of guided wave via tunable meta-elements & Bias to meta-atoms (e.g., PIN diodes) & Low-cost, low-power analog beamforming & Limited bandwidth, coupling challenges & \cite{Shlezinger2019Dynamic, Bjornson2024Towards} \\
\hline
\textbf{Parasitic} & Mutual Coupling (coupled elements) & Reactance loading of parasitic elements & Tunable capacitors/varactors & Single \ac{rf} chain, very low cost & Limited scanning range, lower efficiency & \cite{Deshpande2025Beamforming} \\
\hline
\textbf{Fluid antenna (FA)} & Aperture Reconfiguration & Shape/position change of conductor & Liquid-based, pixel implementation & Flexibility in shape/position, improving system performance & Mainly suitable for slowly-varying channels due to movement speed & \cite{Wong2020Fluid, Wong2023FluidIII} \\
\hline
\textbf{Movable antenna (MA)} & Antenna Element Relocation & Moving antenna ports & Micro-electromechanical systems (MEMS), step motors & Improving channel condition and communication performance & Mechanism-dependent speed, mechanical reliability & \cite{Zhu2023Movable, Ning2024Movable} \\
\hline
\textbf{FIM} & \ac{3d} Surface Shaping & Dynamic shaping of the entire surface & Micro-actuators per element & Optimizes near-field channels directly & Extreme fabrication complexity & \cite{An2025Flexible} \\
\hline
\textbf{SIM} & \ac{em} Wave Computation & Multi-layer wave manipulation via meta-atoms & Per-layer phase profile configuration & In-hardware precoding, ultra-low latency & Fixed function, design complexity & \cite{An2023Stacked, Liu2025Stacked} \\
\hline
\end{tabular}
\end{table*}

\section{Physically consistent modeling for emerging architectures }
\label{Sec:III}

\subsection{Physics-based modeling frameworks}
This subsection presents four complementary physics-based modeling frameworks that form the foundation for physically consistent channel characterization. We begin with Maxwell's equations, then present the Fourier plane-wave representation, multiport network theory, and computational \ac{em} simulation tools.

\subsubsection{Maxwell’s equations}

Classical \ac{em} theory provides the foundation for analyzing and designing emerging architectures. Maxwell's equations describe the spatio-temporal evolution of electric and magnetic fields in the presence of charges and currents, where $\mathbf{r}$ denotes the spatial position vector and $t$ denotes time. In a general macroscopic medium characterized by spatially varying permittivity $\varepsilon(\mathbf{r})$, permeability $\mu(\mathbf{r})$, and free charge and current densities $\rho(\mathbf{r},t)$ and $\mathbf{J}(\mathbf{r},t)$, the electric flux density $\mathbf D$ and magnetic flux density $\mathbf B$ are related to the electric field intensity $\mathbf{E}(\mathbf{r},t)$ and magnetic field intensity $\boldsymbol{\mathbf{H}}(\mathbf{r},t)$ by the constitutive relations
\begin{equation}  
\mathbf D(\mathbf r,t) = \varepsilon(\mathbf r)\mathbf E(\mathbf r,t), \qquad
\mathbf B(\mathbf r,t) = \mu(\mathbf r)\boldsymbol{\mathbf{H}}(\mathbf r,t).
\end{equation}

The time-domain Maxwell's equations then take the form
\begin{subequations}\label{eq:maxwell}
\begin{align}
\nabla \cdot \mathbf D(\mathbf r,t) &= \rho(\mathbf r,t), \label{eq:maxwell_a}\\
\nabla \cdot \mathbf B(\mathbf r,t) &= 0, \label{eq:maxwell_b}\\
\nabla \times \mathbf E(\mathbf r,t) &= -\frac{\partial \mathbf B(\mathbf r,t)}{\partial t}, \label{eq:maxwell_c}\\
\nabla \times \mathbf H(\mathbf r,t) &= \mathbf J(\mathbf r,t) + \frac{\partial \mathbf D(\mathbf r,t)}{\partial t}. \label{eq:maxwell_d}
\end{align}
\end{subequations}

For a linear medium (in which the material parameters $\varepsilon$ and $\mu$ are independent of the field magnitudes), Maxwell’s equations form a linear system, enabling superposition and linear operator techniques.
Under the time-harmonic assumption (all fields vary sinusoidally at a single angular frequency $\omega$), we have $\mathbf E(\mathbf r,t) = \mathbf E(\mathbf r)e^{-j\omega t}$. For a homogeneous, isotropic, and lossless medium where $\varepsilon(\mathbf r) = \varepsilon$ and $\mu(\mathbf r) = \mu$, Maxwell’s equations reduce to the vector Helmholtz equation,
\begin{equation}
\nabla \times \nabla \times \mathbf E(\mathbf r) - \kappa^2 \mathbf E(\mathbf r)
= j\omega\mu \mathbf J(\mathbf r),
\qquad
\kappa^2 = \omega^2 \mu \varepsilon,
\label{eq:vectorwave}
\end{equation}
which governs time-harmonic \ac{em} wave propagation.

The solution to \eqref{eq:vectorwave} can be represented using the dyadic Green's function $\mathbf G(\mathbf r,\mathbf s)$, where $\mathbf r, \mathbf s \in \mathbb{R}^3$ are the three-dimensional position vectors of the observation point and the source point, respectively. Physically, the Green's function gives the electric field at observation point $\mathbf r$ due to an infinitesimal point source located at $\mathbf s$. It satisfies
\begin{equation}
\mathcal{L}_0 \mathbf G(\mathbf r,\mathbf s) = \mathbf I\delta(\mathbf r - \mathbf s),
\qquad
\mathcal{L}_0 = \nabla \times \nabla \times - k_0^2 \mathbf I,
\end{equation}
where $\mathbf I$ denotes the three-dimensional identity tensor (dyadic), $\delta(\cdot)$ represents the three-dimensional Dirac delta function, and $\kappa_0 = \omega\sqrt{\mu_0 \varepsilon_0}$ is the free-space wavenumber. The total electric field is then obtained by integrating the source distribution against the Green’s function:
\begin{equation}
\mathbf E(\mathbf r) = j\omega\mu_0 \int_{V_s} \mathbf G(\mathbf r,\mathbf s) \mathbf J(\mathbf s)\, d \mathbf s.
\end{equation} 

where $V_s$ denotes the volume containing the source current distribution $\mathbf J(\mathbf s)$.

In free space, the scalar Green's function takes the familiar spherical-wave form, which arises as the outgoing causal solution to the homogeneous scalar Helmholtz equation,
\begin{equation}
g(\mathbf{r},\mathbf{s}) = \frac{e^{-j k_0 \|\mathbf{r} - \mathbf{s}\|}}{4\pi \|\mathbf{r} - \mathbf{s}\|},
\end{equation}
and the dyadic Green’s function is derived from it as~\cite{Chew1999Waves}
\begin{equation}
\mathbf{G}(\mathbf{r},\mathbf{s}) =  \left[ \mathbf{I} + \frac{1}{\kappa_0^2} \nabla_{\mathbf{r}} \nabla_{\mathbf{r}} \right] g(\mathbf{r},\mathbf{s}).
\label{eq:green}
\end{equation}

In an inhomogeneous or scattering medium, where $\varepsilon(\mathbf r)$ or $\mu(\mathbf r)$ vary spatially~\cite{Chew1999Waves}, the Green’s function must account for multiple scattering and reflections. The total field can then be expressed as the sum of the direct field and the scattered field~\cite{Wan2024Near},
\begin{equation}
\begin{split}
\mathbf E(\mathbf r) &= j\omega \mu_0 \int_{V_s} \mathbf G(\mathbf r,\mathbf s)\mathbf J(\mathbf s)\, d \mathbf s \\
&\quad + \int_V \mathbf G(\mathbf r,\mathbf r')[\kappa^2(\mathbf r') - \kappa_0^2]\mathbf E(\mathbf r')\, d \mathbf r',
\end{split}
\end{equation}

where $V_s$ denotes the source region, $V$ the domain containing material inhomogeneities, $\mathbf r'$ is the position vector within the scattering volume $V$, $\mathbf E(\mathbf r')$ is the induced electric field in the inhomogeneous regions, and $\kappa^2(\mathbf r') = \omega^2 \mu(\mathbf r') \varepsilon(\mathbf r')$. The first integral represents the direct field from the source currents, while the second captures the scattered field from material inhomogeneities.

\subsubsection{Wavenumber representation}
\label{subsec:fourier_repr}

In a homogeneous, source-free, linear, and isotropic medium, the electric field satisfies the vector Helmholtz equation derived from Maxwell’s equations (see Eq.~\eqref{eq:vectorwave} with $\mathbf{J} = \mathbf{0}$ and constant $\varepsilon$, $\mu$). Under these conditions, the Cartesian components of the electric field decouple, and each component $E_i(\mathbf{r})$ ($i = x,y,z$) satisfies the scalar Helmholtz equation
\begin{equation} 
\label{eq:helmholtz}
(\nabla^2 + \kappa^2)E_i(\mathbf{r}) = 0,
\end{equation}
where $\kappa = \omega\sqrt{\mu\varepsilon}$. This forms the basis for the wave number domain channel representation in~\cite{Pizzo2022Fourier}. The general solution to the Helmholtz equation can be expressed as a superposition of plane waves of the form 
\( e^{i(\kappa_x x + \kappa_y y + \kappa_z z)} \) satisfying \(\kappa_x^2 + \kappa_y^2 + \kappa_z^2 = \kappa^2\). 
For propagating components (\(\kappa_z \in \mathbb{R}\)), \((\kappa_x , \kappa_y) \) is limited to the support 
\begin{equation} 
\label{eq:wavenumber_support}
\mathcal{D}(\kappa) = \{(\kappa_x,\kappa_y)\in\mathbb{R}^2 : \kappa_x^2 + \kappa_y^2 \le \kappa^2\}.
\end{equation}
The total field is represented as an integral over \(\mathcal{D}(\kappa)\) of upgoing and downgoing plane waves with random amplitudes \(H_\pm(\kappa_x,\kappa_y)\), as given in~\cite[Eq.~(7)]{Pizzo2020Spatially}.

In the wavenumber domain, the field’s power spectral density (PSD) is concentrated on the spherical shell \(\kappa_x^2 + \kappa_y^2 + \kappa_z^2 = \kappa^2\),
\begin{equation}
\label{eq:psd}
S_h(\kappa_x,\kappa_y,\kappa_z) = A_h^2(\kappa_x,\kappa_y,\kappa_z)\, \delta(\kappa_x^2 + \kappa_y^2 + \kappa_z^2 - \kappa^2),
\end{equation}
where \(A_h(\cdot)\) determines the angular scattering distribution. 
This formulation links spatial field statistics directly to directional power distributions.

For finite arrays, the continuous representation in~\cite[Eq.~(4)]{Pizzo2022Fourier} is discretized over wavenumber lattices 
\(\mathcal{E}_s\) and \(\mathcal{E}_r\) (see~\cite[Eq.~(17)–(18)]{Pizzo2022Fourier}). 
The wireless channel matrix \(\mathbf{H}_w\in\mathbb{C}^{N_r\times N_s}\) admits a discrete Fourier expansion:
\begin{equation}
\begin{split}
\label{eq:discrete_channel}
\mathbf{H}_w = \sqrt{N_r N_s} 
\sum_{(\ell_x,\ell_y)\in \mathcal{E}_r}
\sum_{(n_x,n_y)\in \mathcal{E}_s}
& H_a(\ell_x,\ell_y,n_x,n_y)\, \\
& \times \mathbf{a}_r(\ell_x,\ell_y)\mathbf{a}_s^\HH(n_x,n_y),
\end{split}
\end{equation}
where \(\mathbf{a}_s(\cdot)\) and \(\mathbf{a}_r(\cdot)\) are the transmit and receive response vectors defined in~\cite[Eq.~(13)-(14)]{Pizzo2022Fourier}. 
The coefficients \(H_a(\ell_x,\ell_y,n_x,n_y)\) characterize the channel in the wavenumber domain\footnote{Under common Rayleigh fading assumptions, these coefficients are often modeled as zero-mean circularly symmetric complex Gaussian random variables with variance determined by the 4D channel PSD \(S(\cdot)\) (see~\cite[Eq.~(23)]{Pizzo2022Fourier}).}.

Building on this wavenumber-domain framework, the effects of mutual coupling between antenna elements can be characterized in a physically consistent manner. The Fourier-domain representation developed for holographic \ac{mimo}~\cite{Pizzo2025Mutual} provides an alternative perspective on mutual coupling beyond the conventional impedance-matrix formulation. In this framework, the transmit coupling kernel for punctiform antennas is given by \(c_{\mathrm{t}}(\mathbf{v}) = \operatorname{sinc}(\|\mathbf{v}\|/\lambda)\)~\cite[Eq.~(20)]{Pizzo2025Mutual}, where \(\mathbf{v} \in \mathbb{R}^2\) denotes the spatial separation vector between two antenna elements on the array aperture. The Fourier spectrum of this kernel, defined over the spatial frequency (wavenumber) vector \(\mathbf{k} \in \mathbb{R}^3\), resides on the wavenumber sphere \(\|\mathbf{k}\| = \kappa\) as \(C_{\mathrm{t}}(\mathbf{k}) = \frac{4\pi^{2}}{\kappa}\delta(\|\mathbf{k}\|^{2} - \kappa^{2})\)~\cite[Eq.~(24)]{Pizzo2025Mutual}. For physical antennas with pattern \(\mathbf{A}_{\mathrm{t}}(\mathbf{k})\), this generalizes to \(C_{\mathrm{t}}(\mathbf{k}) = \frac{4\pi^{2}}{\kappa} \|\mathbf{A}_{\mathrm{t}}(\mathbf{k})\|^{2} \delta(\|\mathbf{k}\|^{2} - \kappa^{2})\)~\cite[Eq.~(32)]{Pizzo2025Mutual}. Comparing these expressions with the fading power spectral density reveals an isomorphism between coupling and fading correlation, but with opposite effects: coupling acts as a \textit{deconvolution} (inverse filter) while fading correlation acts as a convolution~\cite[Eq.~(81)]{Pizzo2025Mutual}. Consequently, the end-to-end channel including transmit coupling becomes \(\mathbf{H} \propto \mathbf{H}_{\text{uncoupled}} \mathbf{C}_{\mathrm{t}}^{-1/2}\)~\cite[Eq.~(48)]{Pizzo2025Mutual}, meaning that coupling can either reinforce or counteract spatial correlation depending on the antenna pattern and the operating \ac{snr}.

This Fourier representation thus provides a physically consistent basis for characterizing spatial correlation and wave propagation in the wavenumber domain.

\subsubsection{Multiport network}
\label{sec:multiport_network}

\ac{em} structures with multiple terminals, such as antennas, transmission lines, and microwave circuits, are commonly modeled as multiport networks. 

At high frequencies (\ac{mmwave} and \ac{thz}) and with densely packed antennas (as in holographic MIMO), conventional voltage–current descriptions become ill‑defined, and mutual coupling significantly affects both signals and noise. Multiport network theory, borrowed from microwave engineering, treats each antenna as a port and describes interactions via impedance, admittance, or scattering parameters, naturally incorporating coupling, impedance mismatch, and correlated noise. The following definitions build toward the end‑to‑end models in Sec.~\ref{subsec:End-to-end models}.

Each port is associated with a pair of conjugate variables: voltage and current \((v_n, i_n)\), or equivalently, incident and reflected wave amplitudes \((a_n, b_n)\). Under the assumptions of linearity, time invariance, and passivity, the network’s behavior is fully characterized by linear relations among these port quantities~\cite{Pozar2011Microwave,Mezghani2023Reincorporating,Ivrlavc2010Toward,ref130}.

Let \(\mathbf{v} = [v_1, v_2, \dots, v_{N_\mathrm{p}}]^{\TT}\) and \(\mathbf{i} = [i_1, i_2, \dots, i_{N_\mathrm{p}}]^{\TT}\) denote the complex phasor vectors of port voltages and currents. The most general linear relationship between them is expressed via the impedance and admittance formulations:
\begin{equation}
\mathbf{v} = \mathbf{Z}\mathbf{i}, \qquad 
\mathbf{i} = \mathbf{Y}\mathbf{v},
\end{equation}
where \(\mathbf{Z} \in \mathbb{C}^{N_\mathrm{p} \times N_\mathrm{p}}\) is the impedance matrix and \(\mathbf{Y} = \mathbf{Z}^{-1}\) is the admittance matrix (assuming \(\mathbf{Z}\) is invertible)~\cite{Pozar2011Microwave}. The element \(Z_{mn}\) represents the open-circuit voltage at port \(m\) due to a unit current injected at port \(n\), with all other ports open-circuited. Conversely, \(Y_{mn}\) is the short-circuit current at port \(m\) resulting from a unit voltage applied at port \(n\), with all other ports shorted~\cite{Balanis2016Antenna}. For reciprocal networks, \(\mathbf{Z}\) and \(\mathbf{Y}\) are symmetric (\(Z_{mn} = Z_{nm}\), \(Y_{mn} = Y_{nm}\)). For lossless networks, \(\mathbf{Z}\) is purely imaginary and satisfies \(\mathbf{Z}^{\HH} = -\mathbf{Z}\) (skew-Hermitian), while \(\mathbf{Y}\) is also purely imaginary and skew-Hermitian. These matrix representations emphasize local voltage–current relationships and are most suitable for lumped or electrically small networks, where these quantities are well-defined.

At high frequencies, direct measurements of voltage and current become impractical due to the distributed nature of \ac{em} fields and the lack of a unique reference for voltage in radiating or waveguide structures. Wave-based variables therefore provide a more suitable framework, as the scattering-matrix formulation describes the linear relationship between incident and reflected power waves, directly capturing reflection, transmission, and power conservation~\cite{Pozar2011Microwave}. 

Defining the normalized incident and reflected waves with respect to a real, positive reference impedance $Z_0$~\cite{Pozar2011Microwave},
\begin{equation}
a_n = \frac{V_n + Z_0 I_n}{2\sqrt{Z_0}}, \qquad
b_n = \frac{V_n - Z_0 I_n}{2\sqrt{Z_0}},
\end{equation}
the behavior of an $N_\mathrm{p}$-port network is concisely expressed via the scattering matrix $\mathbf{S} \in \mathbb{C}^{N_\mathrm{p} \times N_\mathrm{p}}$ as
\begin{equation}
\mathbf{b} = \mathbf{S}\mathbf{a},
\end{equation}
where $\mathbf{a} = [a_1, \dots, a_{N_\mathrm{p}}]^{\TT}$ is the incident waves vector and $\mathbf{b} = [b_1, \dots, b_{N_\mathrm{p}}]^{\TT}$ is the reflected waves vector. The element $S_{mn}$ represents the complex reflection coefficient at port $n$ when $m = n$, and the transmission coefficient from port $n$ to port $m$ when $m \ne n$. 

For reciprocal networks, $\mathbf{S}$ is symmetric ($\mathbf{S} = \mathbf{S}^{\TT}$). For lossless networks, conservation of power implies that $\mathbf{S}$ is unitary, i.e., $\mathbf{S}^{\HH} \mathbf{S} = \mathbf{I}$.

The impedance, admittance, and scattering matrices are mathematically equivalent and are connected through exact transformations defined by the reference impedance \(Z_0\). The relationships are given by~\cite{Pozar2011Microwave,Ivrlavc2010Toward}
\begin{subequations}\label{eq:SYZrelations}
\begin{align}
\mathbf{S} &= (\mathbf{Z} - Z_0 \mathbf{I})(\mathbf{Z} + Z_0 \mathbf{I})^{-1}\!, 
\!&\!\mathbf{Z} &= Z_0(\mathbf{I} + \mathbf{S})(\mathbf{I} - \mathbf{S})^{-1}, \label{eq:SYZrelations_Z}\\[4pt]
\mathbf{S} &= (\mathbf{I} - Z_0 \mathbf{Y})(\mathbf{I} + Z_0 \mathbf{Y})^{-1}, 
& \mathbf{Y} &= \frac{1}{Z_0}(\mathbf{I} - \mathbf{S})(\mathbf{I} + \mathbf{S})^{-1}. \label{eq:SYZrelations_Y}
\end{align}
\end{subequations}

A recent overview of multiport network theory applied to reconfigurable metasurfaces and \ac{ris}-aided wireless communications, covering impedance and scattering parameter formulations, validation through full-wave simulations, and optimization methodologies, is presented in~\cite{Di2025Multiport}. This multiport framework provides the circuit-theoretic foundation used in the end-to-end models developed in Subsection~\ref{subsec:End-to-end models}.

\subsubsection{Numerical EM modeling}

Physically consistent modeling ultimately requires numerical \ac{em} solvers when the interaction between waves and detailed device geometries cannot be captured accurately by analytical or circuit-level abstractions alone. This is particularly challenging for reconfigurable communication environments: structures such as \acp{ris} contain sub-wavelength dielectric and conducting features, while the surrounding propagation environment can span many wavelengths. Direct full-wave simulation of the complete environment is therefore computationally prohibitive, motivating numerical methods that accurately resolve the device-scale \ac{em} interactions while providing scalable representations for system- and environment-level propagation.

\ac{em} interactions on composite objects, such as nearly-passive \acp{ris} composed of piecewise homogeneous dielectric and/or conducting bodies, can be simulated by solving surface integral equations (SIEs)~\cite{ ref100}. For example, the electric field and Poggio-Miller-Chang-Harrington-Wu-Tsai integral equations (EFIE-PMCHWT) can be coupled~\cite{ref101, ref103}. Discretizing this coupled system of first-kind SIEs using Rao-Wilton-Glisson (RWG) functions~\cite{ref104} results in an ill-conditioned matrix equation, requiring a large number of iterations to solve, particularly for electrically large surfaces~\cite{ref105}.

The convergence rate of iterations can be improved by using a second-kind SIE, such as the combined field integral equation (CFIE) or the electric-magnetic field CFIE (JMCFIE)~\cite{ ref107, ref108, ref109}. RWG-based discretization of second-kind SIEs, however, can still yield inaccurate solutions. To apply SIE solvers to multi-scale structures, several approaches have been developed~\cite{ref110, ref111, ref112, ref113, ref114, ref115, ref116, ref117}. Among these, multi-trace SIEs (MT-SIEs)~\cite{ref110}, the contact-region modeling method (CRM)~\cite{ref111}, and domain decomposition method (DDM)-based schemes~\cite{ref112, ref113, ref114, ref115, ref116, ref117} partition the scatterer into subdomains and introduce two sets of SIEs with unknown equivalent currents on the interfaces between subdomains. Upon discretization, MT formulations result in a larger matrix system; however, the easier meshing process and the improved matrix conditioning often justify the increased matrix dimension. DDM typically employs CFIE or JMCFIE as the governing equation to ensure a well-conditioned matrix system~\cite{ref112, ref113, ref114, ref115, ref116, ref117}. CRM, which uses EFIE-PMCHWT, provides more accurate solutions but suffers from worsening matrix system conditioning as the object size increases~\cite{ref110}.

A robust SIE solver can address these issues and simulate structures such as \acp{ris} without loss of accuracy or efficiency. This is achieved using (1) a fine-tuned combination of MT-DDMs and new discretization techniques, and (2) advanced preconditioning methods. In the first part, the locally-coupled multi-trace domain decomposition method (LCMT-DDM) decomposes RISs into subdomains: the exterior region (free space), dielectric bodies (e.g., dielectric substrates), and conducting bodies and surfaces (e.g., conducting ink traces)~\cite{ref113, ref117, ref118}. EFIE and the magnetic field integral equation (MFIE) are used as governing equations for unknown equivalent currents in each subdomain, with these currents on subdomain surfaces locally coupled via Robin transmission conditions (RTCs)~\cite{ref113, ref118}. For perfect conductors, the governing equations are replaced by a special form of RTCs (excluding scattering terms) enforced on their surfaces. For finite conductivity, the governing equations are replaced by impedance or resistive boundary conditions.

The RWG-based discretization of this locally coupled system of equations yields a matrix system that avoids ill-conditioning, except when using dense meshes required for fine geometric details. The matrix blocks corresponding to conducting subdomains and RTCs (coupling between subdomains) remain sparse and well-conditioned. For low and moderately dense meshes, this matrix system can be solved very efficiently using an iterative solver. Additionally, the sparsity of the coupling matrices reduces memory usage. When dense meshes are employed to accurately resolve \ac{ris} array elements in each subdomain, however, matrix blocks corresponding to EFIE/CFIE start to become ill-conditioned, impairing iterative solution efficiency. In the second part, Calderon preconditioning and mixed discretization methods are used to address this issue. Matrices resulting from RWG-based discretization of EFIE (and CFIE, to a lesser extent due to the self-term) become ill-conditioned with dense meshes~\cite{ref120}. Similarly, matrix blocks in LCMT-DDM subdomains involving dense meshes also become ill-conditioned, reducing the efficiency of the iterative solution for the entire matrix system.

To address the ill-conditioning of EFIE, its self-regularizing property is used by preconditioning EFIE with itself, implemented as a multiplicative Calderon preconditioner (CMP)~\cite{ref120}. This method requires using Buffa-Christiansen (BC) basis functions to discretize EFIE~\cite{ref121}. The same basis change is applied to MFIE to ensure that CFIE is derived from a “consistent” combination of EFIE and MFIE. This basis change is implemented using mixed discretization approaches, which combine RWG and BC functions~\cite{ref122, ref123, ref124, ref125}.

The integration of CMP and mixed discretization within DDM-based frameworks remains comparatively unexplored. Their incorporation into LCMT-DDM therefore warrants systematic assessment in terms of accuracy and iterative convergence. Alternative governing formulations, including PMCHWT~\cite{ref100, ref101} and JMCFIE~\cite{ref105, ref107, ref108}, can likewise be considered to assess their interaction with Calderon preconditioning and mixed discretization~\cite{ref126}.

To bridge device-scale and environment-scale simulations, the \ac{ris} can instead be represented by an equivalent surface model in indoor/outdoor \ac{em} propagation simulations. This equivalent model uses generalized sheet transition conditions (GSTCs)~\cite{ref127, ref128, Holloway2012Overview}. The GSTCs replace the \ac{ris}'s geometrical details with an effective homogenized field/wave transformation function defined on the surface occupied by the \ac{ris}. For simplicity, assume that the \ac{ris} is planar and lies in the xy-plane with a surface normal along the z-direction. In this case, the electric and magnetic polarization densities \(\mathbf{P}\) and \(\mathbf{M}\) are expressed in terms of the \ac{em} fields on the two sides of the \ac{ris}~\cite{ref130, ref131}:
\begin{subequations}
\label{eq:pol}
\begin{align}
\mathbf{P}&=\varepsilon \bar{\chi}_{\text{ce}}\frac{\mathbf{E}_1+\mathbf{E}_2}{2}+\bar{\chi}_{\text{em}}\frac{\mathbf{H}_1+\mathbf{H}_2}{2},\\
\mathbf{M}&=\bar{\chi}_{\text{mm}}\frac{\mathbf{H}_1+\mathbf{H}_2}{2}+\bar{\chi}_{\text{me}}\sqrt{\frac{\varepsilon}{\mu}}\frac{\mathbf{E}_1+\mathbf{E}_2}{2},
\end{align}
\end{subequations}
where the factor $\sqrt{\frac{\varepsilon}{\mu}}$ corresponds to the inverse of the wave impedance of the surrounding medium. This scaling normalizes the electric-field contribution in the expression for the magnetic polarization density, ensuring dimensional consistency with the magnetic-field term. 
The surface susceptibilities \(\bar{\chi}_{\text{uv}}\) are extracted from the \ac{ris} reflection and transmission coefficients, either computed using the \ac{em} solver or measured experimentally~\cite{ref132, ref133, ref134}. Incorporating GSTCs into SIE solvers enables efficient indoor/outdoor \ac{em} propagation simulations without explicitly resolving the \ac{ris} geometrical details. Together, detailed SIE-based device modeling and GSTC-based equivalent representations provide a path toward connecting device-scale \ac{em} behavior with environment-scale propagation. However, leveraging such solvers for physically consistent communication modeling remains challenging because of the large disparity between device and environment scales, the computational and memory costs of electrically large problems, numerical conditioning under dense discretization, and the need to preserve configuration-dependent \ac{em} behavior when transitioning from detailed full-wave models to reduced-order surface representations. Addressing these challenges is essential for integrating numerical \ac{em} solvers with scalable channel and system-level simulations.

\subsection{End-to-end models for reconfigurable architectures}
\label{subsec:End-to-end models}

We now specialize the unified system model of \eqref{eq:unified_system_model} to each reconfigurable architecture. The goal is to establish explicit mathematical representations for the effective channel \(\mathbf{H}_{\mathrm{eff},k}\) that capture the distinct physical operating principles of each antenna type.

\subsubsection{Metasurface-based arrays}
\label{subsec:models_metasurface}

The effective channel of the metasurface-based transmitter described in Sec.~\ref{subsubsec:metasurface_antennas} can be characterized through three physical layers: (i) feed-to-surface coupling, modeled by the matrix \(\mathbf{W}\), which governs how guided \ac{em} waves are launched and propagate along the surface, (ii) element-level control, represented by \(\boldsymbol{\Gamma}\), which adjusts the local scattering or radiation characteristics of the meta-atoms, and (iii) free-space propagation, captured by the wireless channel \(\mathbf{H}_{w,k}\). This decomposition aligns with established models for \ac{dma} and \acp{rhs} in the literature~\cite{Wang2019DMA,Shlezinger2021Dynamic,Gong2024Holographic}. Let $N_{\text{fed}}$ denote the number of guided feeds (e.g., microstrip or waveguide ports) and $N_\mathrm{A}$ the total number of radiating elements. For metasurface-based architectures where the EM layer is independent of the propagation channel, the effective channel takes the form
\begin{equation}    
\mathbf{H}_{\mathrm{eff},k}(\mathbf{F}_{\mathrm{EM},k}) = \mathbf{H}_{w,k} \mathbf{F}_{\mathrm{EM},k},
\end{equation}
where \(\mathbf{H}_{w,k} \in \mathbb{C}^{U \times N_{\mathrm{A}}}\) is the wireless propagation channel and \(\mathbf{F}_{\mathrm{EM},k} \in \mathbb{C}^{N_{\mathrm{A}} \times N_{\mathrm{fed}}}\) is the EM-domain processing matrix.

The guided-wave excitation from the feeds to the surface elements is modeled by the coupling matrix $\mathbf{W}\in\mathbb{C}^{N_\mathrm{A}\times N_{\text{fed}}}$, which provides an effective mapping between the input ports and the field amplitudes at the meta-atoms, abstracting the physical propagation of the guided waves.

\underline{For DMAs}, $\mathbf{W}$ captures the guided propagation inside one or multiple waveguides, including attenuation and phase shift as the signal travels along each waveguide~\cite{Shlezinger2019Dynamic,Wang2019DMA}. A key characteristic of DMAs is their frequency-selective response: each radiating element has a resonant frequency, and operation away from resonance leads to magnitude degradation~\cite{Deshpande2026Frequency}. This frequency selectivity can be exploited for frequency-selective beamforming and beam training, but it also imposes constraints on wideband operation~\cite{Deshpande2026Frequency}.

\underline{For RHSs}, $\mathbf{W}$ represents the reference \ac{em} wave propagating across the holographic surface~\cite{Deng2022Reconfigurable,Zhang2025Holographic}. Unlike DMAs, which rely on guided-wave propagation through a waveguide structure, RHSs operate on the holographic principle: the superposition of the reference wave and a desired object wave creates an interference pattern that determines the radiation from each meta-atom. RHSs also exhibit frequency-selective behavior due to the resonant nature of their metamaterial elements, which poses challenges for ultra-wideband operation.

Each element exhibits a tunable complex response depending on its control state (e.g., varactor bias or resonance frequency). Collecting these responses yields
\begin{equation}
\mathbf{\Gamma}=\operatorname{diag}(\gamma_1,\dots,\gamma_{N_\mathrm{A}}),
\end{equation}
where $\gamma_n$ represents the complex amplitude and phase weighting applied by the $n$-th element.

\underline{In DMAs}, the entries of $\mathbf{\Gamma}$ correspond to the tunable impedance or scattering coefficients of each metamaterial element, typically controlled via voltage-biased varactors or PIN diodes that modify the local LC resonance~\cite{Shlezinger2021Dynamic,Bjornson2024Towards}. The Lorentzian model captures the resonant behavior of the DMA elements, where the phase and magnitude of the magnetic polarizability are coupled and cannot be tuned independently~\cite{Deshpande2026Frequency}.

\underline{In RHSs}, $\mathbf{\Gamma}$ represents the amplitude-control mechanism used to realize holographic beamforming. Each element adjusts its radiated amplitude (and possibly phase) so that the superposition of all leaky waves forms the desired far-field beam~\cite{Gong2024Holographic,Deng2022Reconfigurable,Zhang2025Holographic}.

The \ac{em}-domain mapping between the feed excitations and the effective aperture field is therefore
\begin{equation}
\mathbf{F}_{\mathrm{EM},k}=\boldsymbol{\Gamma}_k \mathbf{W}_k,
\end{equation}
where the subcarrier dependence captures the frequency-selective nature of the metasurface response.

For metasurface-based architectures, the radiated power is typically normalized with respect to the incident power at each element. For DMAs, the radiated power is given by
\begin{equation}
P_{k}(\mathbf{F}_{\mathrm{EM},k},\mathbf{F}_{\mathrm{RF},k},\mathbf{F}_{\mathrm{BB},k}) = \sum_{n_{\text{fed}}} P_{\mathrm{in},k}\left(1 - |S_{12,n_{\text{fed}}}^{\mathrm{all}}|^2\right),
\label{eq:dma_radiated_power}
\end{equation}
where \( P_{\mathrm{in},k} = \| \mathbf{F}_{\mathrm{BB},k}\mathbf{F}_{\mathrm{RF},k}\|_{\mathrm{F}}^{2} \) is the input power into DMA waveguide $n_{\text{fed}}$ and \( S_{12,n_{\text{fed}}}^{\mathrm{all}} \) accounts for the forward scattered electromagnetic fields
after the last DMA element for waveguide $n_{\text{fed}}$~\cite{Vikas2026Signal,Heath2025Tri}. For RHSs, the radiated power is determined by the amplitude-control mechanism and the reference wave propagation; the normalization typically ensures that the sum of squared radiated amplitudes equals the number of elements for a lossless surface~\cite{Gong2024Holographic}.

The wireless propagation is represented by the channel $\mathbf{H}_{w,k}\in\mathbb{C}^{ N_\mathrm{A}\times U}$, which includes path loss, phase delay, element radiation patterns, and near- or far-field coupling~\cite{Gong2024Holographic}. To model the propagation channel for DMAs and RHSs, different \ac{em} formulations have been used in the literature. 

Circuit-theory-based models offer a compact yet accurate representation of the \ac{em} behavior by describing each meta-atom through equivalent impedance or scattering networks that capture inter-element coupling and dissipation effects, as already established in the multiport network formulation of Sec.~\ref{sec:multiport_network}. Prior work~\cite{Williams2022Electromagnetic} formulated the DMA as a multiport admittance network, where each \ac{rf} chain, radiating element, and user terminal is represented as a port in the composite system. 
The \ac{em} coupling among these ports is fully characterized by the network admittance matrix,
\begin{equation}
\begin{bmatrix}
\mathbf{v}_t\\[2pt]
\mathbf{v}_s\\[2pt]
\mathbf{v}_r
\end{bmatrix}
=
\begin{bmatrix}
\mathbf{Y}_{tt} & \mathbf{Y}_{st}^\TT & \mathbf{Y}_{rt}^\TT\\
\mathbf{Y}_{st} & \mathbf{Y}_{ss} & \mathbf{Y}_{rs}^\TT\\
\mathbf{Y}_{rt} & \mathbf{Y}_{rs} & \mathbf{Y}_{rr}
\end{bmatrix}
\!
\begin{bmatrix}
\mathbf{j}_t\\[2pt]
\mathbf{j}_s\\[2pt]
\mathbf{j}_r
\end{bmatrix},
\label{eq:DMA_Ymatrix}
\end{equation}
where $\mathbf{v}_t \in \mathbb{C}^{N_{\mathrm{fed}} \times 1}$, $\mathbf{v}_s \in \mathbb{C}^{N_\mathrm{A} \times 1}$, and $\mathbf{v}_r \in \mathbb{C}^{U \times 1}$ denote the magnetic voltages at the transmitter ports, DMA elements, and receivers, respectively. The vectors $\mathbf{j}_t$, $\mathbf{j}_s$, and $\mathbf{j}_r$ denote the corresponding magnetic currents at these ports. 
Each submatrix of $\mathbf{Y}$ captures a specific physical interaction: $\mathbf{Y}_{st}$ represents feed coupling, $\mathbf{Y}_{ss}$ models mutual coupling among elements, and $\mathbf{Y}_{rs}$ describes the wireless propagation channel between the DMA and the users. 
Applying Ohm’s law and network reduction yields the input-output relationship~\cite{Williams2022Electromagnetic,Ramirez2022Performance},

\begin{equation}
\mathbf{j}_r \approx 
(\mathbf{Y}_r + \mathbf{Y}_{rr})^{-1}
\mathbf{Y}_{rs}(\mathbf{Y}_s + \mathbf{Y}_{ss})^{-1}\mathbf{Y}_{st}\,\mathbf{j}_t.
\label{eq:DMA_IOrelation}
\end{equation}

Comparing \eqref{eq:DMA_IOrelation} with the unified system model in \eqref{eq:unified_system_model} identifies the end-to-end effective channel as \(\mathbf{H}_{\mathrm{eff},k} = (\mathbf{Y}_r + \mathbf{Y}_{rr})^{-1} \mathbf{Y}_{rs}(\mathbf{Y}_s + \mathbf{Y}_{ss})^{-1}\mathbf{Y}_{st}\). Although this expression relates input and output magnetic currents (rather than voltages or complex gains), it is directly equivalent to the complex channel gains because the magnetic currents are proportional to the port voltages through the network admittance parameters, and the load term \((\mathbf{Y}_r + \mathbf{Y}_{rr})^{-1}\) converts the induced magnetic currents at the receivers to measurable output quantities. This formulation is therefore fully consistent with the standard complex channel gain representation. This approximation assumes weak mutual coupling between the transmitter, DMA, and receiver sections, which is valid when the transmitter and receiver are well separated, and the DMA elements are designed to minimize cross-coupling.

\begin{figure*}[ht]
 \centering
 \subfloat[Without mutual coupling ]{\label{fig:dma_corr_NoMC} \includegraphics[width=0.48\linewidth]{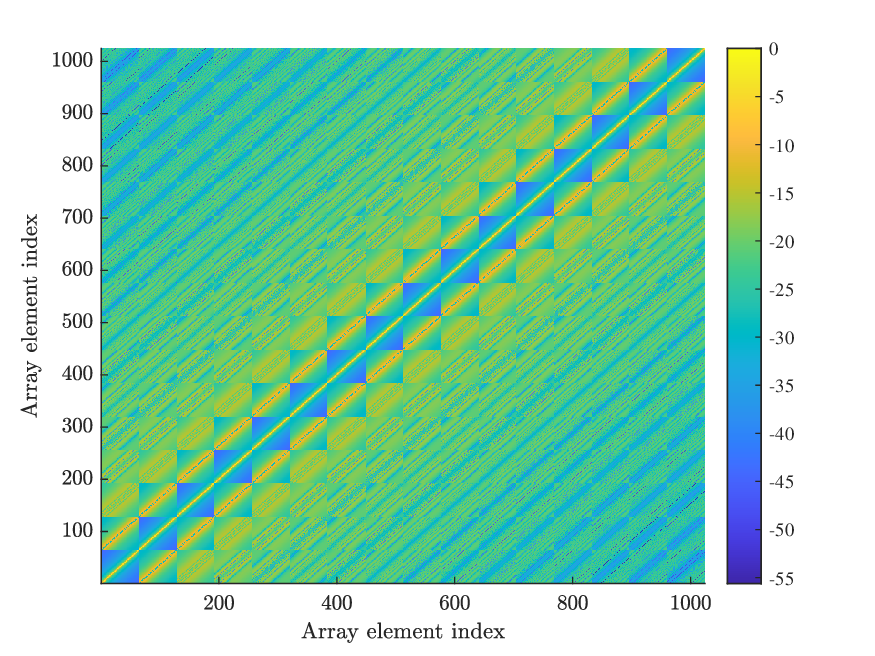}}%
 \hfill
 \subfloat[With mutual coupling ]{\label{fig:dma_corr_MC} \includegraphics[width=0.48\linewidth]{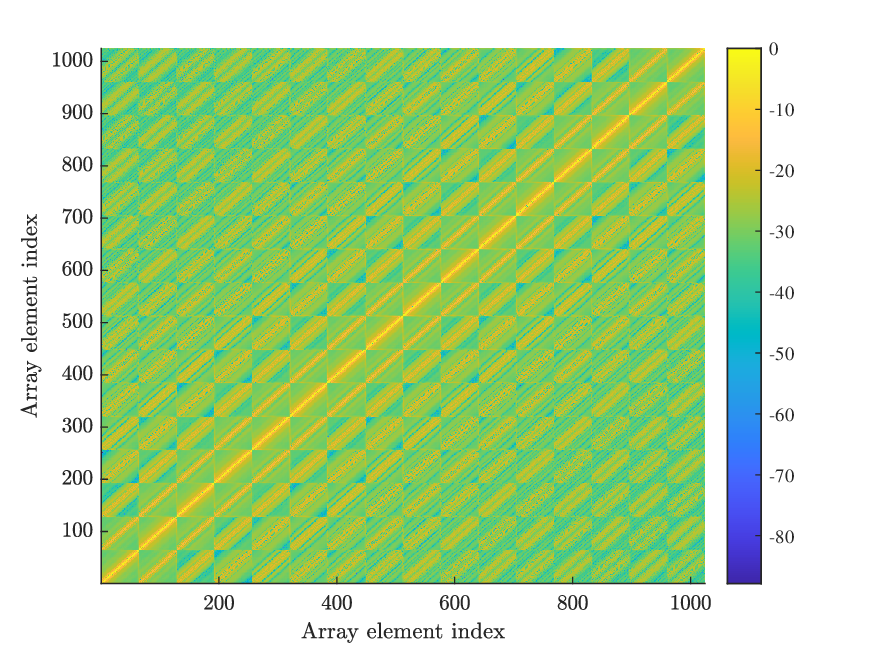}}%
\caption{Spatial correlation magnitude of the DMA effective channel. (a) Without mutual coupling, the matrix is nearly identity, showing no structure. (b) With mutual coupling, significant correlations appear, influencing system design.}
 \label{fig:DMA_SpCo}
 \vspace{-5mm}
\end{figure*}

The impact of mutual coupling, represented by the off-diagonal elements of $\mathbf{Y}_{ss}$ in Eq.~\eqref{eq:DMA_Ymatrix}, is illustrated in Fig.~\ref{fig:DMA_SpCo}. These figures show the magnitude of the spatial correlation matrix \(\mathbf{R}_{\mathrm{DMA}} = \mathbb{E}[\mathbf{H}_{\mathrm{eff}}\mathbf{H}_{\mathrm{eff}}^\HH]\) of the DMA effective channel from Eq.~\eqref{eq:DMA_IOrelation}. Fig.~\ref{fig:dma_corr_NoMC} depicts the idealized case without mutual coupling ($\mathbf{Y}_{ss}$ diagonal), yielding an approximately identity-scaled correlation matrix, indicating weak interaction among $N_{\text{fed}} = 16$ waveguides, each containing $N_w = 64$ meta-atoms. In contrast, Fig.~\ref{fig:dma_corr_MC} includes mutual coupling ($\mathbf{Y}_{ss} \neq 0$), revealing significant correlations, especially among elements on the same or adjacent waveguides. This coupling substantially modifies the channel’s spatial characteristics, affecting channel capacity, channel estimation, and beamforming design. Neglecting it, as in Fig.~\ref{fig:dma_corr_NoMC}, results in overly optimistic and physically inconsistent performance predictions.

Unlike the conventional electric-domain definition \(\mathbf{I} = \mathbf{Y}\mathbf{V}\) (where \(\mathbf{I}\) and \(\mathbf{V}\) are the electric current and voltage vectors, and \(\mathbf{Y}\) is the admittance matrix, as introduced in Sec.~\ref{sec:multiport_network}), the formulation in~\cite{Williams2022Electromagnetic} uses magnetic variables and writes \(\mathbf{v} = \mathbf{Y}\mathbf{j}\). This choice is deliberate: the DMA elements are modeled as magnetic dipoles, and adopting magnetic voltages and currents simplifies the analytical derivations while remaining fully equivalent to the standard electric current formulation.

Moreover, since both DMAs and RHSs can be regarded as holographic surfaces that approximate continuous apertures, their channels can also be modeled in the wavenumber (Fourier) domain. In this view, the surface field is expanded as a superposition of plane waves whose spectral components determine the radiated field distribution. This Fourier plane-wave channel model, adopted in recent \ac{hmimo} studies such as \cite{Gong2024Holographic,Zhang2025Holographic}, captures the spatial \acp{dof} and the transition between near- and far-field regions. The mathematical details of this representation are provided in Sec.~\ref{subsec:fourier_repr}.

Another model in~\cite{Yang2024Extremely} used a spherical-wave geometric channel model for XL-DMAs. Let $P$ denote the number of propagation paths, $\alpha_p$ the complex gain of the $p$-th path, and $\mathbf{b}(\theta_p, \phi_p, r_p)$ the near-field array response vector determined by spherical-wave geometry. The wireless channel vector is then expressed as
\begin{equation}
\mathbf{h}_{w,k} = \sqrt{\frac{N_w N_{\text{fed}}}{P}} \sum_{p=1}^{P} \alpha_p\, \mathbf{b}(\theta_p, \phi_p, r_p).
\end{equation}
Each element of $\mathbf{b}(\theta_p, \phi_p, r_p)$ is given by
\begin{equation}
b_{m,n}(\theta_p, \phi_p, r_p) = \frac{1}{\sqrt{N_w N_{\text{fed}}}}
e^{-j\frac{2\pi}{\lambda}\left(r_{m,n} - r_p\right)},
\end{equation}
where $r_{m,n}$ is the distance between the $(m,n)$-th element and the $p$-th scatterer.

\subsubsection{Spatially reconfigurable arrays}
\label{sec:models_ssra}
This section models a wireless communication system employing spatially reconfigurable antennas, where the radiating elements can dynamically modify their spatial configuration in two- or three-dimensional space. The instantaneous spatial geometry of the antenna array is defined as
\begin{equation}
\mathcal{R} = \{\mathbf{r}_1, \mathbf{r}_2, \ldots, \mathbf{r}_{N_{\mathrm{in}}}\} \subset \mathbb{R}^3,
\end{equation}
where $\mathbf{r}_n$ denotes the position vector of the $n$-th radiating element and $N_{\mathrm{in}}$ is the number of active/selected ports.

In the unified model of \eqref{eq:unified_system_model}, the reconfigurable parameters are the element positions (or geometry). The EM processing is inherently embedded in the geometry, so the effective channel is directly
\(
\mathbf{H}_{\mathrm{eff},k} = \mathbf{H}_{w,k}(\mathcal{R}),
\)
where $\mathbf{H}_{w,k}(\mathcal{R}) \in \mathbb{C}^{U \times N_{\mathrm{in}}}$ is the geometry-dependent propagation matrix. 

The work in~\cite{New2024Tutorial} summarizes several modeling approaches for \ac{fas}, including correlation-based and geometry-based formulations. Furthermore, a geometry-based multipath channel model introduced in~\cite[Eq.~(4)]{Qin2024Antenna} describes the channel characteristics as a function of the antenna element positions, thereby capturing the impact of array geometry on the propagation environment.

Recent work in \cite{New2024Channel} extends the Fourier plane-wave framework of Sec.~\ref{subsec:fourier_repr} to \ac{fas} by enforcing the \ac{em}-compliant channel constraint in \eqref{eq:helmholtz}. Using the same Fourier expansion in \eqref{eq:wavenumber_support}–\eqref{eq:discrete_channel} and the wavenumber-domain PSD in \eqref{eq:psd}, the \ac{fas} effective channel is represented as a superposition of up- and down-going plane waves over \(\mathcal{D}(\kappa)\).

This representation is discretized similarly to the finite-array case in \eqref{eq:discrete_channel}, enabling practical channel sampling and reconstruction. Because the \ac{fas} observes the channel only over a finite spatial aperture, the channel becomes non-band-limited, violating the ideal Nyquist condition implied by \(\mathcal{D}(\kappa)\). Oversampling beyond half-wavelength spacing is therefore required to reduce reconstruction error.

In~\cite{Ramirez2025Metasurface}, \ac{fas} was implemented using a \ac{dma} modeled as a multiport network based on the admittance framework of Sec.~\ref{sec:multiport_network}. Each waveguide port, radiating element, and receiving device is treated as a network port characterized by magnetic voltages and currents. Specifically, $\mathbf{v}_t,\mathbf{j}_t \in \mathbb{C}^{N_\text{fed}\times1}$ denote the \ac{rf} chain interface variables; $\mathbf{v}_s,\mathbf{j}_s \in \mathbb{C}^{N_\mathrm{A}\times1}$, with $N_\mathrm{A} = N_w\times N_\text{fed}$, correspond to the radiating elements; and $\mathbf{v}_r,\mathbf{j}_r \in \mathbb{C}^{U\times1}$ describe the single-antenna users. The mutual admittance matrices $\mathbf{Y}_{mn}$ capture both guided-wave and radiative coupling among all ports~\cite[Eq.~(1)]{Ramirez2025Metasurface}.  

The effective channel from the \ac{rf} input to the device currents is  
\begin{equation}\label{eq:end_to_end}
\mathbf{j}_r = (\mathbf{Y}_r \mathbf{I}_M + \mathbf{Y}_{rr})^{-1} 
\mathbf{Y}_{rs} (\mathbf{Y}_s + \mathbf{Y}_{ss})^{-1} 
\mathbf{Y}_{st} \mathbf{j}_t,
\end{equation}
where $\mathbf{Y}_{st}$ and $\mathbf{Y}_{ss}$ are defined in~\cite[Eq.~(3)--(4)]{Ramirez2025Metasurface}.

The corresponding radiated \ac{em} field, useful for characterizing patterns~\cite[Eq.~(31)]{Ramirez2025Metasurface}, is obtained as a superposition of contributions from each element, given by  
\begin{equation}\label{eq:hrad}
\mathbf{h}_\text{rad}(\mathbf{r}) = -j\,2\,l_m\,\omega\epsilon_0 
\sum_{i=1}^{N_w} \mathbf{G}(\mathbf{r},\mathbf{r}_i)\hat{\mathbf{z}}(\mathbf{j}_s)_i,
\end{equation}
where $\mathbf{G}(\mathbf{r},\mathbf{r}_i)$ is the Green’s function in~(Eq.~\eqref{eq:green}), $\hat{\mathbf{z}}$ is the unitary vector along the z-direction, and $l_m$ is the effective dipole length~\cite[Eq.~(31)--(34)]{Ramirez2025Metasurface}.

For fluid and movable antenna systems, the radiated power is typically normalized with respect to the number of active ports or the position-dependent channel gain. A common normalization assumes equal power allocation across the selected ports,
\(
\sum_{k} \| \mathbf{F}_{\mathrm{BB},k}\|_F^2 \leq P_{\mathrm{max}},
\)
where antenna movement does not introduce additional insertion losses beyond the hardware implementation~\cite{Wong2020Fluid,Zhu2023Movable}. For FIM, a geometry-driven representation is adopted in~\cite{An2025Flexible}, where each radiating element is independently adjusted along its normal direction, and the surface shape is characterized by a deformation vector representing the reversible unilateral deformation of the array. A different implementation is presented in~\cite{Ding2025Flexible}, which employs multiple pinching antennas activated on a single waveguide. The channel is modeled using a spherical-wave formulation. Due to the shared waveguide, the signal radiated by each pinching antenna is a phase-shifted version of the common feed signal, with the phase shift determined by the antenna's position along the waveguide and the waveguide wavelength. Consequently, the overall transmit signal vector incorporates these phase relationships, distinguishing the pinching-antenna configuration from conventional MISO arrays.

\subsubsection{Parasitic arrays}
\label{Parasitic_model}

Parasitic arrays are antenna systems in which only a subset of elements is actively driven, while the remaining elements are passive and excited via mutual coupling. The passive elements are terminated with tunable reactive loads that control the induced currents and, consequently, shape the radiation pattern~\cite{Nikkhah2014Theory}.

For a parasitic array at the transmitter with \( N_\mathrm{A} \) total antenna elements, \( N_{\mathrm{in}} \) elements are actively fed and the remaining elements are parasitic. Let \( \mathbf{Z} \in \mathbb{C}^{N_\mathrm{A} \times N_\mathrm{A}} \) denote the mutual coupling matrix, which depends on the element spacing and geometry, and let \( \operatorname{diag}(\mathbf{z}_l) \) denote the diagonal matrix of tunable load impedances. Using the generalized Ohm's law, the vector of currents at the antenna ports is
\begin{equation}
\mathbf{i} = (\mathbf{Z} + \operatorname{diag}(\mathbf{z}_l))^{-1} \mathbf{v},
\label{eq:i_parasitic}
\end{equation}
where \( \mathbf{v} \in \mathbb{C}^{N_\mathrm{A} \times 1} \) is the excitation voltage vector (nonzero only at driven ports). The EM-domain processing matrix is
\begin{equation}
\mathbf{F}_{\mathrm{EM}}(\mathbf{z}_l) = (\mathbf{Z} + \operatorname{diag}(\mathbf{z}_l))^{-1},
\label{eq:F_EM_parasitic}
\end{equation}
which maps the \( N_{\mathrm{in}} \) driven ports to the \( N_\mathrm{A} \) antenna currents. Substituting into the unified model of \eqref{eq:unified_system_model}, the effective channel is
\(
\mathbf{H}_{\mathrm{eff},k} = \mathbf{H}_{w,k} \mathbf{F}_{\mathrm{EM}}(\mathbf{z}_l),
\)
where \( \mathbf{H}_{w,k} \in \mathbb{C}^{U \times N_{\mathrm{A}}} \) is the wireless propagation channel. The wireless propagation channel can be modeled using Maxwell's equations, circuit-theoretic formulations, or geometric multipath models, as in \cite[eqs. (5)--(7)]{Li2017MIMO} and \cite[Eq.~(5)]{Alexandr2014Precoding}.

\begin{figure*}[ht]
 \centering
 \subfloat[Spectral efficiency vs. number of parasitic elements for different spacings $\Delta$.]{%
   \label{fig:parasitic_spacing}%
   \includegraphics[width=0.48\linewidth]{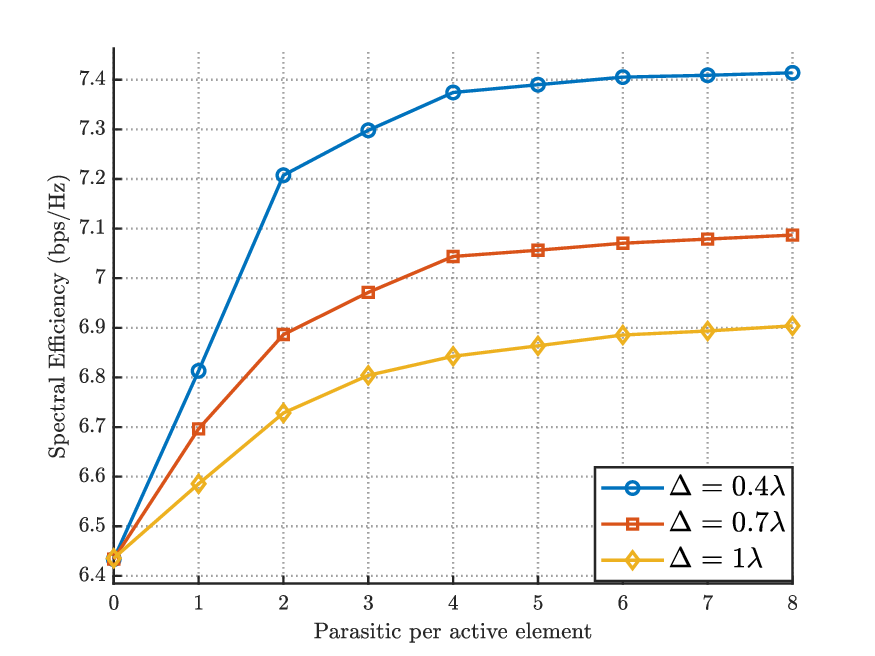}%
 }%
 \hfill
 \subfloat[Spectral efficiency vs. number of parasitic elements for different total aperture sizes $D_x$.]{%
   \label{fig:parasitic_fixapertures}%
   \includegraphics[width=0.48\linewidth]{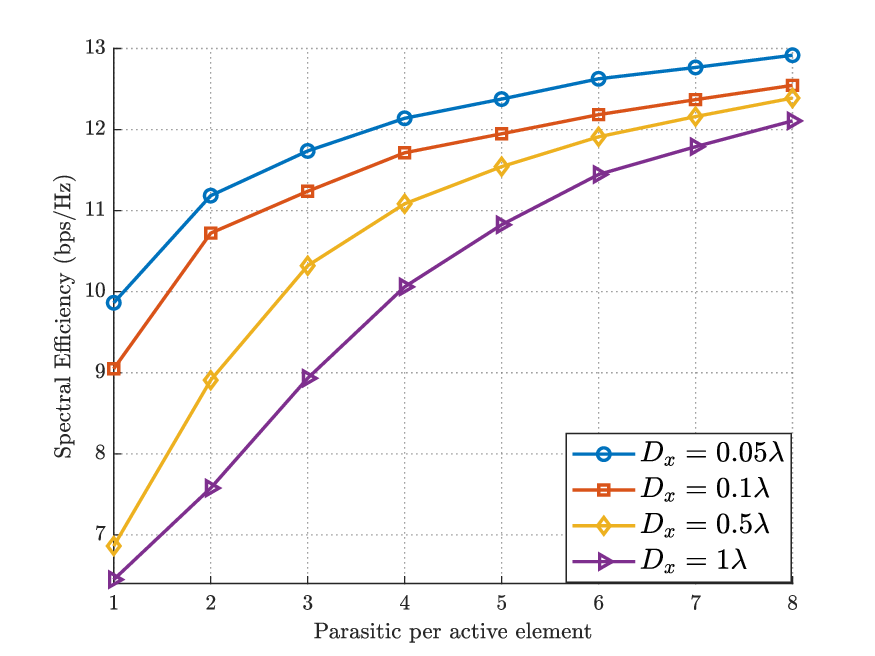}%
 }%
 \caption{Parasitic array performance with optimal reactive loading.}
 \label{fig:Parasitic_Perf}
 \vspace{-2mm}
\end{figure*}

The impact of design choices on system performance is shown in Fig.~\ref{fig:Parasitic_Perf}. The spectral efficiency versus the number of passive elements per active feed is depicted in Fig.~\ref{fig:parasitic_spacing} for various fixed inter-element spacings \( \Delta \). For a given spacing, adding parasitic elements initially improves performance due to enhanced beamforming from optimally tuned reactive loads, but gains diminish as the number of elements increases. The benefit is greatest at smaller spacings (e.g., \( \Delta = 0.4\lambda \)), where strong mutual coupling, captured by the off-diagonal entries of \( \mathbf{Z} \) in \eqref{eq:i_parasitic}, enables superdirective effects. Fig.~\ref{fig:parasitic_fixapertures} considers spectral efficiency versus the number of parasitic elements for different total array widths \( D_x = (N_{\mathrm{A}} - 1)\Delta \) along the x-axis. Efficiency increases monotonically with the number of parasitics for all apertures, with diminishing returns at higher numbers. Smaller apertures (\( D_x = 0.05\lambda, 0.1\lambda \)) achieve higher efficiency than larger apertures, reflecting the enhanced beamforming possible when closely spaced elements exploit mutual coupling. These results highlight the importance of co-designing antenna geometry and tunable loads, as assumptions of weak coupling can lead to inaccurate predictions.

\subsubsection{Multi-layer reconfigurable arrays}
\label{sec:models_stacked}

Multi-layer reconfigurable antennas, such as \ac{sim}~\cite{An2023Stacked,An2024Stacked,Liu2025Stacked} and stacked flexible intelligent metasurfaces (Stacked-FIM)~\cite{Magbool2025Stacked}, employ multiple closely spaced programmable layers to manipulate \ac{em} waves directly in the wave domain. Each layer performs a controllable \ac{em} transformation, typically a phase or amplitude modulation, on the impinging field, enabling near-instantaneous precoding and combining at the speed of light.

The generic structure consists of $L$ programmable metasurface layers, each containing $N_\ell$ meta-atoms. The \ac{em} propagation between adjacent layers $\ell$ and $\ell+1$ is characterized by the channel coefficient matrix $\mathbf{W}_{\ell,\ell+1} \in \mathbb{C}^{N_{\ell+1} \times N_{\ell}}$, where $N_0 = N_{\mathrm{in}}$ denotes the number of input ports and $N_L$ is the number of elements in the last layer.

The matrix $\mathbf{W}_{0,1}$ was discussed in Sec.~\ref{subsec:models_metasurface}, while $\mathbf{W}_{\ell-1,\ell}$ for $\ell \in \{2, \dots, L\}$ is typically defined using Rayleigh-Sommerfeld's diffraction theory~\cite{Lin2018All,Yao2024Channel}. Each layer applies an element-wise programmable phase and amplitude response, modeled as
\begin{equation}
\boldsymbol{\Gamma}_{\ell} = \operatorname{diag}\!\left(\gamma_{\ell,1}, \gamma_{\ell,2}, \dots, \gamma_{\ell,N_\ell}\right),
\end{equation}
where 
\begin{equation}
\gamma_{\ell,n} = \alpha_{\ell,n} e^{j \phi_{\ell,n}}, \quad 
\phi_{\ell,n} \in [0,2\pi),
\end{equation}
represents the \ac{em} transmission coefficient of the $n$-th meta-atom in the $\ell$-th metasurface layer.

The \ac{em}-domain mapping from the input ports to the radiating aperture can therefore be expressed recursively as
\begin{equation}
\mathbf{F}_{\mathrm{EM},k} = \boldsymbol{\Gamma}_{L} \mathbf{W}_{L-1,L} \cdots \boldsymbol{\Gamma}_{1} \mathbf{W}_{0,1},
\end{equation}
where the subcarrier dependence is captured implicitly through the frequency-selective nature of the meta-atom responses.

The wireless propagation from the outermost layer to the receivers is represented by \(\mathbf{H}_{w,k} \in \mathbb{C}^{U \times N_L}\). In the literature, this channel has been modeled in various ways: \cite{Yao2024Channel,An2023Stacked,Papazafeiropoulos2025Performance} adopt correlation-based channels, while \cite{Jia2024Stacked} considers a near-field geometric model. For a single-antenna receiver, this reduces to \(\mathbf{H}_{w,k} \in \mathbb{C}^{N_L \times 1}\). Substituting into the effective channel definition yields
\(
\mathbf{H}_{\mathrm{eff},k}(\mathbf{F}_{\mathrm{EM},k}) = \mathbf{H}_{w,k} \mathbf{F}_{\mathrm{EM},k}.
\)

Beyond wave-optics models, \ac{sim} can be characterized using circuit and multiport network theory~\cite{Nerini2024Physically}. In particular, the scattering-parameter formulation in~\cite{Yahya2025T} models the \ac{sim} as a cascaded network of $L$ \ac{ris} layers with $N$ cells each. For this configuration, the incident and reflected waves at the input and output ports are related through the global scattering matrix $\mathbf{S}_I \in \mathbb{C}^{2N_\mathrm{p} \times 2N_\mathrm{p}}$ as

\begin{equation}
\begin{bmatrix}
\mathbf{b}_{i1} \\ \mathbf{b}_{oL}
\end{bmatrix} = \mathbf{S}_I 
\begin{bmatrix}
\mathbf{a}_{i1} \\ \mathbf{a}_{oL}
\end{bmatrix},
\end{equation}
where $\mathbf{b}_{i1}$ and $\mathbf{b}_{oL}$ represent the reflected waves at the input and output ports, and $\mathbf{a}_{i1}$ and $\mathbf{a}_{oL}$ represent the corresponding incident waves~\cite[Eq.~(1)]{Yahya2025T}. The equivalent S-parameters of the cascaded network can be computed recursively using the non-linear operator $S(\cdot)$~\cite[Eq.~(2)]{Yahya2025T}, yielding the effective channel,
\begin{equation}
\mathbf{H}_{\mathrm{eff},k}(\mathbf{F}_{\mathrm{EM},k}) = \mathbf{H}_{w,k} \mathbf{S}_{I,21} \mathbf{W}_{0,1},
\end{equation}
where $\mathbf{S}_{I,21}$ is the $(2,1)$-th block of $\mathbf{S}_I$, and $\mathbf{H}_{w,k}$, $\mathbf{W}_{0,1}$ denote the effective transmission matrices from the last \ac{sim} layer to the receiver and from the transmitter to the first \ac{sim} layer, respectively~\cite[Eq.~(6)]{Yahya2025T}. To improve tractability, a T-parameters formulation is introduced, where the effective channel is expressed as
\begin{equation}
\mathbf{H}_{\mathrm{eff},k}(\mathbf{F}_{\mathrm{EM},k}) = \mathbf{H}_{w,k} \mathbf{T}_{I,22}^{-1} \mathbf{W}_{0,1},
\end{equation}
with $\mathbf{T}_{I,22}$ the $(2,2)$-th block of the T-parameters matrix $\mathbf{T}_I$, which is computed recursively from the T-matrices of each \ac{ris} layer and the transmission medium~\cite[eqs.~(11)]{Yahya2025T}. Compared to the S-parameters approach, the T-parameters formulation is more compact and requires only a single matrix inversion, making the optimization of \ac{sim} layers simpler and computationally efficient~\cite[Eq.~(14)]{Yahya2025T}.

The radiated power for \ac{sim} is given by
\begin{equation}
P_{k}(\mathbf{F}_{\mathrm{EM},k},\mathbf{F}_{\mathrm{RF},k},\mathbf{F}_{\mathrm{BB},k}) = \| \mathbf{F}_{\mathrm{EM},k}\mathbf{W}_{0,1}\mathbf{F}_{\mathrm{RF},k}\mathbf{F}_{\mathrm{BB},k}\|_{\mathrm{F}}^{2},
\label{eq:sim_radiated_power}
\end{equation}
which couples the \ac{sim} configuration to the power constraint through the EM-domain transmission response \( \mathbf{F}_{\mathrm{EM},k} \)~\cite{An2023Stacked,Heath2025Tri}.

A model for reconfigurable electromagnetic structures (REMSs), such as reflectarrays and \acp{ris}, which accounts for real-world effects including inter-antenna coupling, polarization, ohmic and matching losses, non-reciprocal materials, and antenna noise, was proposed in~\cite{Stutz2025Efficient}. This approach enables efficient computation of the full far-field pattern for arbitrary configurations of the reconfigurable elements, providing a physically consistent framework for REMS design and control.

In this model, the radiating structure acts as the interface between the matching network, reconfigurable elements, and the far-field. It relates the outgoing power waves, denoted by $\mathbf{b}_R$ and $\mathbf{a}_F$, to the incoming waves $\mathbf{a}_R$ and $\mathbf{b}_F$ through a scattering operator $\mathbb{S}$:

\begin{equation}
\begin{bmatrix}
\mathbf{b}_R \\
\mathbf{a}_F
\end{bmatrix}
=
\underbrace{
\begin{bmatrix}
\mathbb{S}_{RR} & \mathbb{S}_{RF} \\
\mathbb{S}_{FR} & \mathbb{S}_{FF}
\end{bmatrix}
}_{\triangleq \mathbb{S}}
\begin{bmatrix}
\mathbf{a}_R \\
\mathbf{b}_F
\end{bmatrix}.
\end{equation}

Each sub-block has a specific physical interpretation:  
$\mathbb{S}_{RR}$ models inter-element coupling,  
$\mathbb{S}_{RF}$ the receiving operator (field to ports),  
$\mathbb{S}_{FR}$ the transmitting operator (ports to field),  
and $\mathbb{S}_{FF}$ the far-field scattering operator. This representation strengthens the link between deterministic \ac{em} theory and system-level modeling,

\subsection{Wideband and frequency-selective considerations}
\label{subsec:wideband}
In wideband operation at \ac{mmwave} and sub-\ac{thz} frequencies, the assumption of frequency-flat
beamforming breaks down. The use of frequency-independent phase shifters causes the
radiation beam to deviate from the intended direction across different subcarriers, an
effect known as beam squint or beam splitting~\cite{Najjar2023Hybrid, 
Zhu2025Flexible}. This results in substantial array gain loss and limits the effective
utilization of available bandwidths~\cite{Najjar2023Hybrid}. More generally, the amplitude, phase, and frequency responses of reconfigurable elements are mutually coupled, an effect referred to as amplitude-phase-frequency coupling (APFC)~\cite{Ozen2025Beam}, which further complicates wideband \ac{ris} beamforming.

A common hardware solution to mitigate beam squint is the use of \ac{ttd} elements, which provide frequency-dependent phase shifts that align the beam
across the entire band~\cite{Domae2025Millimeter, Li2024Holographic}. \ac{ttd} lines
enable controllable beam-squint and beam-split ranges, and have been shown to
effectively restore beamforming gain in wideband \ac{ris}-aided systems~\cite{Li2024Holographic,
Li2024User}. However, \ac{ttd} units often suffer from high power consumption, insertion
loss, and hardware complexity, motivating the development of alternative mitigation
techniques such as subband-based phase design, delay-phase precoding (DPP), and active
\ac{ris} architectures~\cite{Najjar2023Hybrid, Butt2025Beam}.

Beyond beam squint, wideband operation also exacerbates frequency-dependent
mutual coupling between closely spaced elements. Conventional circuit-theoretic models
with frequency-independent coupling matrices become inaccurate, necessitating a
coupling-aware wideband framework that captures frequency-dependent mutual coupling
across multiple subbands~\cite{Ibrahim2026Stacked, Gueuning2025Broadband}. Recent
work has proposed a circuit-based, coupling-aware framework for \ac{sim} that maintains a consistent phase response across frequencies,
enabling significant performance gains over single-layer \ac{ris} in wideband multiuser
MIMO systems~\cite{Ibrahim2026Stacked}. More generally, the interaction between beam
squint, frequency-dependent mutual coupling, and the frequency-selective response of
meta-atoms represents a key challenge for physically consistent wideband modeling,
and remains an active research area.

\subsection{Measurement validation and prototype studies}

Physically consistent channel models must be validated against real-world measurements to be useful for system design. Recent measurement campaigns have revealed significant gaps between idealized theoretical models and practical \ac{ris} behavior. For example, Weinberger et al.~\cite{Weinberger2024Validating} deployed an \ac{ris} prototype on a turntable in an anechoic chamber and measured the angle-dependent reflection coefficient. They found that non-perpendicular incidence can introduce additional attenuation of up to 14.5\,dB and phase deviations that vary per element–effects not captured by standard point-scatterer models. The same study proposed a refined model with angle-dependent coefficients (approximated by Lorentzian magnitude and polynomial phase functions) that aligns much more closely with measurements~\cite{Weinberger2024Validating}.

Other prototype studies have validated specific architectural claims: an active liquid-crystal \ac{ris} at 27\,GHz demonstrated significant \ac{snr} improvements in both laboratory and corridor scenarios~\cite{Guirado2025}; a transmissive $16\times16$ \ac{ris} achieved 7\,dB received power gain when deployed to overcome obstacles~\cite{Zhang2024Design}; and the first experimental \ac{bdris} prototype confirmed that inter-element connections are beneficial, but also that hardware constraints (e.g., limited load states) strongly influence performance~\cite{Tapie2025Beyond}. The
\ac{mc} in \ac{ris} has been experimentally characterized using a 1‑bit mm‑Wave prototype, where a scattering‑matrix model trained from a single full‑wave simulation significantly improved prediction accuracy over coupling‑unaware models~\cite{Zheng2024Mutual}.
For channel modeling specifically, a recent study systematically compared \ac{em} simulations, ray-tracing simulations, and over-the-air measurements for a lossy anomalous reflector, finding good agreement and validating the consistency of the macroscopic \ac{ris} model~\cite{Hao2025Modeling}. Beyond RIS, a survey of reconfigurable antenna prototypes for 6G systems has documented a liquid‑metal‑based pattern‑reconfigurable antenna array prototype that demonstrates substantial gains in array gain and received signal power in over‑the‑air experiments, while also discussing current hardware limitations such as slow actuation speed\cite{Zheng2025Reconfigurable}. Over-the-air functional testing methodologies have also been developed for \ac{ris} at 28\,GHz, enabling rapid verification of continuous amplitude and phase control~\cite{Krasov2025ris}.

Beyond \ac{ris}, recent experimental efforts have extended to other reconfigurable architectures. For \ac{dma}, a complete K-band end-to-end wireless prototype has been validated with over-the-air experiments, demonstrating the ability to simultaneously steer a beam toward a desired transmitter and a null toward an interfering jammer, achieving up to 43\,dB discrimination by leveraging strong inter-element mutual coupling~\cite{Yven2025DMA}. Using software-defined radios transmitting QPSK-OFDM waveforms, this platform bridges theory and practice, revealing algorithmic and technological challenges for future real-time \ac{dma}-based systems. Complementary \ac{dma} prototypes include a 32-element X-band leaky-wave design that experimentally generated seven distinct beam patterns via PIN-diode control, with a wide beam coverage of 136° and an average gain of 6.3\,dBi, while also demonstrating direction-of-arrival estimation for one or two simultaneous sources with an error as low as 1°~\cite{Jain2026DMA}. Another experimental work has applied programmable coded-metasurfaces for computational-imaging-driven channel characterization at microwave frequencies, demonstrating direction-of-arrival estimation within a computational imaging framework~\cite{Yurduseven2025DMA}. A comprehensive 2D waveguide-fed metasurface model has further been validated against full-wave \ac{em} simulations, confirming its ability to accurately characterize both electric and magnetic dipole responses in the near- and far-field regimes~\cite{Gavriilidis2026Modeling}.

\ac{sim} have also seen significant modeling and experimental progress. A multiport network model based on scattering parameters has been developed and validated against \ac{em} simulations, capturing realistic circuit characteristics such as multiport coupling, circuit losses, and non-ideal hardware behavior~\cite{Pettanice2026SIM}. This framework has been shown to provide accurate performance predictions for both communication and sensing scenarios. On the experimental front, a dedicated \ac{sim} hardware platform has been designed and built to evaluate integrated sensing and communication tasks, where the \ac{sim} is jointly optimized to minimize the Cramér–Rao bound for target estimation under signal-to-interference-plus-noise ratio constraints while maintaining multi-user communication quality~\cite{Wang2024SIM}. Additionally, early \ac{sim} prototypes include a programmable diffractive deep neural network operating at microwave frequencies, where each meta-atom is integrated with amplifier chips to act as an active artificial neuron, successfully performing inference tasks such as image classification, mobile communication coding–decoding, and real-time multi-beam focusing directly in the wave domain~\cite{Liu2022Programmable}. A more recent study has explored functional coding and data-enhanced deep unfolding detection for \ac{sim}-enabled transceivers~\cite{Bohrium2026SIM}.

\ac{fas} have transitioned from theoretical concepts to multiple validated physical prototypes. A comprehensive survey of early-stage \ac{fas} hardware has documented liquid-metal fluid antennas, mechanically movable elements, pixel-reconfigurable antennas, and meta-fluid designs, with initial measurements confirming potential diversity and outage-probability advantages~\cite{Tong2025FAS}. A more recent advancement is the electromagnetically reconfigurable fluid antenna system, which uses liquid metal to dynamically adjust the radiation pattern without physical movement or rotation; this work covers the full development cycle from antenna design and channel modeling to optimization algorithms, prototype fabrication, and experimental measurements~\cite{Wang2025ERFAS}. A parallel effort has developed a broadband L-probe-fed pixel-reconfigurable antenna capable of $\mu$s-range switching speeds, specifically designed for high-speed \ac{fas} applications~\cite{Zhang2025Broadband}.

Progress has also extended to simultaneous transmitting and reflecting \ac{ris}. The first experimental STAR-RIS designed for integrated communication, computing, and sensing was fabricated and tested at 26\,GHz, demonstrating simultaneous reflection-based communication and transmission-based Fourier transforms for angle-of-arrival estimation within a single aperture~\cite{Omam2025STAR}.

Finally, for pinching-antenna systems, the literature has focused on system modeling and performance analysis, but recent comprehensive tutorials and foundational modeling studies have accelerated the path toward practical validation. A detailed tutorial on PASS has outlined signal models, hardware models, power radiation models, and pinching antenna activation methods, establishing the theoretical groundwork for experimental prototyping~\cite{Liu2025PASS}. Physics-based hardware models have further refined the understanding of PASS, where pinching antennas are modeled as open-ended directional couplers using coupled-mode theory, providing a foundation for future prototype development and experimental verification~\cite{Wang2025Modeling}. An overview of emerging \ac{sim} technologies has also surveyed existing prototypes and their applications~\cite{An2025Emerging}.

These studies collectively show that while physically consistent models are necessary, they are not sufficient; measurement-informed refinements (angle-dependent coefficients, insertion losses, mutual coupling) are essential for bridging the gap between theory and practice.  The rapid expansion of experimental work across DMA, \ac{sim}, FAS, STAR-RIS, and PASS indicates that the field is maturing toward prototypes that can validate the theoretical promises of these reconfigurable architectures. Future work should extend such validation to emerging hybrid architectures, where experimental data remain scarce.

\section{Signal processing for emerging architectures}
\label{Sec:IV}

We now address the signal processing algorithms for reconfigurable wireless systems, building on the physically consistent channel models of Sec.~\ref{Sec:III}. The constraints imposed by mutual coupling, near-field wavefront curvature, and hardware impairments fundamentally reshape classical approaches to estimation, beamforming, and detection. Accordingly, this section covers these three areas, highlighting the role of physical consistency in algorithm design and the resulting performance trade-offs.

\subsection{Channel estimation}
Channel estimation for reconfigurable architectures must account for high dimensionality, hardware constraints, and physically consistent channel structure. We organize estimation techniques into three categories: linear, sparse, and AI-aided approaches.

\subsubsection{Linear estimation}
Conventional linear estimators such as \ac{ls} and \ac{lmmse} serve as baselines, but they are poorly suited for reconfigurable large-array systems: \ac{ls} requires orthogonal pilots across all ports (prohibitive overhead), while \ac{lmmse} needs second-order statistics that are hard to acquire and its complexity scales badly with array dimension~\cite{Jensen2020Optimal}.

\begin{figure*}[ht]
 \centering
 \subfloat[Effect of neglecting mutual coupling]{%
   \label{fig:nmse_dma}%
   \includegraphics[width=0.48\linewidth]{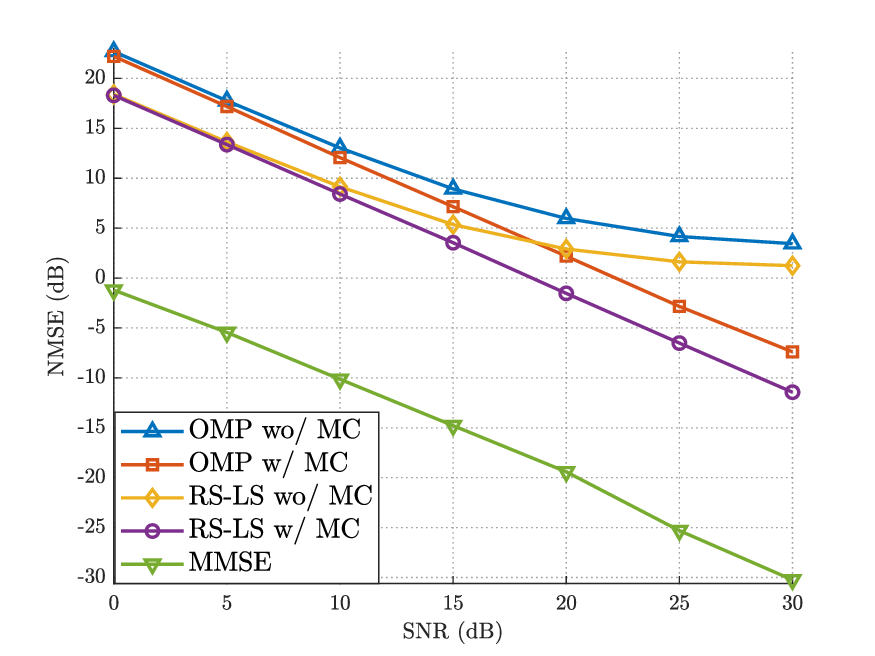}%
 }%
 \hfill
 \subfloat[Effect of neglecting spherical‑wavefront curvature]{%
   \label{fig:nf_vs_ff_mismatch}%
   \includegraphics[width=0.48\linewidth]{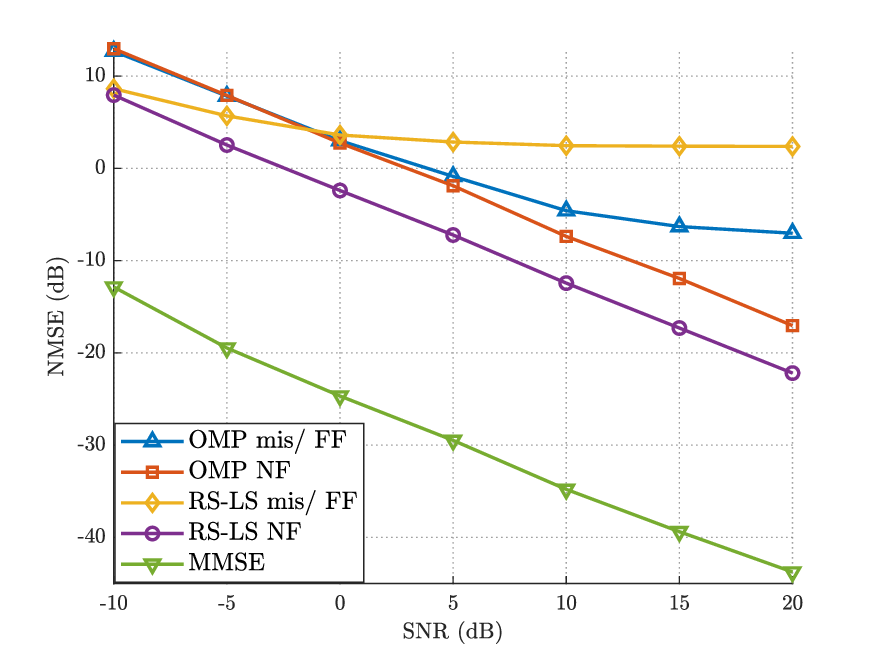}%
 }%
 \caption{The \ac{nmse} versus \ac{snr} for channel estimation in a DMA‑based system.}
 \label{fig:estimation}
 \vspace{-2mm}
\end{figure*}

For reconfigurable architectures, more efficient linear estimators exploit physical structure. The \ac{rsls} method leverages the low-rank nature of spatial correlation matrices, operating on a subspace that spans all plausible correlations. This reduces pilot overhead and complexity while approaching \ac{lmmse} performance without full statistical knowledge~\cite{Demir2022Exploiting}.

Physics-aware enhancements further improve estimation. Incorporating mutual coupling into the subspace model refines channel accuracy, as demonstrated in our prior works on \ac{ris}-aided near-field scenarios~\cite{Dkhan2025Nearfield,Dkhan2025Thz}. Similar ideas have been applied to \ac{dma} architectures~\cite{Rezvani2023Channel}, \ac{sim} architectures~\cite{Yao2024Channel,An2024Hybrid}, and \ac{fas} systems~\cite{Wang2023Estimation,Skouroumounis2022Fluid}. In \ac{sim}, hybrid wave–digital estimators achieve near full-digital \ac{lmmse} performance with far fewer \ac{rf} chains~\cite{An2024Hybrid}. In \ac{fas}, sequential \ac{lmmse} with port selection reduces pilot overhead~\cite{Skouroumounis2022Fluid}.

While linear estimators provide a solid baseline, they do not exploit the inherent sparsity of high‑frequency channels, motivating the sparse estimation techniques discussed next.

\subsubsection{Sparse estimation}

Sparse channel estimation exploits the inherent sparsity of mmWave and \ac{thz} channels, where only a few dominant multipath components exist in a suitable transform domain. Compressive sensing (CS) techniques reduce the number of required pilot measurements while preserving estimation accuracy. Methods are broadly categorized as gridless or on-grid, based on assumptions about the underlying parameter space.

Gridless methods, such as \ac{anm}, \ac{sbl}, and perturbation-based refinements, estimate continuous-valued parameters without discretization, thereby avoiding basis mismatch issues~\cite{Ma2023Compressed,Zhang2024Bayesian}. These methods have been applied in near-field and hardware-constrained scenarios; for example, \ac{sbl} has been successfully applied to \ac{sim}-based \ac{mmwave} near-field communications, where a low-complexity polar-domain SBL algorithm was proposed to address the underdetermined channel estimation problem arising from the large number of meta-atoms per layer and the limited RF chains at the base station~\cite{An2026Sparse}.

On-grid methods discretize angular, delay, or frequency parameters to construct dictionaries. Classical algorithms include \ac{omp}~\cite{Tropp2007Signal} and its variants. Dictionary design is crucial and depends on the propagation domain: angular-domain DFT bases suit far-field planar wavefronts~\cite{Alkhateeb2014Channel}, while polar-domain dictionaries account for near-field spherical propagation~\cite{Cui2022Channel}. Hybrid-field environments are addressed with cross-field dictionaries or polar-domain \ac{omp} variants, enabling unified estimation of near- and far-field paths. This includes our prior work on cross-field dictionaries~\cite{Tarboush2024Cross,Tarboush2023Compressive}, approaches for hybrid-field scenarios~\cite{Wei2021Channel,Yue2024Hybrid}, and a hidden Markov model that processes pairwise power differences between subarrays within an array-of-subarrays architecture to determine the correct field region before channel estimation~\cite{Tarboush2024Near}.

For \ac{hmimo}, the wave-number domain provides a physically consistent representation (see Sec.~\ref{subsec:fourier_repr}), capturing fine spatial variations and sub-wavelength effects. Algorithms exploiting wave-number domain sparsity include \ac{omp} variants, tensorized CoSaMP, graph-cut swap expansion, and ellipse fitting~\cite{Guo2024Wavenumber, Chen2025Unified, Du2024Tensor, Di2024Electromagnetic}. Prior knowledge and structural information further enhance performance. \ac{em}-aware array manifolds improve dictionary construction~\cite{Castellanos2023Electromagnetic}. Environmental factors, including beam squint, polarization diversity, spatial non-stationarity, and near-field curvature, must be considered to ensure robust estimation~\cite{Myers2022Near, Tang2024Spatial, Amiri2018Extremely}.

Emerging architectures require tailored sparse estimation strategies. In \ac{sim}-based systems, wave-domain and Bayesian methods effectively capture multilayer \ac{em} interactions~\cite{An2023Stacked, An2024Hybrid, Liu2025Stacked, Liu2025Stacked}. For \ac{ris}-assisted systems, estimators must handle mutual coupling \cite{Zheng2024MutualTSP}, switching noise, and quantized phase resolution~\cite{Zheng2024On, Akrout2023Super}, relying on \ac{em}-consistent models for reliable performance~\cite{Chian2024Novel}. In \ac{fas} and \ac{dma} systems, sparsity is exploited at the port level using algorithms such as low-sample-size sparse channel reconstruction (L3SCR) and successive transmitter-receiver
compressed sensing (STRCS)~\cite{Xu2023Channel, Ma2023Compressed}, while tensor decomposition-based methods have been proposed for \ac{dma}-based \ac{mmwave} massive \ac{mimo} to extract multipath parameters and reconstruct high-dimensional channels with reduced training overhead~\cite{Miao2025Channel}; adaptive strategies further address dynamic or time-varying sparsity patterns~\cite{Yang2024Extremely}.

For a \ac{dma} with \(N_{\mathrm{fed}}=8\) feeds, \(N_{\mathrm{w}}=32\) elements per feed, waveguide spacing \(\lambda\), element spacing \(\lambda/16\), and \(M=1\) user, Fig.~\ref{fig:estimation} compares the \ac{nmse} of \ac{omp} and \ac{rsls} estimators versus \ac{snr}. Fig.~\ref{fig:nmse_dma} shows that neglecting \ac{mc}, i.e., omitting the admittance matrix \(\mathbf{Y}_{ss}\) in~\eqref{eq:DMA_IOrelation}, significantly degrades accuracy. \ac{mc} introduces additional spatial correlations and parasitic interactions, increasing the effective channel dimensionality and violating the sparsity assumptions of both \ac{omp} and \ac{rsls}. Fig.~\ref{fig:nf_vs_ff_mismatch} quantifies the penalty of using a far‑field (planar‑wave) dictionary in a near‑field spherical‑wave channel. A matched near‑field dictionary allows both algorithms to approach \ac{mmse} performance, whereas a mismatched far‑field dictionary yields an irreducible error floor due to its inability to model wavefront curvature. These results demonstrate that physically consistent models, capturing both mutual coupling and the correct wavefront geometry, are essential for channel estimation in DMA‑based near‑field massive \ac{mimo} systems.

Finally, low-complexity beamspace methods such as beamspace channel estimation (BEACHES) leverage sparsity with adaptive soft-thresholding and Stein’s unbiased risk estimator (SURE)-based denoising, achieving performance comparable to \ac{anm} and NOMP at significantly lower complexity~\cite{Ghods2019Beaches}. While sparse estimators exploit channel sparsity, they rely on predefined dictionaries or parametric models that may mismatch real hardware impairments; AI‑aided estimation learns the channel structure directly from data, offering robustness to such imperfections.

\subsubsection{AI-aided estimation}

Both conventional linear or sparse techniques often struggle under high dimensionality, limited pilot resources, and complex spatial correlations. To address these challenges, AI-based approaches have emerged as a promising alternative, using data-driven models to learn underlying channel structures and recover complete \ac{csi} from partial or noisy observations~\cite{Alhammadi2024Artificial}. More recently, the limitations of purely data-driven methods have motivated physics-aware AI paradigms for wireless estimation, where \ac{em}-grounded wireless foundation models (WFMs) embed physical principles into self-supervised pre-training to improve physical consistency, generalization, and data efficiency for scalable channel estimation~\cite{Xiao2025Wireless}.

For example, in near-field channels, \cite{Zhang2023Near} formulated channel estimation as a compressed sensing problem and proposed the sparsifying dictionary learning-learning iterative shrinkage and thresholding algorithm (SDL-LISTA) algorithm, which integrates dictionary learning into the network architecture. This method outperforms conventional LISTA and other benchmarks in both estimation accuracy and computational efficiency.

In the context of \ac{dma}-based architectures, prior work~\cite{Zhang2023Channel} formulated channel estimation as a compressed sensing problem and proposed a model-based learning approach based on LISTA, where the sparse channel parameters are efficiently recovered under analog compression; furthermore, a sensing-matrix-aware extension (LISTA-SMO) jointly optimizes the \ac{dma} weighting matrix and the recovery network using a self-supervised training strategy.

In rich-scattering environments, neural-network-based models can exploit spatial dependencies across antennas to improve \ac{csi} recovery. The unified asymmetric masked autoencoder (UAMA)~\cite{Wang2022Dynamixer,Lee2022Fnet,He2016Deep} extrapolates missing \ac{csi} at non-sampled ports, while attention-based networks~\cite{Gao2021Attention}, multilayer perceptrons (MLPs)~\cite{Belgiovine2021Deep,Balevi2020Massive}, convolutional neural networks (CNNs)~\cite{He2018Deep}, and long short-term memory (LSTM) architectures~\cite{Chai2022Port} have demonstrated robust performance across diverse propagation conditions. These AI-driven techniques reduce pilot overhead and enhance estimation reliability in scenarios where traditional methods are limited.
Multi‑user channel estimation for hybrid far‑ and near‑field \ac{thz} UM‑MIMO systems with non‑orthogonal pilots has been addressed using a model‑driven deep learning framework that integrates a UNet neural network into the orthogonal approximate message passing (OAMP) algorithm~\cite{Cao2026Multi}

\subsection{Beamforming}

In reconfigurable wireless systems, beamforming must be designed under hardware constraints that are absent in fully digital arrays: a limited number of \ac{rf} chains, constant‑modulus phase shifters for analog beamforming, programmable responses of metasurface elements, and \ac{mc} among densely spaced antennas. Building on the reconfigurable architectures in~\ref{sec:reconfigurable_transceiver_architectures}, this subsection examines how hardware constraints and available processing domains shape beamforming objectives, formulations, and algorithms. For emerging architectures, such as hybrid digital-analog arrays, \ac{em} domain processors, and tri-hybrid systems, beamforming design must explicitly account for these constraints while maintaining high spectral and energy efficiency. Fundamental insights into beamforming architectures and their trade-offs between hardware complexity, energy consumption, and signal-processing flexibility are discussed in~\cite{Love2003Grassmannian,Elayach2014Spatially,Love2008Overview}.
A summary of beamforming paradigms across different transceiver architectures, including their constraints, update timescales, and computational complexity, is provided in Table~\ref{tab:BF_arch}.

\begin{figure}
    \centering
    \includegraphics[width=0.98\linewidth]{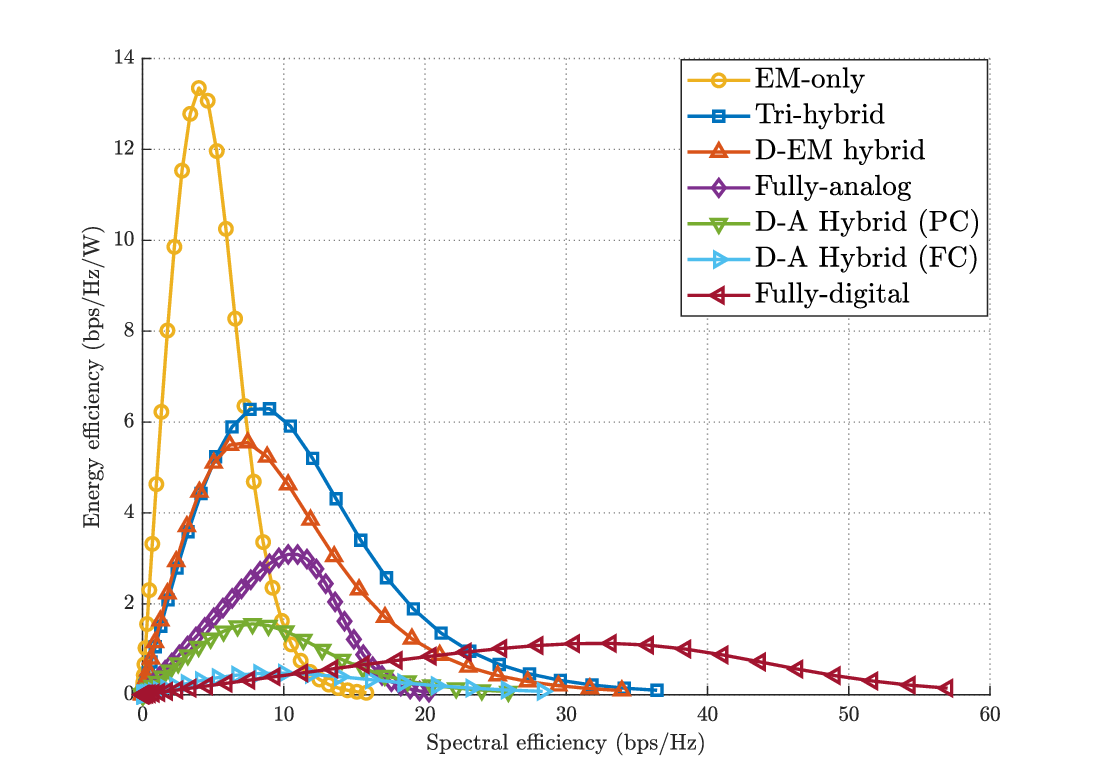}
    \caption{Energy efficiency versus spectral efficiency for different transceiver architectures.}
    \label{fig:arch_eff}
\end{figure}

Figure~\ref{fig:arch_eff} quantifies this trade‑off by comparing the energy efficiency versus spectral efficiency for seven transceiver architectures in a wideband \ac{mmwave} MIMO system. The simulation parameters are: \(N_\textit{RF} = 4\), \(N_{\mathrm{A}}=128\), \(N_{\mathrm{r}}=64\) receive antennas, \(K=64\) subcarriers, \(P=4\) multipath components, and a total transmit power varying from \(-20\) to \(50\) dBm. The architectures include fully digital, fully analog, fully‑connected hybrid (FC), partially‑connected hybrid (PC), DMA‑only (EM‑only), DMA‑digital hybrid (D‑EM hybrid), and the tri‑hybrid.

The tri‑hybrid architecture achieves a balance in energy‑efficiency–spectral‑efficiency. While the EM‑only architecture attains the highest energy efficiency at low spectral efficiencies (e.g., \(11\) bps/Hz/W at \(6\) bps/Hz), its spectral efficiency is limited and degrades beyond \(8\) bps/Hz. Fully digital systems reach very high spectral efficiencies (up to \(55\) bps/Hz) but with extremely low energy efficiency (\(<0.2\) bps/Hz/W). The tri‑hybrid bridges this gap: it maintains competitive efficiency across a wide range, outperforming both conventional hybrid (fully‑ and partially‑connected) and DMA‑digital hybrid architectures. Intuitively, the EM layer provides low‑power analog beam shaping with few \ac{rf} chains, the analog layer refines the beam with moderate complexity, and the digital layer handles multi‑stream interference, These results validate the tri‑hybrid architecture as a promising solution for energy‑efficient massive \ac{mimo} in future generations, complementing the qualitative comparisons in Table~\ref{tab:BF_arch}.

For a multi-user downlink system with \(K\) subcarriers serving \(U\) single-antenna users, the received signal at user \(u\) and subcarrier \(k\) is modeled using the unified system model of \eqref{eq:unified_system_model},
\begin{equation}
y_{u,k} = \mathbf{h}_{\mathrm{eff},u,k}^\HH \, \mathbf{F}_k \, \mathbf{s}_k + n_{u,k},
\end{equation}
where \(\mathbf{h}_{\mathrm{eff},u,k}^\HH(\mathbf{F}_{\mathrm{EM},k}) \in \mathbb{C}^{1 \times N_{\mathrm{in}}}\) is the \(u\)-th row of the effective channel matrix \(\mathbf{H}_{\mathrm{eff},k}(\mathbf{F}_{\mathrm{EM},k})\), \(\mathbf{F}_k \in \mathbb{C}^{N_{\mathrm{in}} \times U}\) is the precoding matrix applied after the EM layer, \(\mathbf{s}_k \in \mathbb{C}^{U}\) contains the data symbols with \(\mathbb{E}[\mathbf{s}_k\mathbf{s}_k^\HH]=\mathbf{I}\), and \(n_{u,k}\sim\mathcal{CN}(0,\sigma^2)\) is the additive noise. Here, \(N_{\mathrm{in}}\) denotes the number of effective input ports to the antenna system (e.g., RF chains, driven elements, or guided feeds), and the specific forms of \(\mathbf{F}_{\mathrm{EM},k}\), \(\mathbf{F}_k\), and \(\mathbf{H}_{\mathrm{eff},k}(\mathbf{F}_{\mathrm{EM},k})\) depend on the architecture as detailed in Section~\ref{subsec:End-to-end models}. For notational simplicity, the subcarrier index is dropped in the sequel, and all equations apply per subcarrier.

Fully digital beamforming allows unconstrained optimization over $\mathbf{F}$, supporting multi-stream transmission and interference suppression~\cite{Sun2014mimo}. Analog beamforming reduces hardware complexity and power consumption by controlling only the phase of each antenna via a single \ac{rf} chain~\cite{Sun2014mimo}. \ac{em}-domain approaches, including \ac{sim} can directly realize multi‑stream precoding operators in the wave domain, effectively performing beamforming matrix operations without per‑symbol digital computation~\cite{An2023Stacked,Nerini2024Physically,Nerini2025Analogcomputing,Nerini2025Capacity}.

A generic beamforming optimization problem can thus be formulated as the joint design of the EM-domain configuration \(\mathbf{F}_{\mathrm{EM}}\) and the remaining precoder \(\mathbf{F}\):

\begin{equation}
\begin{aligned}
& \underset{\mathbf{F}_{\mathrm{EM}} \in \mathcal{F}_{\mathrm{EM}}, \, \mathbf{F} \in \mathcal{F}_{\text{arch}}}{\text{max}}
& & \mathcal{U}\left(\mathbf{H}_{\mathrm{eff}}(\mathbf{F}_{\mathrm{EM}}), \mathbf{F}\right) \\
& \text{s.t.}
&& \|\mathbf{F}_{\mathrm{EM}} \mathbf{F}\|_F^2 \leq P_{\text{tot}},
\end{aligned}
\end{equation}
where \(P_{\text{tot}}\) denotes the total transmit power budget, and \(\mathcal{U}(\cdot)\) denotes a system utility function such as sum-rate, \ac{sinr}, or \ac{mse}. The feasible set \(\mathcal{F}_{\mathrm{EM}}\) enforces the physical constraints of the EM layer (e.g., passivity, reciprocity, and wave-physics constraints), while \(\mathcal{F}_{\text{arch}}\) captures the hardware constraints of the remaining precoder (e.g., constant-modulus phase shifters and limited \ac{rf} chains). The power constraint \(\|\mathbf{F}_{\mathrm{EM}} \mathbf{F}\|_F^2 \leq P_{\text{tot}}\) limits the total power radiated by the antenna array, where \(\mathbf{F}_{\mathrm{EM}} \mathbf{F}\) is the overall mapping from data streams to the radiating elements.

Hybrid digital-analog architectures have been studied extensively~\cite{Elayach2014Spatially,Sohrabi2016Hybrid,Molisch2017Hybrid,Ahmed2018Survey}, while \ac{ris}- and DMA-assisted \ac{em} beamforming exploit passive reconfigurable elements to shape the effective channel without increasing the number of \ac{rf} chains~\cite{Di2020Hybrid,You2022Deploy,Huang2023Integrating,Jamali2020Intelligent,Mishra2023Transmitter,Shlezinger2019Dynamic,Shlezinger2021Dynamic,Zhang2022Beam}. Tri-hybrid systems further integrate these domains, combining digital, analog, and \ac{em}-layer beamforming to optimize spectral and energy efficiency~\cite{Castellanos2023Energy,Castellanos2025Embracing,Liu2025Tri,Zheng2025Tri,Li2025Tri}.

\subsubsection{Beamforming objectives}

We can naturally categorize beamforming strategies by their optimization objectives, which remain consistent across architectures, even though their realizations differ.

\paragraph{Signal power maximization}
In noise‑limited scenarios, beamforming aims to maximize the received signal power. The \ac{mrt} approach aligns the transmit beamformer with the effective channel vector to maximize the post‑combining \ac{snr}~\cite{Lo1999},
\begin{equation}
\mathbf{f}_{\text{MRT}} = \sqrt{\frac{P_{\max}}{\|\mathbf{h}_{\mathrm{eff},u}\|^2}}\,\mathbf{h}_{\mathrm{eff},u}.
\end{equation}
\ac{mrt} is a classical linear precoding method that maximizes signal power in fully digital systems~\cite{Albreem2021Overview}, and it does not require channel inversion. Hybrid and constrained implementations approximate \ac{mrt} by projecting the unconstrained digital solution onto the feasible set enforced by analog phase‑shifters or metasurface elements~\cite{Ioushua2019Family,Molisch2017Hybrid}.

\paragraph{Interference suppression}
In multi‑user settings, \ac{zf} beamforming suppresses inter‑user interference by enforcing orthogonality among users' spatial channels through matrix pseudo‑inversion~\cite{Spencer2004Zero,Albreem2021Overview},
\begin{equation}
\mathbf{F}_{\text{ZF}} = \sqrt{\beta}\,\mathbf{H}_{\mathrm{eff}}^\HH\left(\mathbf{H}_{\mathrm{eff}}\mathbf{H}_{\mathrm{eff}}^\HH\right)^{-1},
\end{equation}
where $\mathbf{H}_{\mathrm{eff}} = [\mathbf{h}_{\mathrm{eff},1}, \ldots, \mathbf{h}_{\mathrm{eff},U}]^\HH$, and $\beta$ is a power normalization factor selected to ensure that the precoder satisfies the total transmit power constraint. This expression requires $U \leq N_{\mathrm{in}}$ (at least as many input ports as users) and that $\mathbf{H}_{\mathrm{eff}}$ has full row rank so that the inverse exists. With perfect \ac{csi} and sufficient spatial \acp{dof}, \ac{zf} achieves exact interference cancellation, albeit at the cost of noise amplification in ill‑conditioned channels. Under hardware constraints inherent in hybrid and reconfigurable systems, exact \ac{zf} becomes infeasible; instead, constrained or regularized \ac{zf} formulations are employed. For example, hybrid analog‑digital beamforming designs combine limited \ac{rf} chain analog processing with digital \ac{zf} or regularized \ac{zf} to mitigate interference while respecting hardware limits~\cite{Yu2022Regularized}. Optimizing hybrid \ac{zf} under partially connected or sub‑array hybrid architectures has been shown to achieve a favorable compromise between interference suppression, hardware complexity, and spectral efficiency~\cite{Su2021Optimal}.

\paragraph{Error minimization}
\ac{mmse} beamforming balances noise enhancement and interference suppression by minimizing the mean square error between transmitted and received signals,
\begin{equation}
\mathbf{F}_{\text{MMSE}} =
\left(\mathbf{H}_{\mathrm{eff}}^\HH\mathbf{H}_{\mathrm{eff}} + \frac{\sigma^2}{P_{\max}}\mathbf{I}\right)^{-1}\mathbf{H}_{\mathrm{eff}}^\HH.
\end{equation}
This formulation provides an optimal linear precoder in the \ac{mmse} sense under Gaussian noise, full CSI, and a sum‑power constraint, balancing the trade‑off between interference cancellation and noise amplification.

\ac{mmse} criteria have been widely adopted in hybrid analog–digital beamforming designs, where the digital part of the precoder is optimized to \ac{mmse} under analog hardware constraints~\cite{Li2022Hybrid,Cong2018Hybrid}.  
Recent research has also investigated the use of analog computing structures such as MiLACs to realize linear estimators like \ac{lmmse} directly in the analog domain with significantly reduced computational complexity, highlighting their potential for scalable \ac{mimo} beamforming and detection~\cite{Nerini2025Analogcomputing}.

\paragraph{Rate maximization}
For heterogeneous traffic demands, beamforming design is often cast as a weighted sum‑rate (WSR) maximization problem,
\begin{equation}
\max_{\mathbf{F}_{\mathrm{EM}}, \mathbf{F}} \sum_{u=1}^U w_u
\log_2\!\left(1+
\frac{|\mathbf{h}_{\mathrm{eff},u}^\HH\mathbf{f}_u|^2}
{\sum_{j\neq u}|\mathbf{h}_{\mathrm{eff},u}^\HH\mathbf{f}_j|^2+\sigma^2}
\right),
\end{equation}
where $w_u$ denotes a priority weight for user $u$. This optimization is inherently non-convex and nondeterministic polynomial-time hard (NP-hard) due to interference coupling across users, and is jointly optimized over the EM-domain configuration \(\mathbf{F}_{\mathrm{EM}}\) and the remaining precoder \(\mathbf{F}\). A widely adopted approach for tackling this problem is the \ac{wmmse} algorithm, which exploits the equivalence between the WSR objective and a suitably weighted mean square error formulation; stationary points of the \ac{wmmse} cost correspond to local optima of the WSR problem under linear precoding~\cite{Christensen2008Weighted}. The \ac{wmmse} framework has seen extensive application and adaptation, including extensions to \ac{ris} and hybrid precoding contexts~\cite{Choi2024Wmmse}.

\subsubsection{Algorithmic implementation techniques}

Architectural constraints introduced by hybrid, \ac{em}-domain, and tri-hybrid transceivers necessitate specialized algorithmic strategies for beamforming realization to accommodate heterogeneous signal processing domains and reconfiguration capabilities.

\paragraph{Alternating optimization}

Alternating optimization addresses the resulting non-convex design problem by iteratively optimizing one component while fixing the others. Specifically, the optimization alternates between the EM-domain configuration \(\mathbf{F}_{\mathrm{EM}}\), the analog RF precoder \(\mathbf{F}_{\mathrm{RF}}\) (if present), and the digital precoder \(\mathbf{F}_{\mathrm{BB}}\), enabling tractable handling of heterogeneous constraints across domains. This approach is widely adopted in hybrid analog-digital beamforming designs~\cite{Sohrabi2016Hybrid,Yu2016Alternating}.

An alternating optimization framework for DMA-assisted radiating near-field wireless power transfer was proposed in~\cite{Zhang2025Near}, where the DMA weights and digital precoder are jointly designed to maximize the weighted sum-harvested energy under spherical-wave propagation and practical hardware constraints (including discrete phase shifts).
This approach is extended to tri-hybrid systems with pattern-reconfigurable antennas in~\cite{Zheng2025Tri}, employing \ac{wmmse}-based block coordinate descent to jointly optimize digital, analog, and antenna-domain precoders under per-antenna power constraints.

\paragraph{Hierarchical and multi-timescale optimization}
In emerging tri-hybrid architectures, alternating optimization naturally evolves into a hierarchical or multi-timescale optimization framework, reflecting the disparate physical control granularities of the beamforming components. Specifically, \ac{em}-domain configurations are adjusted at a slow reconfiguration timescale based on long-term channel statistics or environmental awareness, \ac{rf} beamformers are updated at an intermediate timescale tracking slow channel variations, and digital precoders operate at the fastest timescale to adapt to instantaneous \ac{csi}~\cite{Castellanos2025Embracing,Liu2025Tri}.

This hierarchy is exemplified in~\cite{Castellanos2023Energy} through sequential optimization of DMA, analog, and digital precoders in a tri-hybrid architecture, first selecting DMA weights via codebook search, then analog phases via singular value decomposition (SVD)-based phase quantization per subarray, and finally digital weights from the dominant singular vectors of the effective channel, achieving higher energy efficiency than conventional fully‑digital and hybrid precoding schemes, despite spectral efficiency suboptimality.

This hierarchical decomposition significantly reduces computational complexity and control signaling overhead while maintaining near-optimal performance. By aligning optimization updates with hardware reconfigurability and channel dynamics, multi-timescale beamforming provides a scalable and implementation-aware solution for large-scale and reconfigurable wireless systems.

\paragraph{Manifold optimization}
Phase-only constraints imposed by analog phase shifters or reconfigurable \ac{em} surfaces define a complex circle manifold,
\begin{equation}
\mathcal{M}=\{\mathbf{F}_{\text{RF}} : |[\mathbf{F}_{\text{RF}}]_{i,j}|=1\}.
\end{equation}
Riemannian manifold optimization methods, such as Riemannian gradient descent and conjugate gradient algorithms, efficiently enforce constant-modulus constraints while navigating the feasible set along the manifold geometry~\cite{Absil2008Optimization}.  

These techniques are particularly effective for hybrid beamforming and \ac{em}-domain precoding, where direct Euclidean optimization would violate hardware constraints. Manifold-based formulations have been successfully applied to hybrid analog-digital precoding~\cite{Sohrabi2016Hybrid} and dynamically reconfigurable \ac{em} surfaces, enabling physics-consistent beam shaping with reduced dimensionality~\cite{Shlezinger2019Dynamic}.

\begin{figure*}[ht]
 \centering
 \subfloat[\ac{nmse} between the optimized channel and the diagonal channel versus the number of transmit \ac{sim} layers.]{\label{fig:mse_layers} \includegraphics[width=0.48\linewidth]{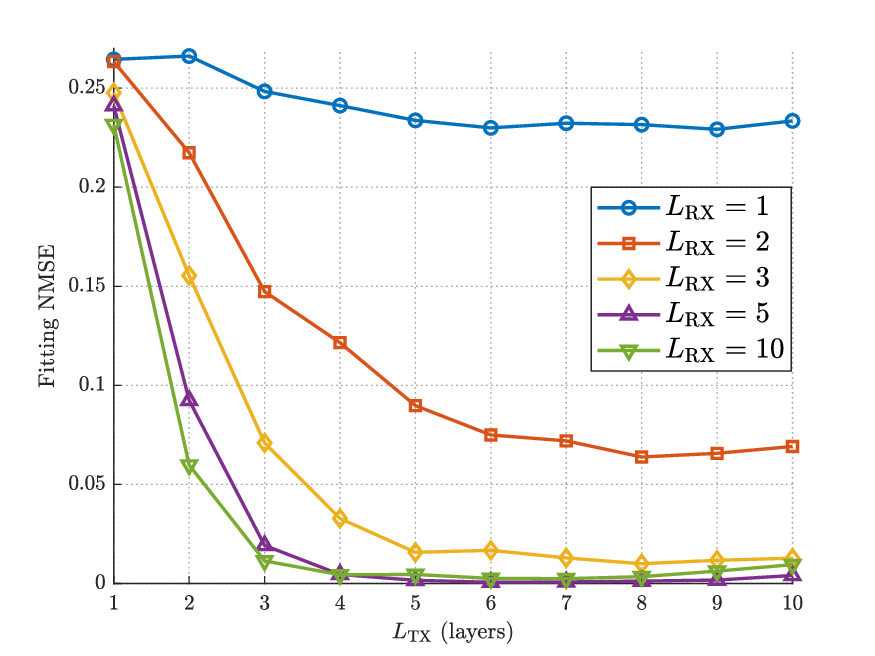}}%
 \hfill
 \subfloat[Spectral efficiency versus number of transmit \ac{sim} layers.]{\label{fig:capacity_layers} \includegraphics[width=0.48\linewidth]{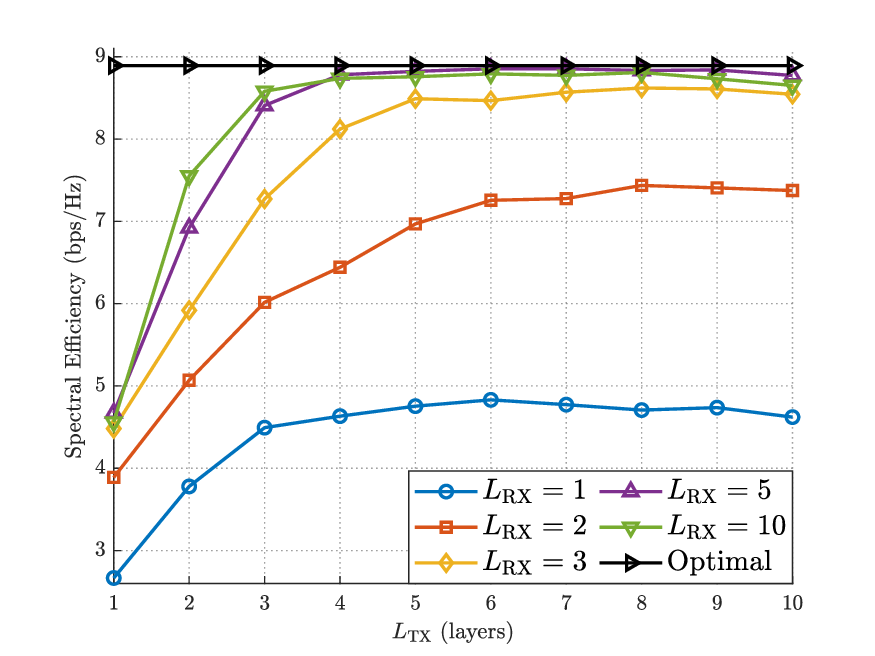}}%
 \caption{Performance of gradient-descent optimization for the \ac{sim} system with multiple metasurface layers. $L_{\text{TX}}$ and $L_{\text{RX}}$ denote the number of programmable layers at the transmitter and receiver, respectively.}
 \label{fig:SIM_phase_opt}
 \vspace{-5mm}
\end{figure*}

Figure~\ref{fig:mse_layers} shows the \ac{nmse} between the achieved channel matrix and the ideal diagonal matrix as a function of the number of transmit metasurface layers. As the number of layers increases, the \ac{nmse} decreases rapidly, indicating that the optimization effectively fits the channel. Fig.~\ref{fig:capacity_layers} shows the corresponding spectral efficiency, which approaches the upper bound of a full‑precision digital precoder/combiner when a sufficient number of layers (approximately 3) is used. These results demonstrate that a moderate number of layers is sufficient to achieve near‑optimal wave‑domain processing.

\begin{table*}[t]
\caption{Beamforming Paradigms Across Transceiver Architectures}
\label{tab:BF_arch}
\centering
\renewcommand{\arraystretch}{1.2}
\begin{tabular}{|p{3.3cm}|p{3.1cm}|p{3.1cm}|p{3.1cm}|p{3.1cm}|}
\hline
\textbf{Aspect} & \textbf{Digital} & \textbf{Hybrid} & \textbf{EM} & \textbf{Tri-hybrid} \\ \hline
\acp{dof} & Channel-limited & RF-chain limited & Physics-limited & Multi-domain \\ \hline
Constraints & None & Constant-modulus, low-rank & Passivity, reciprocity & Mixed constraints \\ \hline
Algorithm Type & Closed-form, convex & Alternating, manifold & Physics-aware, analog & Hierarchical \\ \hline
Update Timescale 
& Fast (symbol-level) 
& Medium (coherence-level) 
& Slow (reconfiguration-level) 
& Multi-timescale \\ \hline
Computational Complexity & High & Moderate & Low--moderate & Moderate \\ \hline
\end{tabular}
\end{table*}

\paragraph{Learning‑based beamforming}
Learning‑based beamforming leverages machine learning models to directly infer precoding matrices or beam selection policies from observed channel or environmental data, thereby addressing the non‑convex optimization challenges inherent in hybrid and constrained architectures. For example, deep reinforcement learning has been applied to joint beam‑user selection and hybrid beamforming design in massive \ac{mimo}‑NOMA systems, where a Q-network agent selects analog beams and digital precoding patterns to improve sum‑rate and energy efficiency over classical methods~\cite{Ahmed2022Deep}. Advanced deep reinforcement learning techniques with continuous action spaces have also been proposed to adapt hybrid precoders in highly dynamic \ac{mmwave} environments using experience replay and policy gradient methods, demonstrating improved spectral efficiency relative to traditional optimization approaches~\cite{Arjoune2022Experience}. In addition, neural network‑based hybrid beamforming frameworks, such as autoencoder and deep learning driven designs, have been shown to reduce beamforming training overhead and approach near‑optimal performance by learning implicit mapping from channel observations to analog/digital beamformers~\cite{Elbir2020Deep,Wang2020Precodernet}.

\subsection{Detection and Decoding}
\label{sec:detection_decoding}

Detection and channel decoding constitute the final stages at which the physical structure of the communication system directly affects the recovered information bits. In conventional \ac{mimo} models, receiver processing is often designed assuming an abstract channel matrix and spatially white additive noise. These assumptions become increasingly inaccurate for large-aperture, electromagnetically dense, and reconfigurable architectures. Near-field wavefront curvature, \ac{mc}, configuration-dependent radiation characteristics, frequency selectivity, and circuit-level impairments jointly modify both the effective channel and the noise statistics observed by the detector. Consequently, physical consistency must propagate beyond channel modeling and beamforming into symbol detection, reliability extraction, and channel decoding. Accordingly, this section discusses physically consistent data detection, soft and pseudo-soft reliability information, reliability-aware and noise-centric decoding, and their joint design under practical performance--complexity constraints.

Following the end-to-end system model in \eqref{eq:unified_system_model}, where $\mathbf{H}_{\mathrm{eff},k}(\mathbf{F}_{\mathrm{EM},k})$ captures the end-to-end propagation and the transformations imposed by a given reconfiguration state, we can assume zero-mean Gaussian thermal and circuit noise, $    \mathbf{n}_k \sim \mathcal{CN}\!\left(\mathbf{0},\mathbf{R}_{n,k}\right)$, where $\mathbf{R}_{n,k}$ captures the resulting noise power and spatial correlation. Thus, physically consistent detection should account for both the configuration-dependent effective channel and the associated noise statistics.

\subsubsection{Physically consistent data detection}

Data detection estimates the transmitted symbols from the received signal using the available effective channel and noise statistics. In general, \ac{ml} detection selects the candidate that maximizes the likelihood induced by the underlying noise distribution. For the important case of spatially colored Gaussian noise, this yields
\begin{equation}
    \hat{\mathbf{x}}_{\mathrm{ML}}
    =
    \operatorname*{arg\,min}_{\mathbf{x}\in\mathcal{X}^{N}}
    \left(
        \mathbf{y}
        -
        \hat{\mathbf{H}}_{\mathrm{eff}}\mathbf{x}
    \right)^{H}
    \hat{\mathbf{R}}_{n}^{-1}
    \left(
        \mathbf{y}
        -
        \hat{\mathbf{H}}_{\mathrm{eff}}\mathbf{x}
    \right),
    \label{eq:colored_ml}
\end{equation}
where $\hat{\mathbf{R}}_{n}$ is the estimated noise covariance matrix. Although \ac{ml} detection is optimal under the assumed statistical model, its complexity grows exponentially with the number of transmitted streams and modulation order, making it impractical for large \ac{mimo} systems~\cite{Cao2016Distributed,Yang2015Fifty}. More generally, non-Gaussian or hardware-induced noise requires a likelihood metric consistent with its actual statistics.

Lower-complexity linear detectors, such as \ac{zf} and \ac{mmse}, are more scalable but can degrade in highly correlated large-\ac{mimo} channels~\cite{Sarieddeen2019Terahertz,Yang2015Fifty}. This makes physically consistent channel and noise modeling particularly important: near-field propagation introduces structured spatial characteristics~\cite{Dong2022Near,Demir2022Channel,Haghshenas2023New,Demir2024Spatial,Demir2022Exploiting}, measurement-based small-scale fading distributions such as Mixture Gamma and $\alpha$-$\mu$ can provide more representative statistical characterizations than simplified Gaussian assumptions~\cite{enad2026symbolerroranalysislinear,11072420,11362947,Jemaa2024Performance}, and \ac{mc}, thermal noise, and circuit-level effects in compact apertures can modify the effective signal response and induce spatially correlated receiver noise~\cite{Akrout2022Achievable,Gradoni2021End,Zhu2024Electromagnetic}. The circuit-theoretic multiport formulation of Sec.~\ref{sec:multiport_network} naturally captures these effects, with the noise covariance derived from the impedance or admittance quantities that also govern signal coupling. Hence, detectors based on uncoupled channels or spatially white noise can become mismatched.

For a generic linear detector $\mathbf{W}$, $\hat{\mathbf{x}} = \mathbf{W}\mathbf{y}$. The post-detection noise covariance is
\begin{equation}
    \mathbf{R}_{\mathrm{post}}
    =
    \mathbf{W}
    \mathbf{R}_{n}
    \mathbf{W}^{H}.
    \label{eq:post_detection_covariance}
\end{equation}
Equation~\eqref{eq:post_detection_covariance} directly links the circuit-level noise model to the noise statistics at the detector output. Thus, neglecting the off-diagonal entries of $\mathbf{R}_{n}$ in electromagnetically dense arrays can affect both symbol detection and the reliability information subsequently supplied to the decoder.

The same physical channel structure can be exploited to reduce detection complexity. Near-field, cross-field, and hybrid-field propagation exhibit spatial and distance-dependent structures that can be leveraged through subspace decompositions~\cite{Dong2022Near,Demir2022Channel,Haghshenas2023New,Demir2024Spatial,Demir2022Exploiting}, while analytical characterizations of \ac{em}-field correlation provide further structure for linear estimation and subspace processing~\cite{Wan2024Near,Zhu2023Can}. Such methods reduce the effective \ac{mimo} detection dimension and enable spatial parallelization~\cite{Sarieddeen2024Bridging,Jemaa2024Leveraging,Jemaa2022THz}. The corresponding performance--complexity trade-offs under realistic \ac{thz} channels, including subspace- and QR-based detectors, are analyzed in~\cite{Jemaa2025Performance}. Sparse beamspace methods such as SPADE (SParsity-ADaptive Equalization) further reduce arithmetic complexity by exploiting sparse inner products and beamspace transformations~\cite{Mirfarshbafan2021Spade}. Nonlinear tree-search detectors can improve performance but scale poorly with system dimension~\cite{Yang2015Fifty}, whereas approximate message-passing algorithms may lose robustness in strongly correlated channels~\cite{Jeon2015Optimality}.

Finally, detector design is inherently architecture dependent. Fully digital arrays retain the complete observation space and provide maximum digital-processing flexibility at high hardware and computational cost. Hybrid analog--digital architectures reduce the observation dimension before detection, making the effective channel and noise statistics dependent on the analog transformation. Digital--\ac{em} and tri-hybrid architectures introduce additional configuration-dependent transformations, while full-\ac{em} architectures can shift part of the conventional spatial processing into the wave domain. Detection should therefore operate on the effective signal and noise statistics available \emph{after} the architecture-dependent transformations.

\subsubsection{Reliability extraction and pseudo-soft information}

Since exact soft-information generation and transmission between baseband blocks can be computationally expensive in large-array systems, \ac{psi}, which is mainly channel-dependent, provides an intermediate solution between hard and fully soft outputs~\cite{Sarieddeen2022GRAND}. For conventional \ac{zf} detection with spatially white noise, the filter based on the effective channel is
\begin{equation}
    \mathbf{W}_{\mathrm{ZF}}
    =
    \left(
        \mathbf{H}_{\mathrm{eff}}^{H}\mathbf{H}_{\mathrm{eff}}
    \right)^{-1}
    \mathbf{H}_{\mathrm{eff}}^{H},
\end{equation}
such that the post-equalization noise variance of the $m$-th data stream, $m\in\{1,\ldots,N_S\}$, is
\begin{equation}
    \sigma_{\mathrm{eff},m}^{2}
    =
    \sigma_n^{2}
    \left[
        \left(
            \mathbf{H}_{\mathrm{eff}}^{H}\mathbf{H}_{\mathrm{eff}}
        \right)^{-1}
    \right]_{m,m},
    \label{eq:zf_eff_noise}
\end{equation}
where $\sigma_n^2$ denotes the spatially white noise variance. The corresponding stream-level \ac{psi} is $ \psi_m = \frac{1}{\sigma_{\mathrm{eff},m}^{2}}$,
where larger $\psi_m$ indicates lower post-equalization noise enhancement and, hence, a more reliable stream estimate. More generally, for spatially correlated noise, the post-detection additive-noise covariance in~\eqref{eq:post_detection_covariance} can be used to define the noise-based reliability proxy,
\begin{equation}
    \psi_m
    =
    \frac{1}
    {\left[\mathbf{R}_{\mathrm{post}}\right]_{m,m}},
    \label{eq:generalized_psi}
\end{equation}
where $\left[\mathbf{R}_{\mathrm{post}}\right]_{m,m}$ is the post-detection noise variance of the $m$-th stream. This generalization allows the physically consistent noise covariance, including spatial correlation, to propagate into the reliability metric. When residual inter-stream interference is present, however, its contribution should also be incorporated into the reliability calculation.

Beyond individual stream reliabilities, the distribution of reliability across bits, streams, and resources can significantly affect decoding performance. Let $\lambda_i$ denote the reliability of the $i$-th coded bit and $N_b$ the number of coded bits in a codeword. The ordered reliability vector is
\begin{equation}
    \boldsymbol{\Lambda}
    =
    \left[
        \lambda_{(1)},
        \lambda_{(2)},
        \ldots,
        \lambda_{(N_b)}
    \right],
    \qquad
    \lambda_{(1)}
    \leq
    \lambda_{(2)}
    \leq
    \cdots
    \leq
    \lambda_{(N_b)},
    \label{eq:ordered_reliability}
\end{equation}
where $\lambda_{(i)}$ denotes the $i$-th smallest bit reliability. The resulting reliability profile conveys information beyond average \ac{snr}: configurations with comparable received power can yield different profiles due to channel conditioning, coupling, propagation geometry, and post-equalization noise enhancement. Weak reliabilities can therefore limit coded-link performance, whereas accurate identification and ordering of unreliable bits can be exploited by reliability-based decoders. Accordingly, \ac{psi} need not reproduce exact likelihoods, but can provide lower-complexity channel-dependent reliability information for decoder ordering.

The accuracy of this reliability information depends on the underlying physical model. Spherical-wave and near-field models capture spatial structures absent under conventional planar-wave assumptions~\cite{Dong2022Near,Demir2022Channel,Haghshenas2023New,Demir2024Spatial,Demir2022Exploiting}, while analytical \ac{em}-field correlation models can support improved \ac{mmse} and subspace processing~\cite{Wan2024Near,Zhu2023Can}. Circuit-theoretic noise models similarly allow \acp{llr} and \ac{psi} metrics to account for correlated noise under strong \ac{mc}~\cite{Akrout2022Achievable,Gradoni2021End,Zhu2024Electromagnetic}. This physical structure can further support bit-to-resource mapping and shorter codewords across parallel baseband processing units, increasing decoding throughput in large-array systems~\cite{Sarieddeen2024Bridging,Jemaa2024Leveraging}.

\subsubsection{Reliability-aware and noise-centric decoding}

Error-correction coding introduces redundancy that enables recovery from detection errors. Modern coded systems include Turbo codes with iterative soft-input soft-output decoding~\cite{Berrou1993}, \ac{ldpc} codes commonly decoded using belief propagation or sum-product algorithms~\cite{ryan2009channel}, and polar codes~\cite{Arikan2009Channel}, typically decoded using \ac{scl} and related algorithms. Physical consistency does not alter the underlying codes, but rather the statistical information supplied to their decoders, making the detector--decoder interface critical to coded-link performance.

Noise-centric decoding provides a natural framework for exploiting this information. Rather than directly searching for the transmitted codeword, \ac{grand} searches for likely noise realizations that could have transformed a valid codeword into the received sequence~\cite{Duffy2019Capacity,Sarieddeen2022GRAND}. This code-agnostic formulation can accommodate structured noise processes, including channels with memory and colored noise. Reliability-aware variants such as ordered reliability bits GRAND (ORBGRAND) use detector-generated reliability information to order candidate error patterns, prioritizing the least reliable bits. Consequently, reliability estimates incorporating channel-state and colored-noise information can improve both decoding efficiency and error-correction performance~\cite{Sarieddeen2022GRAND}. In physically consistent systems, inaccurate channel or noise models can therefore lead to mismatched reliability ordering and decoding.

Noise-centric processing can also account for correlation rather than treating it solely as an impairment. Noise recycling exploits correlations between decoding errors and subsequent decoder inputs to improve decoding efficiency~\cite{Riaz2023Noise}, while ordered-statistic decoding (OSD) and related reliability-based approaches provide alternative performance--complexity trade-offs~\cite{Yue2021Revisit}. \ac{grand}-based \ac{mimo} detection and decoding with antenna selection has demonstrated complexity reductions exceeding $80\%$ relative to bit-level \ac{grand}, while maintaining near-\ac{ml} performance~\cite{Allahkaram2023Symbol}.

Joint detection and decoding further extends reliability-aware processing by iteratively exchanging detector reliability and decoder extrinsic information. In large \ac{mimo} and \ac{thz} systems, the resulting performance gains must be balanced against latency, memory access, and hardware complexity. Parallelizable detection, shorter codewords, spatial subspace decomposition, and noise-centric decoding provide complementary means of mitigating these bottlenecks~\cite{enad2025theoretical,dkhan2026physically,Sarieddeen2024Bridging,Jemaa2024Leveraging,Jemaa2022THz}. Physically representative simulation frameworks are essential for evaluating such receiver designs. For example, TeraMIMO captures \ac{thz}-specific effects, including spherical-wave propagation and wideband beam split~\cite{Tarboush2021Teramimo}, while circuit-level noise and hardware impairments should also be considered to assess performance under realistic transceiver conditions~\cite{Zhu2024Electromagnetic}.

\begin{figure*}[ht]
 \centering
 \subfloat[BLER versus \ac{snr} for decoding schemes under \ac{mc} effects.]{%
 \label{fig:dec_MC}
 \includegraphics[width=0.48\linewidth]{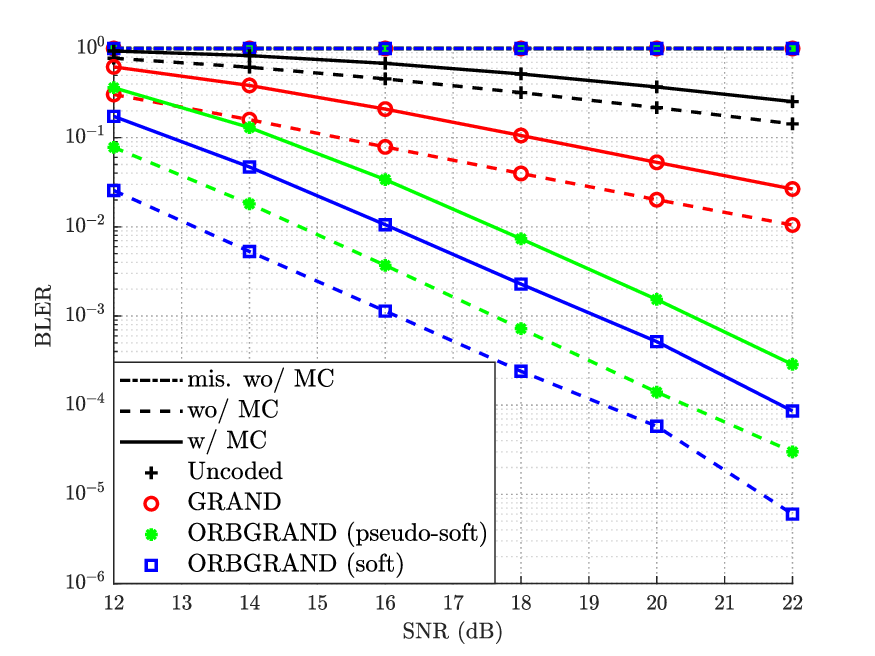}}%
 \hfill
 \subfloat[BLER versus \ac{snr} for decoding schemes under near-field effects.]{%
 \label{fig:dec_NF}
 \includegraphics[width=0.48\linewidth]{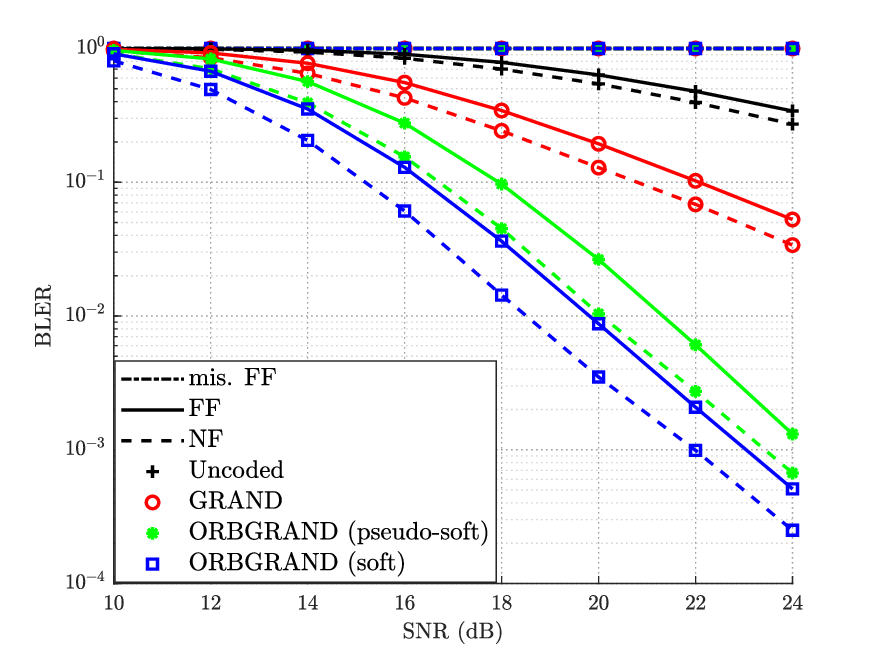}}%
 \caption{BLER versus \ac{snr} under BCH $(127,113)$ coding and \ac{zf} equalization: uncoded hard-decision detection, hard GRAND, ORBGRAND with PSI, and ORBGRAND with soft information. (a) Impact of MC and model mismatch; (b) Impact of near-field propagation and planar-wave model mismatch.}
 \label{fig:decoding}
 \vspace{-5mm}
\end{figure*}

Fig.~\ref{fig:decoding} illustrates how physical-model mismatch propagates through detection and reliability generation to coded-link performance for a $(127,113)$ BCH code with \ac{zf} equalization, comparing uncoded hard-decision detection, GRAND, ORBGRAND using \ac{psi}, and ORBGRAND using soft information. Fig.~\ref{fig:dec_MC} considers a coupled channel with coupling-aware processing, an uncoupled reference, and a mismatched receiver that neglects \ac{mc}. While \ac{mc} modifies the effective signal and noise characteristics, neglecting it during receiver processing produces a substantially higher error floor, highlighting the impact of model mismatch on the post-detection statistics and reliability information. Error correction under \ac{mc} can be further improved by accounting for noise correlation during noise-pattern guessing~\cite{ORBGRANDAI}. Fig.~\ref{fig:dec_NF} compares near- and far-field propagation and the mismatch caused by applying a planar-wave model in the near field. The resulting degradation highlights the importance of accurate wavefront modeling. For the considered configuration, reliability-aware decoding provides a larger gain in the near field, consistent with the structured post-detection reliabilities obtained when spherical-wave propagation is accurately represented rather than approximated by a far-field planar-wave model.

\subsubsection{Toward joint architecture--detection--decoding design}

The preceding discussion suggests that detection and decoding should not be treated independently of the physical architecture. The \ac{em}-domain configuration $\mathbf{F}_{\mathrm{EM}}$ shapes the effective channel and associated noise statistics, which in turn determine the post-detection reliability and coded-link performance. This dependence can be summarized as
\begin{equation}
    \mathbf{F}_{\mathrm{EM}}
    \longrightarrow
    \left\{
        \mathbf{H}_{\mathrm{eff}}(\mathbf{F}_{\mathrm{EM}}),
        \mathbf{R}_{n}
    \right\}
    \longrightarrow
    \boldsymbol{\Lambda}
    \longrightarrow
    \text{BLER},
    \label{eq:architecture_reliability_chain}
\end{equation}
where $\boldsymbol{\Lambda}$ provides an intermediate link between physical configuration, detector behavior, and decoder performance.

This motivates \emph{decoder-aware reconfiguration}. Conventional beamforming and reconfiguration typically optimize channel-centric metrics such as received power, \ac{snr}, spectral efficiency, or sum rate; however, the configuration optimizing these metrics need not yield the reliability profile most favorable to the decoder. A decoder-aware design can instead consider
\begin{equation}
    \mathbf{F}_{\mathrm{EM}}^{\star}
    =
    \operatorname*{arg\,max}_{\mathbf{F}_{\mathrm{EM}}}
    \mathcal{J}
    \left(
        \boldsymbol{\Lambda}(\mathbf{F}_{\mathrm{EM}})
    \right),
    \label{eq:decoder_aware_reconfiguration}
\end{equation}
where $\mathcal{J}(\cdot)$ denotes a decoder-aware measure of the post-detection reliability profile. Low-complexity surrogates derived from \ac{psi} can facilitate such optimization without repeatedly executing the complete decoder, which is particularly attractive for large reconfigurable systems.

This perspective extends to antenna and frequency-resource selection, where configurations can be chosen to shape the reliability distribution rather than solely maximize channel gain. Since reliability-aware decoders can exploit both the magnitude and ordering of the resulting reliabilities, maximizing average reliability alone may not be optimal. The physical architecture, channel, detector, and decoder should therefore be viewed as coupled components of an end-to-end design that balances physical-model fidelity, reliability, decoding performance, and implementation complexity.

\section{Conclusion and future research directions}
\label{sec:conclusion}

This tutorial develops a unified framework for physical consistency in next-generation reconfigurable wireless systems, connecting emerging architectures with their underlying \ac{em} and circuit-level models and the corresponding advances in signal processing. As reconfigurability extends beyond the digital and analog domains into the antenna and propagation domains, the physical architecture, channel, and communication algorithms become increasingly intertwined. The central premise is that the relevant physical properties of these systems should be preserved in the effective channel and noise models and propagated throughout the processing chain, from channel estimation and beamforming to detection and decoding. This perspective unifies the diverse reconfigurable architectures considered in this tutorial and provides a basis for understanding their physical constraints, exploitable degrees of freedom, and associated performance--complexity trade-offs. The key insights can be synthesized as follows:
\begin{itemize}

    \item \textbf{Reconfigurability expands the physical design space:}
    Emerging architectures introduce controllable degrees of freedom across digital, analog, and \ac{em} domains, giving rise to bi-hybrid, tri-hybrid, and full-\ac{em} architectures. These include reconfigurable apertures, antenna positions, and multi-layer wave-domain processors, with fundamental trade-offs among hardware complexity, power consumption, and beamforming flexibility. Realizing these degrees of freedom therefore requires co-optimization across domains~\cite{Bjornson2024Towards,Gong2024Holographic,You2025Next,Castellanos2025Embracing,Heath2025Tri}.

    \item \textbf{Physical consistency connects architecture to the communication model:}
    Maxwell's equations, dyadic Green's functions, Fourier plane-wave representations, and multiport network theory provide complementary tools for translating the physical architecture and propagation environment into effective channel and noise models. The required level of physical fidelity depends on the architecture and operating regime, particularly when near-field propagation, MC, spatial non-stationarity, polarization, structural scattering, or circuit-level effects materially affect system behavior. For example, neglecting structural scattering in multi-\ac{ris} systems can substantially misrepresent the achievable physical gain~\cite{Shen2021Modeling,Wu2024Intelligent,Li2023Beyond}.

    \item \textbf{Physical consistency must propagate into signal processing:}
    Once physical effects modify the effective channel and noise statistics, they must also be reflected in channel estimation, beamforming, detection, reliability extraction, and decoding. Reconfigurable systems exhibit exploitable structures, including the low-rank behavior of \ac{sim} channels, port selectivity in FAS, and coupling-dependent behavior of parasitic and dense arrays. Channel estimation can exploit these structures through reduced-subspace, sparse, and physics-aware learning methods~\cite{Dkhan2025Nearfield,Rezvani2023Channel,Guo2024Sparse,Zhu2023Can}, while beamforming, detection, and decoding must account for hardware constraints and correlated noise arising from coupling and circuit impairments~\cite{11078009,Akrout2022Achievable,Akrout2023Super,Yu2016Alternating,Sarieddeen2024Bridging,Jemaa2024Leveraging,Sarieddeen2022GRAND}. Physical consistency therefore does not merely add modeling complexity; it can reveal structure that enables lower-dimensional processing, improved reliability, and joint optimization of the architecture and communication algorithms.

\end{itemize}

\subsection{Future research directions}

Despite substantial progress, translating this paradigm into scalable and experimentally validated communication systems requires advances across modeling, algorithms, and hardware. The central challenge is to retain the physical information that materially affects communication performance without sacrificing the tractability required for system-level design.

\subsubsection{Advancing physically consistent modeling and simulation}

\begin{itemize}

    \item \textbf{Scalable computational \ac{em} for large structures:}
    A major bottleneck is the development of computationally efficient yet accurate \ac{em} solvers capable of resolving electrically large reconfigurable structures with sub-wavelength details. Promising approaches such as LCMT-DDM~\cite{ref117} require further advances in scalability, preconditioning, including Calderón-based techniques, and integration with communication-level simulators. Bridging detailed full-wave descriptions with reduced-order representations is particularly important for extending device-scale accuracy to environment-scale propagation. Such capabilities can ultimately support physics-based digital twins and real-time optimization of reconfigurable propagation environments~\cite{Stutz2025Efficient}.

    \item \textbf{Refined end-to-end channel models:}
    Channel models should increasingly integrate propagation physics with circuit-theoretic descriptions of transceivers and \acp{ris}. Existing abstractions often separate the antenna array from the propagation channel, whereas physically consistent models should capture the interplay among feed networks, tunable elements, mutual coupling, free-space propagation, structural scattering, and specular reflections~\cite{Gradoni2021End,Shen2021Modeling,Wu2024Intelligent}. A key open question is how much of this physical detail must be retained to preserve accuracy while remaining tractable for communication-system optimization.

    \item \textbf{Modeling emerging physical \acp{dof}:}
    New forms of reconfigurability require corresponding channel abstractions. Examples include FAS models that capture continuous antenna-position variations~\cite{Wong2020Fluid,Wong2023FluidIII} and \ac{sim} models that accurately describe wave-domain transformations across multiple metasurface layers~\cite{An2024Stacked,An2023Stacked,Liu2025Stacked}. More generally, future models should expose new physical \acp{dof} to communication algorithms without discarding the constraints that govern their realizability.

\end{itemize}

\subsubsection{Enabling advanced signal processing and system optimization}

\begin{itemize}

    \item \textbf{Physics-aware and AI-driven channel estimation:}
    The dimensionality and physical structure of channels in reconfigurable systems challenge conventional estimation techniques. Future algorithms should exploit sparsity while explicitly incorporating \ac{em} constraints, such as wavenumber-domain support and MC models~\cite{Dkhan2025Nearfield,Rezvani2023Channel}. At the same time, wireless foundation models (WFMs) offer a path toward task-agnostic, multimodal representations that can generalize across environments and tasks with limited retraining and reduce pilot overhead. WavesFM demonstrates the potential of shared architectures across tasks including channel estimation and positioning~\cite{Aboulfotouh2025WavesFM}, while transformer-based large wireless models (LWMs) provide contextualized channel representations~\cite{Alikhani2025LWM}. Broader developments in multimodal foundation models for wireless prediction and control are surveyed in~\cite{Zhang2026MultiModal}. An important direction is therefore to embed physical constraints and priors into such learned representations so that scalability and generalization do not come at the expense of physical consistency.

    \item \textbf{Holistic multi-domain beamforming and resource allocation:}
    Tri-hybrid and related architectures introduce hierarchical optimization across digital, analog, and \ac{em} domains. New methods are needed for their joint design under practical constraints including power budgets, discrete phase states, tunable-load constraints, and latency. Promising directions include manifold optimization, hierarchical optimization across different timescales, and learning-based mappings from environmental observations to feasible configurations~\cite{Castellanos2025Embracing,Castellanos2023Energy,Yu2016Alternating}. For wideband systems, beam squint and frequency-dependent coupling must also be incorporated. More broadly, optimization should account for physical realizability so that configurations that are optimal under abstract channel models remain feasible when losses, coupling, dispersion, and hardware constraints are included.

   \item \textbf{Physically aware detection and decoding:}
    A key research direction is to extend physical consistency beyond channel modeling and beamforming to the complete receiver chain. Reconfigurable architectures can jointly shape the effective channel, noise statistics, and post-detection reliability, motivating detection and decoding strategies that exploit these dependencies. Subspace-based detection can reduce dimensionality and enable spatial parallelization~\cite{Sarieddeen2024Bridging,Jemaa2024Leveraging}. On the decoding side, conventional code-specific decoders can be adapted to exploit physically consistent soft or reliability information, while noise-centric paradigms such as \ac{grand} provide an alternative capable of directly exploiting structured noise and reliability information with highly parallelizable decoding~\cite{Sarieddeen2022GRAND}. More broadly, joint architecture--detection--decoding design should optimize physical configurations and resource allocation according to their impact on post-detection reliability and coded-link performance, rather than solely through channel-centric metrics such as received power or \ac{snr}.

\end{itemize}

\subsubsection{Bridging theory and practice: hardware, prototyping, and standardization}

\begin{itemize}

    \item \textbf{Closing the loop with hardware prototyping and testbeds:}
    A critical gap remains between physically consistent theoretical models and real hardware. Advanced prototypes and testbeds for reconfigurable architectures, including DMA, \ac{sim}, and FAS, are needed to validate both modeling assumptions and algorithmic performance under realistic operating conditions. Field trials demonstrating signal and capacity gains with \ac{ris} provide valuable evidence for directing theoretical research toward practically relevant regimes~\cite{Di2020Smart,Wu2024Intelligent}. Hardware-oriented studies and active-\ac{ris} calibration efforts further expose system-level considerations for mm-Wave \ac{ris} deployment, including wideband and three-dimensional implementations~\cite{Wang2026Millimeter}. Such measurements are essential for determining which physical effects must be retained and which can be safely abstracted.

    \item \textbf{Co-design of hardware and algorithms:}
    Once physical models are experimentally validated, the corresponding degrees of freedom should be co-designed with the algorithms that exploit them. This requires closer integration among antenna engineering, circuit design, electromagnetics, and communication theory. For example, tunable meta-atoms should be designed with the beamforming and reconfiguration algorithms they support in mind, while those algorithms must respect the attainable hardware states and associated losses. Information metasurfaces capable of computation directly in the wave domain exemplify this increasingly tight integration of hardware and signal processing~\cite{Cui2014Coding,An2024Stacked,Nerini2025Analogcomputing}.

    \item \textbf{Standardization and reproducible research:}
    Translation from research prototypes to deployable technologies requires common physical models, interfaces, benchmarks, and validation procedures. Existing tutorials have highlighted the path toward \ac{ris} specifications and associated standardization challenges~\cite{Di2020Smart}. Future efforts should define meaningful key performance indicators, standardized over-the-air testing and certification methodologies, reference geometries and measurement procedures, and open-source reproducible modeling and simulation frameworks. Of particular importance are standardized interfaces between \ac{em}, channel, link-, and system-level models, enabling physical consistency to be preserved across simulation layers. Such efforts can establish a common basis for evaluating reconfigurable technologies and facilitate their incorporation into future standards.

\end{itemize}

In conclusion, reconfigurability is progressively moving communication-system design beyond the digital domain and into the physical domain, where antennas, circuits, propagation, and signal processing become increasingly coupled. Realizing its potential therefore requires more than new architectures or more detailed channel models in isolation: it requires a physically consistent methodology that connects realizable hardware degrees of freedom to effective channel and noise models and, ultimately, to the algorithms that estimate, optimize, detect, and decode the transmitted information. Achieving this vision demands continued integration of electromagnetics, circuit theory, communications, signal processing, and experimental validation. Such architecture--physics--algorithm co-design provides a foundation for reconfigurable wireless systems whose physical environment is not merely modeled as part of the channel, but actively incorporated into the communication design.

\bibliographystyle{IEEEtran}
\bibliography{IEEEabrv,my_bibliography}

\vfill

\end{document}